\documentclass[a4,10pt]{article}
\usepackage{array}
\usepackage{xcolor,colortbl}
\usepackage{csquotes}
\usepackage{tabularx}
\usepackage{enumitem}
\usepackage{arxiv}
\usepackage[backend=bibtex]{biblatex}

\usepackage{placeins}
\usepackage{outlines}

\usepackage[utf8]{inputenc} 
\usepackage[T1]{fontenc}    
\usepackage{url}            
   
\usepackage{nicefrac}       
\usepackage{microtype}      
\usepackage{lipsum}
\usepackage{graphicx}
\graphicspath{ {./images/} }

\usepackage{booktabs}
\usepackage{amsmath,amssymb,amsfonts}
\usepackage{xcolor}
\definecolor{myorange}{RGB}{100, 50, 0}
\definecolor{myblub}{RGB}{34, 52, 168}
\definecolor{dg}{RGB}{64,64,64}

\usepackage[colorlinks = true,
		citecolor = blue,
		urlcolor= blue,
		linkcolor=dg,]{hyperref}
\usepackage{algorithmic}
\usepackage{textcomp}
\usepackage{xcolor}
\def\BibTeX{{\rm B\kern-.05em{\sc i\kern-.025em b}\kern-.08em
    T\kern-.1667em\lower.7ex\hbox{E}\kern-.125emX}}	

\title{A Trust-Centric Approach To Quantifying Maturity and Security in Internet Voting Protocols}

\author{
Stanis\l{}aw Bara\'{n}ski\\
\href{mailto:stanislaw.baranski@pg.edu.pl}{stanislaw.baranski@pg.edu.pl}\\
Faculty of Electronics\\
Telecommunications and Informatics \\
Gdansk University of Technology\\
Gdansk, Poland \\
\href{https://orcid.org/0000-0001-7181-8860}{0000-0001-7181-8860}\\
\And
Ben Biedermann\\
\href{mailto:bb@acurraent.com}{bb@acurraent.com}\\
Islands and Small States Institute \\
University of Malta\\
Msida, Malta \\
\href{https://orcid.org/0000-0003-1331-6517}{0000-0003-1331-6517}\\
\And
Joshua Ellul\\
\href{mailto:joshua.ellul@um.edu.mt}{joshua.ellul@um.edu.mt}\\
Centre for DLT \\
University of Malta\\
Msida, Malta \\
\href{https://orcid.org/0000-0002-4796-5665}{0000-0002-4796-5665}\\
  }

\begin{document}
\maketitle

\begin{abstract}
Voting is a cornerstone of collective participatory decision-making in contexts ranging from political elections to decentralized autonomous organizations (DAOs). Despite the proliferation of internet voting protocols promising enhanced accessibility and efficiency, their evaluation and comparison are complicated by a lack of standardized criteria and unified definitions of security and maturity. Furthermore, socio-technical requirements by decision makers are not structurally taken into consideration when comparing internet voting systems. This paper addresses this gap by introducing a trust-centric maturity scoring framework to quantify the security and maturity of seventeen internet voting systems. A comprehensive trust model analysis is conducted for selected internet voting protocols, examining their security properties, trust assumptions, technical complexity, and practical usability. In this paper we propose the Internet Voting Maturity Framework (IVMF) which supports nuanced assessment that reflects real-world deployment concerns and aids decision-makers in selecting appropriate systems tailored to their specific use-case requirements. The framework is general enough to be applied to other systems, where the aspects of decentralization, trust, and security are crucial, such as digital identity, Ethereum layer-two scaling solutions, and federated data infrastructures. Its objective is to provide an extendable toolkit for policy makers and technology experts alike that normalizes technical and non-technical requirements on a univariate scale.
\end{abstract}
 

\maketitle
\section{Introduction}\label{sec_1}
Voting is a fundamental mechanism for collective decision-making, used across a broad spectrum of contexts, from political elections and corporate boards to decentralized autonomous organizations (DAOs) and cryptocurrency governance. While an increasing number of local communities and governments start experimenting with internet voting protocols, for online communities internet voting is a crucial aspect of participatory governance. Thus, internet voting promises to increase the accessibility of governance processes and democratic institutions, ultimately contributing to their resilience in times of crisis and contexts of remoteness. In this regard, voting protocols that are more accessible, faster, and safer can mitigate societal risks highlighted by the COVID-19 pandemic and far-reaching distributed island communities of archipelagic states~\cite{hobbis_voter_2017}, where traditional or in-person voting posed significant risks.

Internet voting offers numerous advantages, including convenience, cost reduction, increased voter turnout, and claims enhanced transparency and verifiability, with the potential to catalyze the further development of modern democracy~\cite{barandiaranDecidimTechnopoliticalNetwork2024,williamsRemoteVotingAge2022,lichtIvoteNotIvote2021, buterinBlockchainVotingOverrated2021,plurality2023}. Its applications range from high-stakes environments like national elections and corporate governance to more routine uses such as social media polls and crowd-sourced data labeling. For example, \textit{CHVote} and \textit{sVote} are Switzerland voting systems used in multiple cantons for large-scale elections and referenda. 
Decidim is widely used for participatory democracy in cities, associations, universities, and NGOs, empowering users to engage in transparent and verifiable decision-making processes across hundreds of public and private organizations~\cite{DecidimUse}.
This wide range of applications is reflected in the diversity of internet voting systems. These systems can be broadly grouped into three categories based on their origin and primary use case. First are state-sponsored systems designed for legally-binding national elections, such as Switzerland's CHVote~\cite{haenniCHVoteProtocolSpecification2017}, sVote~\cite{larraiaSVoteControlComponents2022}, and the Estonian i-voting system (IVXV)\cite{GeneralFrameworkElectronic}, which are built to meet stringent security and operational requirements. Second are academically-rooted protocols like Helios\cite{adidaHeliosWebbasedOpenAudit2008} and Belenios~\cite{cortierBeleniosSimplePrivate2019}, which pioneered web-based end-to-end verifiability and provide strong, provable security guarantees. A third stream of innovation has emerged from the blockchain space, with protocols such as Votem~\cite{mattbeckerProofVoteEndtoend2018}, Agora~\cite{AgoraBringingOur}, Snapshot~\cite{WelcomeSnapshotDocs2024}, and MACI~\cite{ethereumfoundationMinimalAntiCollusionInfrastructure2022} tailored for the governance of online communities and Decentralized Autonomous Organizations (DAOs).

The introduction of blockchain technology has sparked a wave of innovation in internet voting protocols, as the addition of several new solutions to the above list highlights over the course of this paper. At their core, these solutions promise to improve security, transparency, verifiability, as well as simplify the whole architecture. However, as discussed in the preliminary interview with David Dueñas-Cid, internet voting researcher highlighted~\cite{interviewer1BlockchainTechnologyNot2024} ``while blockchain-based protocols might bring new perspectives to internet voting due to their decentralized, integral and transparent nature, they often simplify or overlook complex requirements that more traditional voting systems address, enlarging the attack surface of the system''. 
For this reason, the security of internet voting systems is not a one-dimensional property that can easily be assessed for any and each internet voting protocol. Conversely, different solutions achieve different levels of security and performance in various ways, depending on their design. Some systems rely on a single trusted authority~\cite{adidaHeliosWebbasedOpenAudit2008}, while others decentralize out as many components as possible~\cite{mattbeckerProofVoteEndtoend2018}. Similarly, some prioritize anonymity and privacy~\cite{williamsRemoteVotingAge2022,haenniCHVoteProtocolSpecification2017}, while others emphasize coercion resistance ~\cite{ethereumfoundationMinimalAntiCollusionInfrastructure2022} or verifiability and integrity~\cite{adidaHeliosWebbasedOpenAudit2008,cortierBeleniosSimplePrivate2019, IntroducingSnapshotsOnchain} instead.

The lack of a unified definition of what constitutes a ``\textit{good}'' internet voting system further complicates comparisons, as use case requirements may conflict with security requirements, which generally vary widely between systems~\cite{bernhardSoKComprehensiveAnalysis2015,cortierSoKVerifiabilityNotions2016,neumannHolisticFrameworkEvaluation2014,canettiUniversallyComposableSecurity2001}. This challenge is particularly pronounced when comparing blockchain-based protocols with non-blockchain systems, where the advantages of decentralization can contrast sharply with the reliance on centralized trust models in more conventional systems. In other words, the rapid development of new internet voting protocols has created a dynamic and complex landscape without a discrete set of properties that would allow for comparison. Internet voting systems may promise improved accessibility and efficiency, however, the diversity of these systems, coupled with the lack of standardized terminology and evaluation criteria, complicates their selection for use cases, which should be based on a fair comparison. This poses significant challenges for decision-makers, who ultimately carry the responsibility for their selection of a solution in this evolving field.

Existing studies do not address this gap, rather they focus on analyzing internet voting systems either purely regarding theoretical security properties, such as those outlined in systematization of knowledge (SoK) reviews~\cite{cortierSoKVerifiabilityNotions2016,bernhardSoKComprehensiveAnalysis2015,khoReviewCryptographicElectronic2022}, or on evaluating the claimed security properties of individual systems~\cite{neumannSecIVoQuantitativeSecurity2016,suwitoEvolutionBulletinBoard2021,haenniCHVoteProtocolSpecification2017}. In the latter case, this is often done through specific case studies~\cite{alonsoEvotingSystemEvaluation2018}. Hence, these studies rarely offer a comprehensive heuristic for effectively assessing and comparing internet voting protocols, particularly when it comes to technical functionality, system complexity, quantified security levels, and the overall maturity of the systems. This contribution aims at closing this gap by responding to the following research questions.

\begin{itemize}
    \item How can the concept of maturity be used to assess the relevance of internet voting protocols?
    \item What are the trust models underpinning these protocols, and how do they impact their practical application?
\end{itemize}

Thus, in the following, selected internet voting systems are analyzed by examining their trust models and scoring them according to maturity levels. This provides a more nuanced view of the trust assumptions underlying the voting protocols, because it abstracts technical complexity by providing a one-dimensional assessment criterion that allows use case-specific selection depending on the stakes involved. By expanding research on internet voting protocols beyond traditional security evaluations, this research provides a holistic framework that reflects real-world implementation concerns, with a specific emphasis on the priorities of decision-makers, such as policymakers, administrators, and system operators. In sum, the study makes four key contributions to the field, each of which is tied to the practical applicability of internet voting protocols:

\begin{enumerate} 
    \item Systematizing the knowledge on internet voting protocols, targeted at decision-makers. 
    \item Conducting a trust model analysis of the most popular internet voting protocols. 
    \item Quantifying the level of trust in security properties of selected voting protocols.
    \item Establishing a maturity score for quantifying the security of internet voting protocols.
\end{enumerate}

The maturity scoring framework not only fulfills the need for decision-making tools that can assist civil society representatives, including non-governmental organizations (NGOs), decentralized autonomous organizations (DAOs), and corporate board members, in selecting an appropriate electronic voting system for their specific use cases. Moreover, it addresses the lacking attention to the trust assumptions that underlie claims over the security of internet voting systems. As a result, this analysis can contribute to overcoming current adoption trends that primarily reflect popularity rather than rigorous security validation, by adding a useful and understandable framework for the selection of internet voting systems to the toolbox of decision-makers both in the private and public sector.

To answer the research questions and substantiate the claimed benefits of a maturity framework, the paper is structured as follows. Section 2 presents the state of the art, summarizing existing research on internet voting protocols. Section 3 outlines our analysis methodology, including the criteria for protocol selection and the evaluation framework. Section 4 describes the selected internet voting protocols. Section 5 discusses the results and implications for decision-makers. Section 6 presents a discussion of the findings. Finally, Section 7 presents the limitations of the research and offers directions for future work, followed by the presentation of conclusions in Section 8.

\section{The State-of-the-Art of Electronic Voting}\label{sec_2}
Drawing from the field of digital identity, when undertaking a state-of-the-art review, some scholars have resorted to sketching a timeline from the very first passport in 746 AD to the emergence of the European Digital Identity Wallet in 2023~\cite{freitag_new_2022,european_parliament_regulation_2024}. Such timelines, however, do not have much explanatory power for the current state-of-the-art of any system because modern digital systems do not have much in common with a passport from ancient times. Now, for internet voting it is still worthwhile acknowledging that postal voting can be traced back to the Roman Empire~\cite{staveleyGreekRomanVoting1972} and is still used in many elections around the world~\cite{krimmerBitsPaperComparing2005}. It highlights the relevance of and the need for remote voting protocols, but does not imply the necessity for internet voting nor its design. In fact, the following analysis will show that internet voting is not limited to remote voting practices, but can and is also used for in-person ballot casting. Consequently, this section focuses on explaining the developments that led to the state of internet voting today, the particularities that define it as such, and pave the way for future development~\cite{barry_understanding_2022}.

The way in which elections are conducted did not change much since the emergence of the nation-state, where elections necessarily are comparatively stable systems, given the high stakes involved. Meanwhile, the development of electronic voting systems only picked up after the invention of asymmetric cryptography and is by and large based on initial theoretical work undertaken between 1985 and 1995~\cite{benalohVerifiableSecretBallotElections1987,fujiokaPracticalSecretVoting1993,benalohReceiptfreeSecretballotElections1994}. More concretely, the first implementation of an internet voting protocol called \textit{Sensus}, was proposed at AT\&T Labs Research in 1997 ~\cite{cranorSensusSecurityconsciousElectronic1997}. It focused on four core properties: accuracy, invulnerability, privacy, and verifiability, and additional properties: convenience, flexibility, and mobility. While some of the terms are still used to describe internet voting protocols today, this study highlights a core problem of research on internet voting in general. Namely, terminology and definitions are given ambiguously or \textit{en passant}. For example, at a first glance \textit{invulnerability}, in the meaning of not exhibiting any vulnerabilities, implies a perfect system, which is not provably possible \cite{ellul2023good}. Even for \textit{Sensus} specifically, where the term refers to the inability of talliers maliciously casting ballots when all voters have cast or invalidated their vote, the honesty of auditors guaranteeing this property is assumed.

For this reason, it is not surprising that the technologies used for all parts of internet voting systems have become a focus of regulators and technological oversight. In the case of the European Union (EU), the Directive 1999/93/EC on Digital Signatures did not only aim at standardizing the usage of digital signatures across EU member states, but also seeded the development of a pan-European trust framework to reduce the reliance on assumed honesty of verifiers~\cite{european_commission_directive_1999}, i.e. auditors. In its ambition, the directive may not have fully fulfilled its objectives, but it set a precedent for organizational safeguards for cryptographic technologies. Against this background, in 2005, Estonia became the first country to hold a legally binding election using internet voting~\cite{maatenRemoteEvotingEstonian2004}. Since then, France has also experimented with allowing citizens abroad to vote via the Internet. Yet, most nation-states, including states within the EU, have not advanced internet and electronic voting beyond an experimentation stage, with Estonia and Switzerland being notable exceptions. This observation is underpinned by pilot results from the EU. When the European Union launched the CyberVote project as early as 2000~\cite{CybervoteProgrammeResearch2003}, which conducted various electronic voting experiments, such as the use of e-voting for nationwide student representative elections in Austria~\cite{harrisonStudyEVotingPractices2023}, the project was discontinued following initial trials. Cryptography experts had raised concerns about the system's security, specifically its integrity, and a significant number of voters encountered technical difficulties during the identification process.

According to~\cite{lichtIvoteNotIvote2021}, the decisions to abandon internet voting (i-voting) projects are driven by fear, reflecting a broader discourse around i-voting that often hinges on arguments related to trust and distrust. These arguments are not necessarily tied to the technology itself but instead are influenced by broader sociopolitical factors and individual perceptions~\cite{duenas-cidTrustDistrustEDemocracy2022}. Dueñas-Cid's interviews with stakeholders, including technology providers, election officials, journalists, politicians, and activists, underscore the chasm for overcoming barriers to adoption~\cite{electrustAnxietiesInternetVoting2021}. In this regard, it is worthwhile addressing the naming conventions that seem to increase the complexity for stakeholders and policymakers to understand internet voting protocols. That is, internet voting and electronic voting are not interchangeable. While electronic voting (e-voting) does not necessarily require the \textit{Internet}, internet voting (i-voting) protocols are by definition electronic. Moreover, both categories of voting protocols may enable remote voting but are not unequivocally doing so. Leaving this distinction implicit increases the perceived complexity of those looking to implement and use either i-voting or e-voting protocols.

Moreover, the distinction between i-voting and e-voting signifies an emergent divergence of voting applications in the state-of-the-art, which is explicitly addressed in the analysis of this contribution. Voting applications appear to either exclusively focus on non-state governance or address high-stakes elections within a nation-state context. In the former case, internet voting increasingly makes use of blockchain technology, also addressing governance needs of DAOs and Web3 organizations~\cite{tan_open_2023}. Meanwhile, electronic voting can encompass both direct-recording electronic voting machines (DREs) and internet voting protocols that predominantly integrate with national digital identity solutions~\cite{tammpuu_estonian_2022}. The most notable implementation of an internet voting protocol, created in 2008 and used as a reference implementation in academia, is Helios~\cite{adidaHeliosWebbasedOpenAudit2008}. Its creation also demarcates the diversification of protocols into use cases, technologies, and definitions. The appeal of Helios is that its architecture is quite simple and focuses on end-to-end verifiability, whereas it arguably was the last i-voting protocol to be created before a trend towards more complex i-voting architectures emerged.

With the advent of blockchain technology, exemplified by the launch of the Bitcoin network~\cite{nakamotoBitcoinPeerpeerElectronic2008} and its generalization through Ethereum~\cite{buterinEthereumWhitePaper2013}, many found parallels between the problem of double-spending in cryptocurrencies and double-voting in internet voting protocols. This comparison highlighted the central role of a component common to most modern voting schemes: the \textit{bulletin board}, an immutable, publicly accessible, and append-only register. For decades, cryptographic research has advanced along two parallel lines: designing secure protocols that assume the existence of a bulletin board to achieve verifiable and secure voting~\cite{juelsCoercionresistantElectronicElections2010,achenbachImprovedCoercionResistantElectronic2015,suwitoEvolutionBulletinBoard2021}, and developing robust implementations of the bulletin board itself to provide a trustworthy foundation for these protocols~\cite{culnanePeeredBulletinBoard2014,kiayiasSecurityPropertiesEVoting2018}.

A public blockchain can thus be understood as a modern, globally-verifiable answer to this implementation challenge. This architectural choice, however, creates a critical tension: how to reconcile the public nature of the board with the strict privacy requirements of voting. The solution lies not in abandoning transparency, but in using it to verify cryptographically protected data, such as encrypted ballots. By publishing transformed data, protocols can ensure that the process integrity is verifiable while the vote's content and its link to a voter's identity remain secret. Nevertheless, this perceived tension remains a significant hurdle for public adoption in high-stakes environments, as highlighted by legal and societal debates in jurisdictions like Germany~\cite{fitzpatrick_high_2022}. Thus, the choice to use a public blockchain versus a more traditional bulletin board implementation has become a key architectural decision, deepening the divergence in trust models and use cases across the landscape of internet voting systems.

Nonetheless, in principle traditional systems and novel blockchain-based systems share the challenge of either overcoming trust assumptions for central servers, which tally cast ballots, or their reliance on public-permissionless systems that do not have the same trust assumptions. Thus, a continuous scale for evaluating the use case-specific trade-offs of digital voting systems, i.e. both i-voting and e-voting systems, appeared as a need from the state-of-the-art review. More concretely, suggestions by system architects, such as Vitalik Buterin, to distribute tallier components over many parties in ``the most secure way [that is] to implement a bulletin board [on] an existing blockchain''~\cite{buterinBlockchainVotingOverrated2021, suwitoEvolutionBulletinBoard2021} are yet to be consistently contextualized with stakeholder requirements.

The need for an overarching theoretical framework, evaluating voting systems from design to use case requirements, is becoming increasingly apparent as many blockchain-based internet voting protocols continue to emerge and their development is accelerating. On one hand, the use of public-permissionless blockchains offers the potential to shift trust assumptions away from a single centralized authority towards a decentralized network of participants~\cite{ethereumfoundationMinimalAntiCollusionInfrastructure2022,mccorryOpenVoteNetwork2023}. This shift, however, introduces its own set of complex trust assumptions tied to crypto-economic incentives, where the security of the network relies on the rational, and often purely financial, behavior of its participants. On the other hand, in some cases, this abstraction has stretched too far. This caused oversimplification and unfulfilled promises of security properties and attracted sustained criticism from the academic e-voting community~\cite{parkGoingBadWorse2021,jeffersonMythSecureBlockchain2019,specterBallotBustedBlockchain2020, leeBlockchainbasedElectionsWould2018, shanklandNoBlockchainIsn2018,mearianWhyBlockchainbasedVoting2019}.

At the same time, thousands of DAOs are using blockchain-based voting protocols in practice~\cite{CaseStudiesMACI,WelcomeSnapshotDocs2024}. Through sustained usage, these protocols are becoming better and more secure~\cite{IntroducingSnapshotsOnchain,FreedomtoolOrg}. While nation-states and offline communities still hesitate and contemplate theoretical security properties, Web3 communities connected through blockchain are putting i-voting to the test in real-world environments. Thus, i-voting is contributing already to sustaining the governance of value, organizations, and people. Although less mature, meaning with fewer contingency measures and less rigorous security considerations, blockchain-based systems offer and make use of the property of decentralization. Blockchain-based systems offer and use the property of decentralization to remove the strong assumption of a trusted third party, which is often a bottleneck in system security. Their integrity and transparency bring new perspectives that require further research to meet the requirements of traditional voting systems~\cite{interviewer1BlockchainTechnologyNot2024}. Particularly, blockchain-based solutions are often dedicated to Web3 users only, requiring cryptocurrency wallets and public-private key authentication. This barrier to entry may simplify the protocol, and provide better security properties, but excludes the majority of the population and bars the way to their broad and large-scale application.

For this reason, the next section outlines the analysis undertaken herein for establishing such a theoretical framework. It draws from the concept of maturity and connects it to the description of voting protocols and security properties. While the focus is on creating the means for evaluating and comparing the characteristics of different i-voting protocols, the need for a holistic understanding of e-voting at the crossroads with internet voting is fulfilled. 

\section{On the Analysis of Internet Voting Protocols}\label{sec_3}
This section establishes the theoretical and practical foundation for analysing i-voting protocols as subcategory of e-voting. The relationship between the categories of e- and i-voting is made explicit through the use of maturity as unifying scale for the evaluation of i-voting systems as part of a larger system of digital voting applications. To that end, drawing from the previous section, first, a brief description of voting protocols in general is given. Second, the meaning and relevance of security properties is explained for the analysis of i-voting systems. Finally, the third subsection introduces the concept of maturity and provides the reader with known implications of using a maturity scoring framework for the analysis of i-voting protocols.

\subsection{Voting Protocol}\label{sec:voting-protocol}
Internet voting protocols vary in their structuring of the voting process. This diversification often divides voting into a set number of stages. Meanwhile, different frameworks and studies propose various models for these stages. For example, Bernhard et al.~\cite{bernhardSoKComprehensiveAnalysis2015} introduce a single-pass model that executes in three phases: (1) setup, (2) voting, and (3) results. Another approach, as described in the General Framework for Electronic Voting~\cite{GeneralFrameworkElectronic}, is to divide the process into four stages: (1) pre-voting, (2) voting , (3) processing, and (4) counting. Similarly, the CHVote protocol~\cite{haenniCHVoteProtocolSpecification2017} employs three general phases with several sub-steps:

\begin{enumerate}
    \item \textbf{Pre-Election (Setup)}: Includes sub-steps such as Key Generation, Election Preparation, and Printing of Voting Cards. 
    \item \textbf{Election (Voting)}: Involves Candidate Selection, Vote Casting, Vote Confirmation, the Benaloh Challenge, and the Generation of Zero-knowledge Proofs. 
    \item \textbf{Post-Election (Counting/Processing/Tally)}: Comprises Mixing (Shuffling), Decryption, Tallying, and Inspection. 
\end{enumerate}

According to these stages, most internet voting protocols must be multi-party systems, involving several distinct roles. For instance, the Helios protocol~\cite{cortierBeleniosSimplePrivate2019} defines roles for the Administrator, Helios Server, Key-holders, and Auditors. Scytl’s protocol specifies a broader array of components, including the Voting Server, Election Administrators, Print Office, Control Components (CCR and CCM), Board Members (Key-Holders), and Auditors. Meanwhile, blockchain-based systems emphasize decentralization. Therefore, the use of blockchain for i-voting protocols introduce even more components that their non-blockchain reliant counterparts. For example, Agora’s architecture includes roles such as the Election Authority, the Private Bulletin Board Blockchain, the Cothority and Cotena networks, Valeda Auditors, and the Bitcoin Blockchain.

Due to this divergence, the structure of each system is analysed in terms of its physically or logically separated components with distinct trust assumptions. It is anticipated that a relationship between the number of components and the distribution of trust models across i-voting systems' security properties can be established. Additionally, the differences between blockchain-based protocols and non-blockchain protocols are analysed in regards of their complexity and the decentralization of their components. In this light, preliminary results and pre-studies revealed that most systems’ roles can be distilled into six key functions:

\begin{enumerate} 
    \item \textbf{Administrator}: Also known as the Election Authority or Organizer. They are responsible for election preparation, including the preparation of eligible voters' lists, candidate lists, and key generation. 
    \item \textbf{Identity Provider}: Also referred to as the Credential Authority, Registration Service, Census Service, or Print Office. This role is responsible for voter authentication and/or authorization, as well as the generation and distribution of authentication codes. 
    \item \textbf{Collector}: Often called the Ballot-box or Bulletin Board, stores ballots securely during the voting process. 
    \item \textbf{Key-Holders}: Also referred to as Trustees or Board Members. They hold shares of decryption keys and perform partial decryption. 
    \item \textbf{Processor}: Performs tasks such as ballot authorization, double-vote prevention, shuffling, mixing, decryption, and tallying. 
    \item \textbf{Auditors}: These entities are granted additional permissions to inspect and validate the correctness of each component's operations and the integrity of the data flow between them throughout all stages of the voting process. 
\end{enumerate}

\subsection{Security Properties}\label{sec:security-properties}
One of the challenges of evaluating internet voting systems is the lack of a standardized set of security properties with universally accepted definitions. Different researchers and practitioners often define and decompose these properties in various ways, leading to inconsistencies in how systems are evaluated.

For example, the property of verifiability is frequently divided into sub-properties such as individual verifiability, universal verifiability, and end-to-end verifiability, with further distinctions made between strong and weak verifiability~\cite{cortierSoKVerifiabilityNotions2016}. Similarly, eligibility verifiability is treated as a distinct aspect only in some frameworks~\cite{cortierBeleniosSimplePrivate2019}. Bribery resistance is another property that is often divided into subcategories, denoting the absence of cast receipts and coercion-resistance, with varying definitions and criteria~\cite{chaidosBeleniosRFNoninteractiveReceiptFree2016}. Some authors consider coercion-resistance to be achieved when voters can overwrite their votes, while others require that there is no discernible way to determine whether the overwrite was successful or not~\cite{clarksonCivitasSecureVoting2008, ethereumfoundationMinimalAntiCollusionInfrastructure2022}. Even privacy and consistency is challenging to define because there are different approaches to prove them~\cite{bernhardSoKComprehensiveAnalysis2015}.   Additionally, terms like anonymity, privacy, and confidentiality are sometimes used interchangeably, further complicating comparisons. Moreover, privacy is defined in spite of its value-rational character and fundamentally ambiguous interpretation. This is highlighted when considering that even the concealment of individual votes does not protect individual votes from being exposed in the case that all voters unanimously vote for the same candidate.

Beyond these basic properties, the literature references numerous other security properties such as everlasting secrecy, auditability, integrity, confidentiality, and censorship-resistance. While all of these categories are important for evaluating the aptitude of a given i-voting protocol, these systems tend to not score well over all when these security measures are not systematically and generally met. Given that it must be assumed that existing analyses may, implicitly or explicitly, favour one or the other solution because of the absence of a single coherent framework for their analysis, a core set of security properties is suggested. These properties were constructed for striking a balance between explaining individual i-voting protocols in-depth and providing a generalizable framework for comparing different i-voting systems with each other. Thus, the condensed list of security properties below is used for a consistent analysis of all i-voting systems irrespective of their specific software implementation design.

\begin{itemize} 
    \item \textbf{Voter Anonymity}: This property addresses the question, ``Who must collude to establish a link between the recorded ballot and the real identity of the voter?'' It concerns the protection of voter identities from exposure. This property ensures that a voter's identity remains anonymous by guaranteeing both the secrecy of their participation in the election and the unlinkability of the voter from their cast ballot~\cite{khoReviewCryptographicElectronic2022, neumannHolisticFrameworkEvaluation2014}.

    \item \textbf{Voting Secrecy}: This property addresses the question, ``Who must collude to decrypt a single vote or all votes before the voting process is complete?'' It ensures that the content of votes remains confidential until the official tallying phase. Also known as \textit{ballot secrecy} or \textit{confidentiality}, this property guarantees that the election system reveals no partial information about the tally before the end of the election, typically achieved through threshold cryptography~\cite{kiayiasSelftallyingElectionsPerfect2002, cortierFeaturesUsageBelenios2022, khoReviewCryptographicElectronic2022} or time-lock puzzles~\cite{glaeserCicadaFrameworkPrivate2023,malavoltaHomomorphicTimeLockPuzzles2019}.

    \item \textbf{Individual Verifiability}: This property addresses the question, ``Who must collude to falsely convince a voter that their ballot has been recorded correctly and included in the final tally?'' It ensures that voters can verify the integrity of their individual votes. As a cornerstone of end-to-end verifiable (E2E-V) systems, this property allows a voter to confirm that their vote was correctly \textit{cast-as-intended}, \textit{recorded-as-cast}, and ultimately \textit{tallied-as-recorded}~\cite{neumannHolisticFrameworkEvaluation2014, williamsRemoteVotingAge2022, suwitoEvolutionBulletinBoard2021}.

    \item \textbf{Universal Verifiability}: This property addresses the question, ``Who must collude to falsely convince observers that the voting procedure was correct and that the final tally accurately represents the collected votes?'' It ensures the process is transparent and trustworthy to the public. This property allows any third-party observer to confirm that the published election outcome is a correct computation of the ballots as recorded on the public bulletin board, ensuring the overall integrity of the tally~\cite{cortierBeleniosSimplePrivate2019, kiayiasSelftallyingElectionsPerfect2002, bokslagEvaluatingEvotingTheory2016}.

    \item \textbf{Eligibility Verifiability}: This property addresses the question, ``Who must collude to falsely convince observers that only eligible voters cast ballots?'' It ensures that only authorized voters participate in the election. This property guarantees two distinct conditions: that every ballot included in the final tally corresponds to a unique, eligible voter (\textit{unicity}), and that eligible voters cannot cast more than one vote (\textit{unreusability}), thereby preventing abuses such as double-voting or "ballot-box stuffing"~\cite{williamsRemoteVotingAge2022, cortierBeleniosSimplePrivate2019, khoReviewCryptographicElectronic2022}.

    \item \textbf{Coercion Resistance}: This property addresses the question, ``Who must collude to give voters the option to convince a briber how they voted?'' It protects voters from being forced to vote against their will. This property is typically realized through \textit{receipt-freeness}, which prevents a voter from obtaining proof of how they voted that could be shown to a coercer~\cite{chaidosBeleniosRFNoninteractiveReceiptFree2016, kiayiasSelftallyingElectionsPerfect2002}. Stronger forms of coercion-resistance allow voters to cast their true vote even under duress, for example, by providing a mechanism to revote or cast a fake vote that is later invalidated~\cite{chaidosBeleniosRFNoninteractiveReceiptFree2016, bokslagEvaluatingEvotingTheory2016}.
\end{itemize}

\subsection{Internet Voting Maturity Framework (IVMF)}\label{sec:EVMI}
Extensive research has been done on token-based governance models~\cite{beck_governance_2018,messias_understanding_2024}, but the evaluation of electronic voting schemes with general relevance to online communities and decentralized autonomous organizations (DAOs) has fallen short of offering practical insight beyond proposing token-based voting. As token-based voting uses blockchain technology, ``every vote could be linked to the identity of the voter, making it difficult or impossible to guarantee anonymous voting''~\cite{beck_governance_2018}. This is particularly problematic as DAOs are structured around blockchain-based systems, where the absence of advanced privacy-preserving electronic voting mechanisms negatively impacts the maturity of their communication~\cite{johansson_roadmap_2019} and governance. Insofar as voting is structured communication of intent and as information systems (IS) research emphasizes “the value of communication” as a measure for maturity~\cite{johansson_roadmap_2019}, DAOs remain unable to mature their governance as long as their voting systems use blockchain technology exclusively~\cite{tan_open_2023}. Thus, the i-voting maturity framework (IVMF) aims at making the maturity of i-voting practices visible and comparable to help mitigate principal-agent dilemmas DAOs face when using immature token-based e-voting schemes that lead to ineffective governance~\cite{beck_governance_2018}.

To this end, the IVMF aggregates multivariate inputs from a qualitative systematization of knowledge into a single maturity metric that adheres to the Council of Europe's recommendations for e-voting~\cite{ministers_deputies_recommendation_2017}. That is universal, free, equal, and secret suffrage. For grounding the IVMF in the literature, the index takes inspiration from the World Bank Group (WBG) Government Technology Maturity Index (GMTI), which assesses digital systems for nation-state governance more broadly~\cite{dener_govtech_2021}. Like the GMTI, the IVMF is proposed to cluster different aspects of the evaluated systems into categories. These categories are then used to provide a topic-specific indication of the systems' performance, while the overall maturity is expressed by the aggregate score of a given system under evaluation. Accordingly, the framework's composite is calculated by multiplying individual indicators with their weight and summing the products to category scores. Furthermore, categories are weighted before being aggregated to the composite maturity score. Thus, the IVMF follows a mixed-method research methodology that is evaluative, but expresses e-voting systems' maturity quantitatively~\cite{creswell_designing_2017}. Hence, the e-voting maturity score follows a \textit{higher-the-better} rating mechanism.

Maturity models have been applied in settings that are as sensitive as e-voting, health being just one example where private and public interests are at play. In this regard, e-voting systems are positioned at the intersection of the public and private sectors. Thus, e-voting is a prime example for advancing maturity models by drawing from previous research in the field of health sciences, where the maturity of health innovation centers~\cite{knosp_research_2018,van_ede_assembling_2024} was the focus. More generally, IS have repeatedly adopted methods from the health sciences, particularly related to technological innovations in hospitals and medical research~\cite{barry_state---art_2022,barry_understanding_2022,kitchenham_guidelines_2007}. This justifies the approach taken by grounding the IVMF in mixed-method interdisciplinary research on the intersections of the public and private sectors, as well as studies on digital innovation for public infrastructures.

\subsubsection{Framework Design}\label{sec_3.3.1}
The appraisal scheme requires evaluators to score programs in the categories and subcategories on a scale from one to ten, where one represents low maturity, five intermediate maturity, and ten high maturity. Note that this scheme was improved throughout the research process and is explained in depth in Section~\ref{sec:trust-model}. Furthermore, evaluators comment with a justification for each score given. To mitigate evaluators' errors, the evaluative part of the IVMF is triangulated~\cite{creswell_designing_2017} through comparing and discussing evaluation results in the research group and substantiating evaluations through the method of document analysis~\cite{bowen_document_2009}. This is performed on the technical documentation of the e-voting systems. For positioning both parts of the IVMF on a shared theoretical foundation, the appraisal framework and the positivist indicators, set out in the following sections, are clustered in three rubrics, i.e., categories. These rubrics are \textit{Practical Usability}, \textit{Trust Model}, and \textit{Complexity Score}, which are discussed in detail in Section~\ref{sec_4}.

\subsubsection{The Concept of Maturity}\label{sec_3.3.2}
Research on information systems (IS) and information technology (IT) generally describes maturity as a staged process, where systems or organizations have to pass through one phase to enter into the next phase~\cite{knosp_research_2018}. Thus, maturity can be seen as a property where structures and processes are improved in stages until they reach a maturity threshold. For evaluating the outputs of outcome-oriented and subsequent iterations in the information communications technology (ICT) sector, maturity models are stated to be “effective tools [for] organizational development [...] prioritization and process improvement”~\cite{knosp_research_2018}. The fact that the WBG approaches the scoring of nation-states according to their government technology (GovTech) readiness through the construction of a quantitative maturity model~\cite{dener_govtech_2021} supports that view.

Now, both e-government solutions procured and deployed by nation-states, which are included in the GTMI, as well as e-voting applications scored by the IVMF are digital technologies. As digital tools, they are outcome-oriented in the sense that they are applied to achieve the best results, while operating under bureaucratic constraints and are limited by their cost. Thus, both the GMTI and the IVMF measure how well specific information systems are adapted to their use case~\cite{nolan_managing_1973}. For e-voting applications in the context of online communities, however, their transformational potential that reconfigures established societal practices, such as research and innovation, into open and participatory processes~\cite{ding_desci_2022} must be taken into account.

Inevitably, the introduction of new technologies for governing social interaction reshapes communities and societies according to the new means of governance. While the principles of openness and decentralization, as paradigms, have resulted in an optimistic vision for the use of these technological means for governance named decentralized science (DeSci), positive outcomes cannot be taken for granted. Thus, increased maturity does not translate into value-rationally better or worse outcomes; it only scores the aptitude of a system to fulfill its stated goal. This is evident, considering that emergent practices, such as DeSci, incorporate key features of established norms and structures, which are not necessarily perceived as desirable. Meanwhile, the DeSci movement still substitutes established governance models through novel organizational structures and decision-making processes, irrespective of underlying epistemological conflicts and paradoxes.

\subsection{Trust Model}\label{sec:trust-model}
To analyze, measure, and compare security properties of complex systems is a challenge. Therefore trust models are used to unify the level of abstraction and normalize the trust assumptions~\cite{hoffmanTrustSecurityExpanded2006,choSurveyTrustModeling2015,shaikhTrustModelMeasuring2015}. Considering that a system is secure as long as some assumptions hold, the security assumptions constitute the ``trust model'' of the system. Trust models captures the combination of assumptions, or weakest link in the system security.

Thus, a system is considered safer when its expected behavior relies on fewer trust assumptions. For example, a system that is safe as long as one central authority behaves correctly is a strong assumption. A system that is safe as long as at least one actor in an independent consortium behaves correctly is a weaker assumption, thus the system is perceived as safer. Systems are considered even saver when they don’t require assumptions about human honesty. Trust reductions are possible through methods like cryptographic proofs and decentralized verification, where system security relies on mathematical guarantees rather than individual actors. This minimizes potential vulnerabilities, making the system more robust and reducing dependency on trusted participants~\cite{ben-sassonAuroraTransparentSuccinct2018}.

For social technologies~\cite{leibetseder_critical_2011}, such as e-voting systems, however, the human factor must always be considered because of their inherently social embeddedness. As trust minimization from a cryptographic perspective remains a desirable quality of any technical system, this paper aims to quantify the balance between different trust models and societal applications that allow for the comparison of systems with different degrees of trust in varying contexts.

In this work, we manually create the trust model for each protocol by analyzing the steps required for one or more actors to compromise the system's security guarantees. We adapt the model proposed by Vitalik Buterin~\cite{buterinTrustModels2020} to evaluate the type of network an e-voting system uses, rather than merely measuring the network's size, as this may vary depending on implementation. In other words, the focus here is on distinguishing between private networks with closed membership and public networks with open membership. The first ordering, which would later be improved but was an important step toward a comprehensive heuristic of classifying the influence of blockchain network type on the trust assumptions for e-voting systems, is defined as follows:
\begin{itemize}
    \item \textbf{N/A:} The protocol does not achieve the property.
    
    \item \textbf{Single Authority (1/1):} The property holds as long as a single authority is honest. For example: Voting Server, Helios Server, Collector, Election Authority, Credential authority.
    
    \item \textbf{One of Closed Set (1/n):} The property holds as long as at least one of participants in a closed set is honest. For example: Talliers, Key-holders, Mix-nodes, Control Components.
    
    \item \textbf{Majority of Closed Network (2/n):} The property holds as long as a majority participants in a closed set are honest. For example: PBFT-based distributed networks, private blockchains, Oracles.
    
    \item \textbf{Majority of Open Network (2/N):} The property holds as long as a majority of participants in an open network are honest. For example: Public blockchains, Public storage networks, IPFS\footnote{https://ipfs.tech/}, DRAND\footnote{https://drand.love/}, Shutter Network\footnote{https://www.shutter.network/}.
    
    \item \textbf{Auditors (1/N):} The property holds as long as at least one party among the public auditors is honest. For example: Auditors, Valeda Auditors, Public Committees.
    
    \item \textbf{Trustless (0/N):} The property is achieved without any assumption of honest parties.
\end{itemize}

Most of the analyzed e-voting systems are complex systems consisting of multiple components, each distributed in its own way, with differing trust models. When applied to the dataset, the above classification proved insufficient in capturing nuanced differences between systems of similar design. Therefore, the classification was extended to reflect complex systems with multiple distributed components that were encountered in the e-voting system selection. The final trust model quantification scheme is outlined in Section \ref{sec:tm_quantification}.

Instead of classifying a whole system by one trust model, the trust model scoring was applied to each of the security properties of an e-voting system, listed in Section~\ref{sec:security-properties}. This provided a higher resolution and a more nuanced view of the similarities and differences of the respective trust models, especially with regard to their maturity. In the following section, the methodology used for aggregating and ranking the different e-voting systems according to, but not limited to, their trust models is explained in detail.

\section{Methodology}\label{sec_4}
After the preceding sections provided a general introduction to the apparent divergence of e-voting systems and presented the theoretical approach, this section outlines the methodology that is used for establishing an assessment framework for i-voting systems specifically. In a mixed-method approach, this contribution foresees responding to the two research questions by applying a phased approach that consists of qualitative protocol and attribute selection, evaluative maturity scoring through multiple researchers, and quantitative data aggregation. Thus, the research falls within the paradigm of sequential mixed-method studies and aligns with the principles of triangulation~\cite{creswell_designing_2011,datta_paradigm_2006}. The methodology section proceeds as follows. First, the selection of protocols and attributes of relevance for this study is outlined. The latter includes an introduction to composite scores, such as the~\textit{complexity quantification} indicator. Finally, the e-voting maturity framework is defined in quantitative terms and its calculation explained.

\subsection{Protocol Selection}\label{sec:protocol-selection}
In this subsection, the methodology is detailed, which was used to select the i-voting protocols included in the dataset. Given the labor-intensive nature of thorough and complete analysis of each and protocols, coupled with the limited number of real-world deployments, a pragmatic approach to protocol selection was chosen that draws from the methodology for state-of-the-art reviews rather than following a conventional Systematic Literature Review (SLR)~\cite{kitchenham_guidelines_2007}. Protocols were included when they are either widely recognized, gained significant attention in the academic and technical communities, or represent the latest developments in the field of i-voting. This focus allowed to include contemporary systems that are not only technically sophisticated but also influential to current trends in the industry and have the potential to shape future directions.

To ensure thorough analysis was undertaken, only protocols were included that had published detailed whitepapers or technical documentation. This criterion was essential as it allowed attention to detail when analyzing the specifics of each protocol's architecture, trust model, and security properties. Protocols that only provided high-level marketing materials or lacked substantial technical detail were excluded from the dataset. Instead, protocols were included when they provided a documented and functional prototype or a working implementation. This criterion was crucial in distinguishing between theoretical designs and practical, deployable systems. Protocols that were still in the conceptual or design phase without any tangible implementation were not considered, as this study is focused on systems with demonstrated functionality.

Applying these criteria led to the exclusion of several notable projects for two primary reasons. First, we targeted only fully functional, end-to-end i-voting platforms. This focus meant excluding valuable but incomplete frameworks, such as Microsoft's~\textit{ElectionGuard}, which is a software development kit (SDK) designed to be integrated into other systems, and~\textit{\href{FreedomTool.org}{FreedomTool.org}}, which primarily serves as an identity provider rather than a complete voting solution. Second, a lack of detailed public technical documentation precluded the rigorous and reproducible analysis our framework requires. For instance, commercial platforms like~\textit{\href{https://electionrunner.com}{electionrunner.com}} and projects such as \textit{Follow My Vote} were not included as their available documentation did not provide the necessary in-depth architectural and security specifications. Therefore, our final selection prioritizes systems that manage the entire voting process—from voter authentication to tallying and verification—and are sufficiently transparent to allow for a practical, in-depth analysis of their trust models and security properties.

\subsection{Attribute Selection}\label{sec:attribute-selection}
This section outlines the methodology for assessing and quantifying the complexity and trust models of the i-voting systems in the dataset. The introduction of a complexity attribute is a noteworthy contribution of this analysis to the academic discourse on i-voting protocols. It describes the number of different set of components and trust models required to define the conditions under which a system may lose its security guarantees. To evaluate the complexity and trust models, available documentation for each voting system was reviewed. This included both explicit descriptions of the trust model, if provided, and an in-depth examination of the system's architecture and data flow to infer the trust model where it was not explicitly stated.

Firstly, the system documentations were reviewed, targeting sections that directly addressed the trust model. In cases where the trust model was explicitly discussed, this information was directly extracted. For example, CHVote provides Protocol Specification~\cite{haenniCHVoteProtocolSpecification2017} with Section ``\textit{6.2. Adversary Model and Trust Assumptions}'', where the trust assumptions and the conditions under which the system maintains its security properties are clearly stated. In instances where the trust model was not explicitly detailed, an architectural analysis of the system was undertaken. This involved examining the flow of information between different components, identifying potential points of vulnerability, and determining who would need to collude or behave maliciously to compromise the system. It was examined how data is processed, transmitted, and validated across the system, focusing on identifying the parties or components whose integrity is crucial for maintaining the system's security.

Using the trust model framework outlined in Section~\ref{sec:trust-model}, each system was categorized based on the trust assumptions required to maintain its security properties. Each security property of the system was individually processed for granular view of the trust model. This classification ranged from systems that require trust in a single party to systems that rely on majority consensus within open or closed networks, and ultimately to systems that are secure based on information-theoretic assumptions without relying on any parties' honesty. It is important to note that this methodology did not involve validating the soundness of the security proofs provided by the systems. Instead, the focus was on analyzing the documented architecture and identifying the trust dependencies as is. Mathematical proofs or the technical correctness of the security claims were not independently verified by the researchers; rather, the information provided by system was relied upon to assess the trust model. A critical part of this approach was to determine the conditions under which a security property could be compromised. It was examined who needs to collude or act dishonestly to break a given security property. This analysis supported accurately classify systems that may have multiple trust dependencies and assign them a more nuanced trust model category. By following this process, trust models of each voting system were systematized in the dataset, providing a clear and comparative understanding of the trust assumptions underlying their security properties.

\subsubsection{Complexity quantification (CMPX)}\label{sec:system-complexity}
The complexity of an i-voting protocol is quantified from the perspective of the election organizer or system deployer. The metric (CMPX) measures the operational and architectural burden required to set up, manage, and secure the system's components. It is intended as a static sub-indicator, where any increase in complexity is expressed through a higher score, providing a baseline expression for system complexity, which is put into context through the aggregation of CMPX(p) with the other sub-indicators to IVMF(p). In sum, a higher score indicates a greater number of distinct, actively managed parts, which can lead to increased costs, coordination overhead, and potential for misconfiguration; lower scores represent less operational burden for the deployer. The sub-complexity of each component is quantified as follows:
\begin{itemize}
    \item \textbf{1: Single dedicated component}: A self-contained unit managed by the organizer. The simplest case.
    \item \textbf{2: Public network}: While technologically complex, using a public network (e.g., Ethereum, IPFS) outsources the infrastructure's maintenance, security, and uptime. The deployer's role is that of a user or developer on the platform, which is operationally simpler than running the network itself.
    \item \textbf{3: A few independent parties}: Requires coordinating with external, human-centric entities like auditors or observers. This introduces logistical and communication overhead.
    \item \textbf{4: Multi-party computation}: Requires the active management and coordination of a closed set of parties to perform sensitive cryptographic operations (e.g., distributed key generation and decryption). This is operationally complex and requires high availability from all participants.
    \item \textbf{5: Dedicated network}: Represents the highest operational burden, as the deployer must build, maintain, secure, and incentivize participation in their own private distributed network (e.g., a private blockchain).
\end{itemize}

The aim of this complexity quantification metric is to provide a simple way to evaluate a protocol's overall architectural complexity. For instance, consider the Estonian e-voting system, which consists of a single Collector, a single Processor, a single Registration Service, a few Talliers (performing MPC), and a few Auditors. Using the CMPX mapping, the complexity values are assigned as follows: 1 (Collector) + 1 (Processor) + 1 (Registration Service) + 4 (Talliers) + 3 (Auditors), resulting in a total CMPX score of 10. This value reflects the system's overall operational complexity and allows for a standardized comparison against other systems with different component configurations.

\subsubsection{Practical usage (PU)}\label{sec:practical_usage}
Even if a protocol claims to have optimal security and performance properties, these attributes are of limited value when the protocol has not been validated through real-world usage. To address this, the Practical Usability (PU) score was introduced This score quantifies the extent to which an internet voting protocol has been deployed and tested in practical scenarios, reflecting its maturity and reliability in real-world contexts. The following values are used to determine the Practical Usability (PU) score:

\begin{itemize}
    \item \textbf{0: Not used in production}: Protocols in this category are in the prototype phase, without any documented instances of deployment in actual election environments.
    \item \textbf{1: Used in low-stakes elections}: These protocols have been deployed in environments with minimal risk, such as elections in academic institutions, student government bodies, or internal organizational decisions.
    \item \textbf{2: Used in medium-stakes elections}: The protocol has been used in elections with moderate significance, such as local government elections, corporate governance voting, or elections in NGOs.
    \item \textbf{3: Used in high-stakes elections}: Protocols that have been implemented in high-stakes environments, such as national or regional government elections. These elections carry substantial political or social consequences.
\end{itemize}

To determine the PU score, publicly available data was collected and reviewed, including: news articles and media coverage of elections or referendums that have adopted internet voting protocols; government or institutional reports documenting the use of these protocols in formal elections; and academic papers, whitepapers, or technical blogs analyzing real-world case studies of these protocols.

\subsubsection{Trust Model Quantification (TM)}\label{sec:tm_quantification}

To systematically compare the trust assumptions underlying each security property, we developed a quantification scale derived directly from our analysis of the sixteen protocols. This process revealed distinct patterns of trust distribution, which we organized into a progressive, ordinal scale based on the principle of trust minimization. A higher score signifies a stronger, more decentralized, and more desirable trust model. The scale is structured into tiers representing qualitative improvements in security, as detailed below.

\begin{enumerate}[series=trustscale]
\addtocounter{enumi}{-1}
    \item \textbf{N/A:} The protocol makes no claim to or fails to achieve the security property.
\end{enumerate}

\bigskip 
\noindent\textbf{Tier 1: Centralized Trust.} Security relies on the presumed honesty of one or a very small, fixed set of named authorities. A compromise of these entities leads to a total failure of the property.
\begin{enumerate}[resume*=trustscale] 
    \item \textbf{Single Authority (1/1):} Represents a single point of failure.
    \item \textbf{A Few Single Authorities (few/1):} Requires collusion among a small, fixed set of trusted parties.
\end{enumerate}

\bigskip
\noindent\textbf{Tier 2: Distributed Trust in Closed Systems.} Trust is distributed among a larger, but still private and permissioned, set of participants (`n`). These models improve resilience but lack public transparency, as their integrity cannot be verified by external observers.
\begin{enumerate}[resume*=trustscale]
    \item \textbf{One or Threshold of Closed Set (1/n or few/n):} The property holds if an honest minority (e.g., at least one mix-node) or a specific threshold is honest.
    \item \textbf{Majority of Closed Network (2/n):} Requires an honest majority of a private network to maintain security.
    \item \textbf{One of Closed Set and a Few Single Authorities (1/n + few/1):} A hybrid model combining distributed and centralized trust within a closed environment.
\end{enumerate}

\bigskip
\noindent\textbf{Tier 3: Publicly Verifiable Trust.} This tier introduces public auditability via an open set of auditors (`N`), where any single honest auditor can detect misbehavior. This transparency makes the system's operation verifiable by anyone.
\begin{enumerate}[resume*=trustscale]
    \item \textbf{Auditors and Single Authority (1/N + 1/1):} A centralized component whose honesty is enforced by public oversight.
    \item \textbf{Auditors and Majority of Closed Network (1/N + 2/n):} A closed distributed system made transparent through public audit.
    \item \textbf{Auditors, One of Closed Set, and a Few Single Authorities (1/N + 1/n + few/1):} A complex hybrid model where all components are subject to public verification.
\end{enumerate}

\bigskip
\noindent\textbf{Tier 4: Permissionless and Cryptographic Trust.} The highest tier removes reliance on trusted parties, instead depending on large-scale economic incentives or pure mathematics.
\begin{enumerate}[resume*=trustscale]
    \item \textbf{Majority of Open Network (2/N):} Trust is placed in the crypto-economic security of a large, public, permissionless network.
    \item \textbf{Trustless (0/N):} The gold standard. The property is guaranteed by cryptographic proofs alone, with no honesty assumptions.
\end{enumerate}

\bigskip 

Rather than assigning a single trust model to the entire system, a score from this scale was assigned to each of the following security properties listed in Section~\ref{sec:security-properties}: SEC(p) (Voting Secy), ANON(p) (Voter Anonymity), IVF(p) (Individual Verifiability), UVF(p) (Universal Verifiability), and EVF(p) (Eligibility Verifiability).

While this general framework applies to most properties, we found that Coercion Resistance (CRES) required a more specialized scale to capture the nuances of its implementation.

\paragraph{Quantifying Coercion Resistance (CRES)}
During our analysis, we observed that Coercion Resistance (CRES) required a more tailored quantification than the other security properties. Applying the general trust model directly resulted in a near-binary outcome, with only two protocols demonstrating any mechanism for strong coercion resistance. To provide a more nuanced and informative comparison, we developed a specific ordinal scale for CRES, derived inductively from the different types of mechanisms observed in the analyzed protocols. This scale distinguishes between protocols with no mechanism, those offering weak resistance via re-voting, systems using strong procedural mitigation via a hybrid model, and finally protocols that implement strong, native cryptographic resistance with trust assumptions. This multi-level approach allows for a more accurate comparison, acknowledging that even partial or procedural solutions offer more protection than none at all. The resulting scale is as follows:
\begin{itemize}[labelindent=0pt, labelwidth=*, leftmargin=!]
    \item \textbf{0: No resistance}: The protocol provides no features to prevent a voter from proving their vote to a coercer.
    \item \textbf{1: Weak resistance}:  The protocol offers plausible deniability, typically through a re-voting mechanism. This is considered weak as its effectiveness depends on the coercer's inability to observe the voter until polls close.
    \item \textbf{2: Hybrid model}: The system mitigates coercion through a robust, external process, such as a hybrid model where an in-person vote can override an electronic one.
    \item \textbf{3: Strong cryptographic resistance with trusted party}: The protocol uses native cryptographic methods to prevent receipt generation, but this guarantee relies on a trusted party (e.g., a coordinator) not colluding with the voter.
    \item \textbf{4: Full receipt-freeness}: The protocol achieves trustless, cryptographically-enforced receipt-freeness, making it impossible for a voter to prove how they voted.
\end{itemize}

This breakdown allows the assessment of protocols in terms of both their individual security features and overall maturity. Some protocols might excel in specific areas, such as Individual Verifiability or Universal Verifiability, while being weaker in areas like Coercion Resistance. By assigning a separate trust score for each property, the trade-offs protocols make are made visible. Nevertheless, trust model scores, e.g., (1) Single Authority (1/1) and (10) Trustless (0/N), are not intended to be interpreted as literal numerical values indicating that ``Trustless'' is ten times better than ``Single Authority''. Instead, these values represent an ordinal ranking, indicating the relative strength of the trust model assumptions. Higher values reflect a more desirable trust model with fewer dependencies on honest parties, but the gap between these values is not necessarily linear. To address this, we normalized the trust scores using the min-max formula when computing the overall maturity score. Normalizing the values allows for a realistic comparison of different trust models on an equal scale.

\subsection{Internet Voting Maturity Framework}\label{sec:def-EVMI}
Drawing from prior art on maturity frameworks for analyzing digital governance applications that make use of blockchains~\cite{biedermann_evaluating_2024}, the maturity score for a protocol, denoted as IVMF(p). Based on the recommendations for composite indicator construction by the Organization for Economic Co-operation and Development (OECD)~\cite{oecd_handbook_2008}, the IVMF(p) is the weighted sum of three~\textit{min-max normalized} indicators, as follows:
\begin{align}
IVMF(p) = w_c\, \text{CMPX}(p) + w_b\, \text{PU}(p)+ w_a\, \text{TM}(p)
\end{align}

The values for each component of IVMF(p), namely, CMPX(p), PU(p), SEC(p), ANON(p), IVF(p), UVF(p), EVF(p), CRES(p), are defined accordingly:
\begin{itemize}
    \item $w_c, w_b,w_a, w_t, w_u, w_v, w_m,$ and $w_n$ are the weights are assigned to all indicators of $IVMF(p)$ and $TM(p)$, reflecting their relative importance.
    \item $\text{CMPX}(p)$, the complexity score, is calculated as described in Section~\ref{sec:system-complexity}, with values detailed in Appendix~\ref{app:cmpx_and_pu}.
    \item $\text{PU}(p)$, the Practical Usability score, is determined as explained in Section~\ref{sec:practical_usage}, with corresponding values presented in Appendix~\ref{app:cmpx_and_pu}.
    \item $\text{TM}(p)$, the trust model score, is composed of the security properties: SEC(p), ANON(p), IVF(p), UVF(p), EVF(p), and CRES(p), which are quantified as outlined in Section~\ref{sec:tm_quantification}. The calculated values are provided in Appendix~\ref{app:trust_models}. $TM(p)$ is a composite of~\textit{min-max normalized} factors constructed using the formula:
\end{itemize}
\begin{align}
TM(p)= w_s\, \text{SEC}(p) + w_t\, \text{ANON}(p) + w_u\, \text{IVF}(p) \nonumber \\ + w_v\, \text{UVF}(p) + w_m\, \text{EVF}(p) + w_n\, \text{CRES}(p)
\end{align}

The weights for computing the IVMF(p) in Table~\ref{tab:IVMF_scores} and the TM(p) sub-indicator were chosen inductively. This was necessary to cover an especially large set of different i-voting systems across various use cases and with a diverse set of trust models. The definition of weights followed the emphasis on practical usage of technology readiness levels (TRLs)~\cite{Mankins2004}. Although TRLs are widely accepted as a measure for maturity~\cite{olechowski_technology_2015}, supporting this approach, sensitivity tests for the weights were performed to increase the robustness and reliability of TM(p) and IVMF(p).
The robustness of the composite indicators to changes in the underlying weights by comparing rankings across alternative weights. The inductively defined \textit{default} weights were as follows: $w_c=-0.5$, $w_b=3, w_a=1, w_s=1.6, w_t=1.8, w_u=2, w_v=1.4$, and $w_u=1.2$. To evaluate the impact of weights on the ranking of i-voting protocols, additional weights were defined for the following scenarios:
\begin{itemize}
    \item For IVMF(p): equal weights, CMPX(p) weighted, TM(p) weighted, and PU(p) weighted.
    \item For TM(p): equal weights, verifiability weighted, and anonymity and security weighted. 
\end{itemize}
Thereafter, for each composite indicator $IVMF(p)$ with weights $w_i$ and $TM(p)$ with weights $w_j$, pairwise correlations of the resulting ranks under different weight specifications were computed~\cite{fieller_tests_1957,oecd_handbook_2008}. For each pair of weight variants $w_\alpha, w_\beta$ ($ i,j \in \alpha, \beta$), Pearson’s product–moment correlation coefficient was calculated and the standard variance $r_{\alpha\beta}^2$ is reported to indicate the relative importance of variance in weighting. Finally, the Pearson’s product–moment correlations were assessed for statistical significance using $t$–test and two-sided $p$–values.

\section{Data}\label{sec_data}
Given the theoretical and methodological framework, this section provides a brief overview of the internet voting protocols that form the dataset. Each protocol was selected based on its relevance to real-world deployments, technical innovations, and its ability to address key challenges in online voting systems. Below, protocols are grouped by their primary purpose, highlighting where each has been deployed and any unique functionalities that distinguish them from other systems.

\subsection{National and Local Government Elections}\label{sec_data.1}

\begin{itemize}
    \item \textbf{Estonian i-voting system (IVXV)}~\cite{GeneralFrameworkElectronic} is one of the most well-known and widely deployed internet voting protocols, used in national elections since 2005. Its primary purpose is to allow citizens to cast votes remotely, offering a convenient alternative to traditional in-person voting. Unique functionalities include the use of national digital ID cards for secure voter authentication and end-to-end verifiability features, ensuring that each vote is correctly tallied while maintaining voter anonymity.
    
    \item \textbf{Scytl's online voting platform}~\cite{larraiaSVoteControlComponents2022} is used in various local and national elections across several countries, including France and Switzerland. It utilizes a microservice architecture for enhanced scalability and security, with Control Components that isolate critical operations like vote mixing and decryption. The system’s use of Choice Return Codes offers end-to-end verifiability without compromising voter privacy.
    
    \item \textbf{CHVote}~\cite{haenniCHVoteProtocolSpecification2017} is an internet voting system developed by the canton of Geneva and used in Swiss cantonal and federal elections. Its key features include strong cryptographic guarantees for both individual and universal verifiability, ensuring transparency without compromising voter privacy. The system was primarily designed for both resident and expatriate voters and has been deployed in multiple trials, with high adoption among Swiss expatriates.
    
    \item \textbf{Agora}~\cite{AgoraBringingOur} is a blockchain-based voting platform designed to provide end-to-end verifiability, transparency, and security for government and institutional elections. The platform uses a custom blockchain architecture, combining a permissioned blockchain with Bitcoin-based immutability for public verifiability. Agora gained international attention in 2018 as an accredited observer in Sierra Leone's presidential election, where it recorded and published verifiable results in real time, days ahead of the official manual tally.

    \item \textbf{Voatz}~\cite{voatzVoatzMobileVoting2019} is a mobile internet voting platform designed for remote ballot return in government elections, primarily targeting military and overseas (UOCAVA) voters. The system uses a smartphone application for identity verification and ballot marking, combined with a permissioned blockchain (Hyperledger Fabric) on the backend for ballot storage. Voatz has been piloted in several U.S. states, including West Virginia and Utah, but has also been the subject of significant public scrutiny regarding its security architecture, particularly the reliance on a centralized backend and the potential for client-side vulnerabilities~\cite{specterBallotBustedBlockchain2020}.
\end{itemize}

\subsection{Organizational, Academic, and Low-Stakes Elections}\label{sec_data.2}

\begin{itemize}
    \item \textbf{Helios}~\cite{adidaHeliosWebbasedOpenAudit2008} is an open-source, web-based voting protocol designed for low-coercion environments, such as academic institutions and organizations. Helios focuses on providing open-audit, end-to-end verifiability while maintaining voter privacy, making it ideal for low-stakes elections. Notably deployed in the University of Louvain’s presidential election in 2009, it has proven its effectiveness in academic settings.
    
    \item \textbf{Belenios}~\cite{cortierBeleniosSimplePrivate2019,cortierFeaturesUsageBelenios2022} is an open-source internet voting system widely used for academic, non-profit, and organizational elections. Built upon Helios, Belenios adds features such as eligibility verifiability and protection against ballot stuffing. It supports various voting methods and has been used in over 1,400 elections yearly.
    
    \item \textbf{Open Vote Network}~\cite{mccorryOpenVoteNetwork2023} is a decentralized, self-tallying voting protocol suited for small-scale elections, such as boardroom voting. It maximizes voter privacy by requiring full collusion among all voters to breach confidentiality. It is primarily suited for environments with lower coercion risks due to the lack of robust coercion resistance mechanisms. 
\end{itemize}

\subsection{DAO Governance and Blockchain-Based Voting}\label{sec_data.3}

\begin{itemize}
    \item \textbf{Snapshot}~\cite{WelcomeSnapshotDocs2024} is a decentralized, off-chain voting platform widely used by DAOs, decentralized finance (DeFi) protocols, and non-fungible token (NFT) communities for governance purposes. Snapshot enables gasless voting and supports flexible voting mechanisms, such as Quadratic Voting and Approval Voting. It has seen widespread adoption, with 96\% of DAOs using it and over 500,000 monthly active users.
    
    \item \textbf{Snapshot X}~\cite{Snapshot2024} is the fully on-chain version of Snapshot, built on Starknet. It integrates trustless execution and on-chain verifiability, improving censorship resistance and security while retaining flexibility for DAOs. Snapshot X is designed for decentralized governance models that require robust decentralization and security.
    
    \item \textbf{Vocdoni}~\cite{williamsRemoteVotingAge2022,VocdoniIntroductionVocdoni} is a decentralized voting protocol designed for large-scale governance in DAOs and blockchain-based communities. The system integrates with Ethereum for added transparency and has been deployed in various contexts, including the Votecaster client for Farcaster's blockchain-based social media protocol.
    
    \item \textbf{Cicada}~\cite{glaeserCicadaFrameworkPrivate2023} is an on-chain voting protocol built on Ethereum. Using time-lock puzzles and homomorphic encryption, Cicada enables private, non-interactive voting, but struggles with providing everlasting privacy. It is designed for decentralized governance and DAOs but has not yet been deployed in practice.
    
    \item \textbf{zkSnap}~\cite{ghangasZkSnapScalableZero} is a novel online voting protocol that uses zero-knowledge proofs and time-lock encryption to ensure privacy and cost-efficiency in blockchain voting. Although zkSnap has not yet been deployed in practice, it is positioned as a solution for privacy-preserving voting in DAOs and decentralized systems.
    
    \item \textbf{Stellot}~\cite{baranskiPracticalIVotingStellar2020} is a blockchain-based i-voting platform built on the Stellar network. It provides a privacy-preserving voting solution using blind signature techniques. Stellot is designed for decentralized governance but has not yet been deployed in practice.
\end{itemize}

\subsection{Participatory Democracy and Public Goods Funding}\label{sec_data.4}

\begin{itemize}
    \item \textbf{Decidim}~\cite{barandiaranDecidimTechnopoliticalNetwork2024,DecidimUse} is an open-source platform for participatory democracy, originally developed by the Barcelona City Council. It supports a wide range of democratic processes, including participatory budgeting and consultations, and is used by over 400 entities globally, including cities like Helsinki and the European Commission.
    
    \item \textbf{Minimal Anti-Collusion Infrastructure (MACI)}~\cite{ethereumfoundationMinimalAntiCollusionInfrastructure2022} is an on-chain voting protocol focused on mitigating collusion and bribery, making it ideal for Quadratic Funding (QF) and decentralized governance. MACI has been successfully deployed in QF rounds, distributing over \$1M to public goods projects via platforms like \url{clr.fund} and the Gitcoin Allo stack.
    
    \item \textbf{Votem’s Proof of Vote}~\cite{mattbeckerProofVoteEndtoend2018} is an end-to-end verifiable voting protocol leveraging blockchain technology, primarily designed for remote and mobile voting in public elections. Votem has been deployed in high-profile elections, for example in the State of Montana and for the Rock and Roll Hall of Fame.
\end{itemize}

\section{Results}\label{sec_6}
Electronic voting systems in general and i-voting applications specifically are used across a variety of use cases, as is demonstrated by the data presented in Section~\ref{sec_data}. While this justified an aggregating approach to compare these systems and to identify new use cases, this sections highlights the utility of the IVMF for narrating the societal implications of using e-voting, as well as explaining their strengths and limitations from an information security perspective. The section proceeds by, first, giving an overview of the i-voting systems, their scores. Thereafter, considerations for the trust models are put forth. Lastly, findings for the usability of i-voting and implications for furthering the research on maturity models in the context of emerging governance technologies in general are outlined.
\subsection{Comparision}\label{sec:comparison}

Overall the results of the quantitative scoring match the qualitative secondary data, anecdotal accounts, and in-depth reviews of i-voting protocols in the literature. Meanwhile, the appropriate use of blockchain was shown to improve the score of centralized systems as long as it does not significantly increase protocol complexity. Snapshot and Snapshot X are pertinent examples for highlighting these improvements. More generally, a threshold of complexity appears to emerge, where an increase in complexity negatively influences the overall performance of an i-voting system. This is of particular importance for blockchain-based systems, as these were suffering the most from disproportionately complex architectures. Thus, in the overall comparison of maturity scores, governmental e-voting platforms outperformed the latest blockchain-based systems with the exception of MACI.

\begin{table*}[ht!]
\footnotesize
\centering
\caption{Comparison of Attribute Scores for Analyzed E-Voting Systems}
\label{tab:comparision}
\begin{tabular}{p{.22\textwidth} p{.07\textwidth} p{.07\textwidth} p{.07\textwidth} p{.07\textwidth} p{.07\textwidth} p{.07\textwidth} p{.07\textwidth} p{.07\textwidth}}
\toprule
Name & CMPX & PU & SEC & ANON & IVF & UVF & EVF & CRES \\
\midrule
Estonian e-voting system & \cellcolor{yellow!30}11 & \cellcolor{green!30}3 & \cellcolor{green!10}5 & \cellcolor{yellow!30}2 & \cellcolor{green!10}6 & \cellcolor{green!10}6 & \cellcolor{red!30}1 & \cellcolor{yellow!30}2 \\
Scytl & \cellcolor{yellow!30}14 & \cellcolor{yellow!30}2 & \cellcolor{green!10}5 & \cellcolor{yellow!30}2 & \cellcolor{green!30}8 & \cellcolor{green!30}8 & \cellcolor{red!30}1 & \cellcolor{red!30}0 \\
CHVote & \cellcolor{green!10}9 & \cellcolor{green!30}3 & \cellcolor{green!10}5 & \cellcolor{green!10}5 & \cellcolor{green!30}8 & \cellcolor{green!30}8 & \cellcolor{red!30}1 & \cellcolor{red!30}0 \\
Belenios & \cellcolor{green!10}10 & \cellcolor{yellow!30}1 & \cellcolor{yellow!30}3 & \cellcolor{yellow!30}2 & \cellcolor{green!10}6 & \cellcolor{green!10}6 & \cellcolor{yellow!30}2 & \cellcolor{red!30}0\footnotemark \\
Helios & \cellcolor{green!10}9 & \cellcolor{yellow!30}1 & \cellcolor{yellow!30}3 & \cellcolor{red!30}1 & \cellcolor{green!10}6 & \cellcolor{green!10}6 & \cellcolor{red!30}1 & \cellcolor{red!30}0 \\
Decidim & \cellcolor{green!10}9 & \cellcolor{yellow!30}1 & \cellcolor{yellow!30}3 & \cellcolor{red!30}1 & \cellcolor{green!10}6 & \cellcolor{yellow!30}4 & \cellcolor{red!30}1 & \cellcolor{yellow!30}1 \\
Votem, Proof of Vote & \cellcolor{red!30}29 & \cellcolor{yellow!30}1 & \cellcolor{yellow!30}3 & \cellcolor{yellow!30}3 & \cellcolor{green!10}7 & \cellcolor{green!10}7 & \cellcolor{yellow!30}3 & \cellcolor{red!30}0 \\
Agora & \cellcolor{red!30}20 & \cellcolor{yellow!30}1 & \cellcolor{yellow!30}3 & \cellcolor{green!10}5 & \cellcolor{green!10}7 & \cellcolor{green!10}7 & \cellcolor{red!30}1 & \cellcolor{red!30}0 \\
Vocdoni & \cellcolor{red!30}28 & \cellcolor{yellow!30}1 & \cellcolor{yellow!30}3 & \cellcolor{green!30}10 & \cellcolor{yellow!30}3 & \cellcolor{yellow!30}3 & \cellcolor{green!30}10 & \cellcolor{yellow!30}1 \\
Voatz & \cellcolor{yellow!30}12 & \cellcolor{yellow!30}2 & \cellcolor{red!30}1 & \cellcolor{red!30}1 & \cellcolor{red!30}1 & \cellcolor{yellow!30}4 & \cellcolor{red!30}1 & \cellcolor{red!30}0 \\
zkSnap & \cellcolor{green!30}6 & \cellcolor{red!30}0 & \cellcolor{red!30}1 & \cellcolor{green!30}10 & \cellcolor{red!30}1 & \cellcolor{red!30}1 & \cellcolor{green!30}10 & \cellcolor{red!30}0 \\
Stellot & \cellcolor{green!30}4 & \cellcolor{red!30}0 & \cellcolor{red!30}1 & \cellcolor{green!30}10 & \cellcolor{green!30}9 & \cellcolor{green!30}9 & \cellcolor{red!30}1 & \cellcolor{red!30}0 \\
MACI & \cellcolor{green!30}4 & \cellcolor{yellow!30}2 & \cellcolor{red!30}1 & \cellcolor{red!30}1 & \cellcolor{green!30}9 & \cellcolor{green!30}9 & \cellcolor{red!30}1 & \cellcolor{yellow!30}3 \\
Cicada & \cellcolor{green!30}4 & \cellcolor{red!30}0 & \cellcolor{red!30}0 & \cellcolor{green!30}10 & \cellcolor{green!30}9 & \cellcolor{green!30}9 & \cellcolor{red!30}1 & \cellcolor{red!30}0 \\
Open Vote Network & \cellcolor{green!30}3 & \cellcolor{red!30}0 & \cellcolor{green!30}10 & \cellcolor{red!30}1 & \cellcolor{green!30}9 & \cellcolor{green!30}9 & \cellcolor{red!30}1 & \cellcolor{red!30}0 \\
Snapshot & \cellcolor{green!30}6 & \cellcolor{yellow!30}2 & \cellcolor{yellow!30}3 & \cellcolor{red!30}0 & \cellcolor{red!30}1 & \cellcolor{red!30}1 & \cellcolor{green!30}10 & \cellcolor{red!30}0 \\
Snapshot X & \cellcolor{green!30}6 & \cellcolor{yellow!30}1 & \cellcolor{yellow!30}3 & \cellcolor{red!30}0 & \cellcolor{green!30}9 & \cellcolor{green!30}9 & \cellcolor{green!30}10 & \cellcolor{red!30}0 \\
\bottomrule
\end{tabular}
\end{table*}
\footnotetext{The standard Belenios protocol is evaluated here, which is not coercion-resistant. A separate variant, BeleniosRF, achieves strong coercion resistance by having the server re-randomize ballots. Under our framework, BeleniosRF would receive a CRES score of 3, as its receipt-freeness relies on a trusted server not colluding with the voter.}

For a general overview, the Table~\ref{tab:comparision} presents the comparison of attribute scores for analyzed e-voting systems. The rationale for each of the values is provided in the Appendix. The table represents the final quantification of Complexity, Practical Usability, and Trust model for each of the security properties. CMPX is Complexity (a lower score is better), PU is Practical Usability (a higher score is better), SEC is Voting secrecy, ANON is Voter anonymity, IVF is Individual Verifiability, UVF is Universal Verifiability, EVF is Eligibility Verifiability, and CRES is Coercion Resistance. Green indicates the best value, yellow is moderate, and red indicates the worst.

\subsection{Trust Models}\label{sec:trust-models}
The investigation revealed that most of the analyzed protocols achieve a high level of both \textbf{Individual} and \textbf{Universal Verifiability}. Individual Verifiability often requires users to perform a ballot spoiling procedure, e.g. the Benaloh Challenge~\cite{benalohSimpleVerifiableElections2006}, which reassures them that their voting device records ballots as intended. Although this step is typically optional, the more users that challenge the system, the greater the overall confidence. Universal Verifiability, however, often assumes that users will verify their vote's inclusion in the final tally, which is also an optional step. A key \textbf{advantage of blockchain-based protocols} lies in their inherent support for both Individual and Universal Verifiability. In these systems, all nodes in the network act as auditors, continuously verifying all processes defined in the smart contracts. Verifiability is thus maintained as long as a majority of the blockchain nodes are correct, follow the protocol, do not censor transactions, and do not reorganize the blockchain. In contrast, non-blockchain systems typically rely on external auditors to verify processes and data flows. Meanwhile, when comparing \textbf{private blockchains} to non-blockchain-based systems, there is little difference in terms of trust models. Although occasional logging to a public blockchain improves the accessibility to the logs compared to a system that is only audited or checked eventually, public logs also need to be checked to be effective. Thus, a private blockchain with a public blockchain component is analogous to a non-blockchain system with external auditors, where the public blockchain plays the role of the auditor.

\textbf{Coercion Resistance} proved to be one of the most difficult properties to achieve, with protocols implementing a wide spectrum of mitigation strategies rather than a single solution. Our analysis, based on the specialized CRES scale (Section~\ref{sec:tm_quantification}), reveals these nuances. Most protocols that address coercion offer only \textbf{Weak resistance} (Score 1), typically through a re-voting mechanism that allows voters to change their vote. This provides plausible deniability but is ineffective if a coercer can monitor the voter until polls close. 

A stronger approach is the \textbf{Hybrid model} (Score 2) employed by the Estonian i-voting system (IVXV), where an in-person paper vote can invalidate a previously cast electronic vote. While this may be a highly effective real-world safeguard, we score it lower than native cryptographic solutions because it relies on a procedure that deviates from the definition of a fully remote e-voting system. The model's success hinges on the additional assumptions and logistical requirements of a secure, parallel paper voting process, placing it outside the direct scope of the protocol's own guarantees.

The most advanced protocols in our analysis achieve \textbf{Strong cryptographic resistance with a trusted party} (Score 3). This is demonstrated by MACI, which uses zero-knowledge proofs and a key-switching mechanism coordinated by a trusted operator to make it cryptographically difficult for a voter to prove their vote. Similarly, a variant of Belenios known as BeleniosRF~\cite{chaidosBeleniosRFNoninteractiveReceiptFree2016} also achieves this level of protection, though its guarantees likewise depend on a trusted server, and it was not included as a separate entry in our analysis. No analyzed protocol achieved the highest level of \textbf{Full receipt-freeness} (Score 4) without some trust assumption, highlighting that while perfect coercion resistance remains elusive, practical and incremental solutions are being successfully deployed.

\textbf{Eligibility Verifiability} was identified as a bottleneck, requiring trust in the organizer responsible for preparing the list of eligible voters or the authentication/registration provider who verifies user identity and grants access to the voting process. This introduces the risk of ballot stuffing, where the authority controlling voter eligibility could add unauthorized ballots without detection. Belenios addresses this problem by introducing a separate party, which must verify and ``sign'' ballots before they are added to the ballot box~\cite{cortierElectionVerifiabilityHelios2014}. The challenge of Eligibility Verifiability becomes more pronounced as the size of the voter base grows. In small elections, any additional votes would likely be noticed by the electorate. In systems where voters are not anonymized, ballot stuffing is also more difficult to execute, as anomalies can be detected and verified. Additionally, systems that ensure individual verifiability allow voters to confirm that their ballot was recorded as intended. In practice, only a small percentage of voters may challenge the system, meaning that in large-scale elections, ballot stuffing could go undetected.

\textbf{Voter Anonymity} is typically achieved through the use of \textit{mix nodes} that shuffle ballots. As long as one \textit{mix node} remains honest, ballots cannot be linked to voter identities. The risk arises when the voter identity is correlated with their actual identity via the authentication provider, organizer, or print office. Systems that do not use mix networks, which are also referred to as \textit{mixnets}~\cite{cortierBeleniosSimplePrivate2019}, rely solely on the trustworthiness of the organizer, who knows the real identities of voters, such as Helios, Belenios, Decidim, OVN, and MACI. Some systems, like \textit{Follow My Vote} and \textit{Stellot}, break the authentication and authorization link using blind signatures. Others, like Cicada~\cite{glaeserCicadaFrameworkPrivate2023}, Vocdoni~\cite{williamsRemoteVotingAge2022}, and zkSnap~\cite{ghangasZkSnapScalableZero}, use cryptographic nullifiers (zero-knowledge proof of inclusion), which allow them to pass authorization without exposing voter identities, relying only on the soundness of the cryptography.

All non-blockchain-based protocols achieve \textbf{Voting Secrecy} by encrypting ballots with a key that is shared among a set of key-holders. Decryption is only possible when a subset of these key-holders participate in a multi-party computation and contribute their key shares. Various voting secrecy approaches are used in the blockchain-based systems. For example, OpenVote Network uses a technique that requires no explicit encryption key; instead, the key is split among voters, and the final result is automatically decrypted once the last vote is cast. This approach has the drawback of low availability, as even one voter can block the voting process by not casting a vote, necessitating a repeat election. In Stellot and MACI, the decryption keys are held by a single trusted third party (TTP), but there seems to be no reason why these protocols could not be extended to distribute keys among multiple key-holders, as other platforms do. Cicada and zkSnap use a time-lock puzzle network that, assuming at least one honest party in a public consortium (such as \textit{\href{https://drand.love/}{drand.love}}), will not reveal the share of encryption key prematurely. Meanwhile, this technique reveals the decryption key to everyone, allowing anyone to decrypt individual ballots, thereby protecting voter confidentiality as long as the anonymity property holds.

\subsection{Usability}\label{sec:usability}
Usability in the context of i-voting systems is a critical factor that encompasses the complexity of the architecture, the number of components involved, and the associated costs. A system's usability can be quantified by tracking the number of parties or components required to achieve the claimed trust models. The more components and parties involved, the less usable the system tends to be, as increased complexity often leads to more work, higher costs, and greater potential for vulnerabilities. Ideally, a system should minimize complexity while maximizing the distribution of trust.

Systems with a high number of components tend to be more complex and less user-friendly. For instance, the \textbf{Estonian e-voting system} involves four primary components, along with multiple talliers (key-holders) and auditors, making it a sophisticated yet complex system. Similarly, \textbf{Scytl}'s system incorporates three primary components, several Control Components, multiple Board Members (key-holders), and auditors, adding layers of complexity. Blockchain-based systems, while offering decentralization, also embody intricate structures. For example, \textbf{Votem}~\cite{mattbeckerProofVoteEndtoend2018}, \textbf{Agora}~\cite{AgoraBringingOur}, and \textbf{Vecdoni}~\cite{williamsRemoteVotingAge2022} combine centralized components with distributed networks, incorporating both public and/ or private blockchains. These systems require a balance between decentralization and operational complexity, which can impact their overall usability.

Simpler systems are generally more accessible and easier to manage. Non-blockchain-based systems like \textbf{Helios}~\cite{adidaHeliosWebbasedOpenAudit2008} and \textbf{Decidim}~\cite{barandiaranDecidimTechnopoliticalNetwork2024} operate with just an organizer, one central component, a set of key-holders, and auditors. These systems are easier to manage and deploy, making them more accessible for smaller-scale or less complex voting scenarios. Among blockchain systems, \textbf{OpenVoteNetwork} stands out as the simplest, being entirely trustless and requiring only the organizer to deploy the contract on the blockchain. Other relatively simple blockchain-based systems like \textbf{Cicida}, \textbf{MACI}, and \textbf{Stellot} require one central component alongside the public blockchain. This simplicity enhances usability, while offering the security benefits of blockchain technology.

Public blockchain-based systems exhibit significant advantages, but also pose challenges related to usability and long-term privacy. One significant advantage is that the security of the network is maintained by the computational power or staked funds of the participating nodes, removing the burden of infrastructure management from the organizers. The network is maintained by nodes, miners, or validators who are compensated through transaction fees. However, this introduces a usability issue: in democratic elections, voters should not be required to pay transaction fees to cast their ballots. To address this, some protocols, like \textbf{Stellot}, introduce a master wallet that covers transaction fees, while systems like \textbf{Snapshot} or \textbf{zkSnap} use off-chain voting with a trusted aggregator. Others implement a hybrid approach with a private blockchain for voting, which helps mitigate costs while preserving some transparency.

A more fundamental disadvantage of using a public blockchain, however, is the risk to everlasting privacy. Publishing encrypted ballots to an immutable public ledger means they are stored permanently and accessibly by anyone. While the encryption may be secure today, advances in cryptanalysis or the advent of quantum computing could allow these ballots to be decrypted in the future. This possibility of retroactive deanonymization poses a severe long-term threat to ballot secrecy, a critical argument that must be weighed when considering public blockchains for storing sensitive electoral data.

\subsection{Internet Voting System Maturity -- Balancing Deployment and Design}\label{sec:results-ivmf}

The aggregate maturity scores were calculated using the normalized, weighted formula of the Internet Voting Maturity Framework (IVMF), following established best practices for the construction of composite indicators~\cite{oecd_handbook_2008}. The results are presented in Table~\ref{tab:IVMF_scores}. The primary ranking uses a weighting scheme $w_1$ explicitly designed to model real-world maturity by heavily rewarding Practical Usability ($w_b=3$), moderately rewarding the Trust Model score ($w_a=1$), and penalizing architectural Complexity ($w_c=-0.5$). The color-coded Practical Usage (PU) column in the table makes the strong influence of this metric visually apparent, as systems with high PU scores (green) dominate the top ranks.

This demonstrates that established, government-backed systems like \textbf{CHVote} (Rank 1, PU=3) and the \textbf{Estonian i-voting system} (Rank 2, PU=3) represent the current pinnacle of maturity. Their extensive use in high-stakes elections, combined with robust operational procedures, sets them apart. Similarly, platforms with high adoption in the blockchain ecosystem, such as \textbf{MACI} (Rank 3, PU=2) and \textbf{Snapshot} (Rank 5, PU=2), also rank highly due to their proven application in real-world governance scenarios.

However, while a high PU score is a strong driver of maturity, the framework's multi-faceted nature reveals a more nuanced picture. The table highlights a vital insight: a higher PU score does not guarantee a higher maturity rank. A prime example is \textbf{Snapshot X} (Rank 6), which, despite a lower Practical Usage score (PU=1, orange), outranks \textbf{Voatz} (Rank 7, PU=2, yellow). This is because Snapshot X's exceptionally strong Trust Model score and low Complexity score compensate for its more limited deployment. Conversely, Voatz's higher PU score is significantly offset by one of the weakest Trust Model scores in the dataset.

This demonstrates that the IVMF does not simply reward popularity; it balances real-world adoption with technical and architectural soundness. Protocols that remain purely theoretical (PU=0, red), such as \textbf{Stellot}, \textbf{Cicada}, and \textbf{zkSnap}, consequently rank at the bottom, as their lack of proven, practical application results in the lowest maturity scores, regardless of their cryptographic strengths.

\begin{table}[htbp]
\centering
\footnotesize
\newcommand{\puhigh}{\cellcolor{green!25}}
\newcommand{\pumed}{\cellcolor{yellow!25}}
\newcommand{\pulow}{\cellcolor{orange!25}}
\newcommand{\puzero}{\cellcolor{red!25}}

\begin{tabular}{lrrrc}
\toprule
Name & \multicolumn{1}{l}{Rank} & \multicolumn{1}{l}{Normalized $IVMF(p)_{w_1}$} & \multicolumn{1}{l}{$IVMF(p)_{w_1}$} & \multicolumn{1}{c}{PU(p)} \\
\midrule
CHVote & 1  & 1.0000 & 3.690 & \puhigh 3 \\
Estonian e-voting system & 2  & 0.9455 & 3.511 & \puhigh 3 \\
MACI & 3  & 0.7677 & 2.927 & \pumed 2 \\
Scytl  & 4  & 0.6334 & 2.487 & \pumed 2 \\
Snapshot & 5  & 0.5144 & 2.096 & \pumed 2 \\
Snapshot X & 6  & 0.4676 & 1.942 & \pulow 1 \\
Voatz  & 7  & 0.4324 & 1.827 & \pumed 2 \\
Belenios & 8  & 0.2847 & 1.342 & \pulow 1 \\
Vocdoni & 9 & 0.2810 & 1.330 & \pulow 1 \\
Agora   & 10 & 0.2803 & 1.328 & \pulow 1 \\
Helios & 11 & 0.2691 & 1.291 & \pulow 1 \\
Decidim & 12 & 0.2623 & 1.269 & \pulow 1 \\
Votem, Proof of Vote  & 13 & 0.2270 & 1.153 & \pulow 1 \\
Stellot & 14 & 0.1746 & 0.981 & \puzero 0 \\
Cicada  & 15 & 0.1678 & 0.958 & \puzero 0 \\
Open Vote Network & 16 & 0.1438 & 0.880 & \puzero 0 \\
zkSnap  & 17 & 0.0000 & 0.408 & \puzero 0 \\
\bottomrule
\end{tabular}
\caption{Internet Voting Systems Ranked by Maturity Score (IVMF), with raw Practical Usage (PU) scores color-coded for clarity. The ranking uses a weighting scheme $w_1$ that heavily prioritizes deployment ($w_b=3$) and discounts complexity ($w_c=-0.5$), reflecting a focus on real-world maturity. The Trust Model score is given a baseline weight ($w_a=1$).}
      \label{tab:IVMF_scores}%
\end{table}

\begin{figure}[htbp]
    \centering
    \includegraphics[scale=0.3]{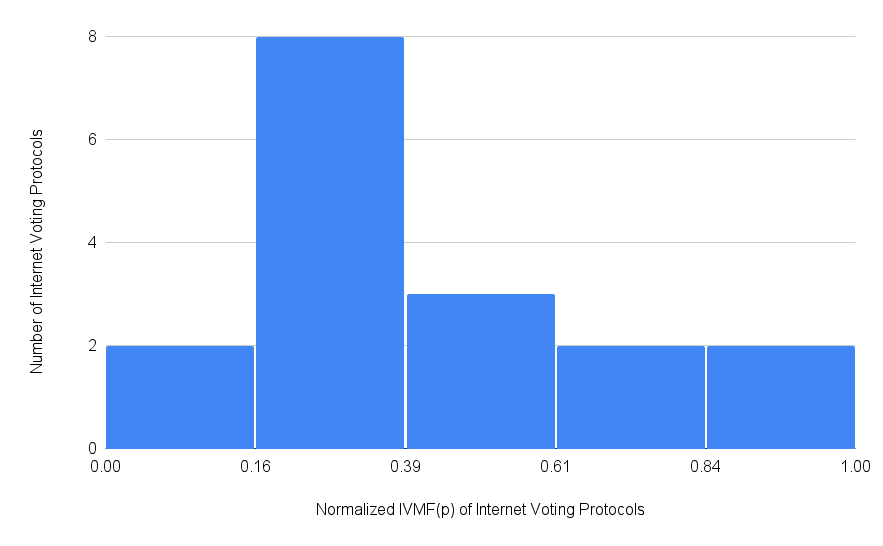}
    \caption{Histogram of normalized Maturity Score of Internet Voting Protocols}
    \label{fig:maturity_score_histogram}
\end{figure}

Figure~\ref{fig:maturity_score_histogram} illustrates the distribution of the normalized maturity scores. The histogram shows a concentration of protocols in the lower-to-mid range (0.16 to 0.61), with a few high-performing outliers. This suggests that while the field is rich with innovative cryptographic ideas, only a select few systems have successfully bridged the gap from theoretical design to mature, practical application. The broad spread indicates a field with distinct clusters of maturity, rather than a gradual continuum, separating the purely academic protocols from those tested in real-world environments.

\FloatBarrier

\subsection{Weights and Sensitivity}\label{sec:sensitivity}
The weights for constructing the IVMF(p) and its TM(p) sub-indicator were chosen inductively. For testing the sensitivity of the composite indicators, rank correlations with the inductive set of weights ($w_{ij,\alpha}$) were calculated for several scenarios ($w_{ij,\beta}$). The rank correlations show that IVMF(p) is usage weighted ($w_{i,4}$) which produced the only statistically significant correlation ($p<0.001$), as opposed to the correlations for trust model weighted ($w_{i,2}$), equal weights ($w_{i,5}$), and complexity weighted ($w_{i,3}$). The correlations are depicted in Table~\ref{tab:ivmf_sensitivity}.

\begin{table}[htbp]
\footnotesize
\centering
\begin{tabular}{lcccc}
\toprule
 & $\mathrm{EVMF(p)}w_{1_2}$ & 
   $\mathrm{EVMF(p)}w_{1,3}$ & 
   $\mathrm{EVMF(p)}w_{1,4}$ & 
   $\mathrm{EVMF(p)}w_{1,5}$ \\
\midrule
PEARSON & 0.316 & 0.520 & 0.887 & 0.561 \\
$R^2$   & 0.100 & 0.270 & 0.787 & 0.315 \\
$t$     & 1.291 & 2.355 & 7.450 & 2.627 \\
$p$     & 0.216 & 0.033 & 0.000 & 0.019 \\
\bottomrule
\end{tabular}
\caption{Rank correlations of $EVMF(p)$ for different weighting scenarios of $TM(p)$, $CMPX(p)$, and $PU(p)$.}
\label{tab:ivmf_sensitivity}
\end{table}
This means that the IVMF(p) favors real-world application over complexity and the trust model, which is in line with the paradigm of TRLs. For the aggregate sub-indicator TM(p) sensitivity to the inductive weighting ($w_{j,1}$) was also measured through rank correlations with anonymity and privacy weighted indicators ($w_{j,2}$), verifiability weighted indicators ($w_{j,3}$), and the equal weights scenario ($w_{j,3}$). All scenarios positively correlate with the inductive weight scenario and are statistically significant at $p<0.001$, which translates into little sensitivity in the tested scenarios.
\begin{table}[htbp]
\footnotesize
\centering
\begin{tabular}{lccc}
\toprule
 & $\mathrm{TM(p)}w_{1,2}$ & 
   $\mathrm{TM(p)}w_{1,3}$ & 
   $\mathrm{TM(p)}w_{1,4}$ \\
\midrule
PEARSON & 0.750 & 0.967 & 0.952 \\
$R^2$   & 0.562 & 0.935 & 0.907 \\
$t$     & 4.391 & 14.642 & 12.100 \\
$p$     & 0.0011 & 0.0000 & 0.0000 \\
\bottomrule
\end{tabular}
\caption{Rank correlations of $TM(p)$ for different weighting scenarios of $SEC(p)$, $IVF(p)$, $UVF(p)$, $EVF(p)$, and $CRES(p)$.}
\label{tab:tm_sensitivity}
\end{table}
Although these rank correlations do not directly address lacking inter-rater reliability, they show the impact of hypothetically different evaluations for the same solution for various sub-indicators and sub-sub-indicators. The low sensitivity to weight changes for TM(p) suggest that the present framework is robust even under deviating scores by other future evaluators. For the IVMF(p), only a significantly different assessment of the practical usage would put the framework at risk of failing to appropriately capture diverging assessments. Given the nature of the indicator for practical usage, significant deviations from the here presented score seem unlikely.

\section{Discussion}\label{sec_7}
This maturity analysis highlights several critical aspects of internet voting systems, particularly concerning their ability to achieve various security properties and the implications of using blockchain technology in the context of social processes. The presented framework explicitly relies on manual assessments because it expresses the relationship between the technical solution and the social context it is deployed in. Thus, formal simulations would not capture the changes to social governance by i-voting systems as \textit{social technologies}~\cite{leibetseder_critical_2011}. In this regard, \textit{maturity} ``evaluates the current level of operational development''~\cite{yatskovskaya_integrated_2018} and ``organisation[al] success''~\cite{kucinska-landwojtowicz_organizational_2023} rather than theoretical computational performance.

While maturity was shown to not rely on any one property, optimizing the use of blockchain technology was identified as the most beneficial architecture improvement that would positively impact the largest subset of e-voting protocols that were analyzed. \textbf{Blockchain-based i-voting protocols offer a significant advantage in terms of common knowledge and verifiability}. In these systems, every node in the network acts as an auditor, validating each message and computation, which inherently supports common knowledge~\cite{halpernKnowledgeCommonKnowledge1990}. As long as the majority of nodes are honest, the blockchain acts as a publicly accessible append-only component often referred to as bulletin board, and as Buterin pointed out, ``the most secure way to implement a bulletin board is to just use an existing blockchain''~\cite{buterinBlockchainVotingOverrated2021}. This property, which blockchain achieves through eventual consistency consensus algorithms, ensures that ``everyone knows that everyone knows'' the same output and that it can not be changed. Unlike traditional systems that rely on a limited number of external auditors -- who often need access to sensitive information for integrity audits, such as the list of eligible voters, identifiers--blockchain systems achieve this layer of trust natively. The process is built into the protocol, eliminating the need for separate implementation and vastly increasing the number of auditors, as every node in the network participates in this role.

\textbf{Achieving universal verifiability is another area where blockchain excels}, as it often depends on the consistent communication between auditors to ensure the integrity of the system. In non-blockchain systems, the central server could potentially provide different outputs to each auditor, leading to undetected inconsistencies. In contrast, blockchain’s native support for common knowledge inherently prevents such discrepancies, as every node verifies the same data independently. This built-in communication and verification process not only ensures common knowledge but also guarantees universal verifiability, significantly enhancing the security and trustworthiness of the election process. McCorry et al.~\cite{mccorrySmartContractBoardroom2017} further highlight how using blockchain as a public bulletin board in e-voting protocols allows for independent verification by all nodes in the network, enhancing security and reducing the need for voters to perform complex verification themselves.

Even if the i-voting protocol is treated as a secure black-box, \textbf{the biggest vulnerability often lies in the process of preparing, printing, and distributing voter credentials}. Ensuring the integrity and confidentiality of these credentials is critical, as any compromise could undermine the entire voting process. This concern is well-documented in the CHVote protocol specification, which highlights two major trust assumptions in internet voting protocols: the integrity of the voting client and the reliability of the printing authority~\cite{haenniCHVoteProtocolSpecification2017}. Specifically, the printing authority (referred to as Identity Provider herein) receives highly sensitive information, such as credentials for vote submission and verification codes for candidates, which, if compromised, could allow votes to be cast on behalf of eligible voters without their knowledge. This situation is particularly problematic for abstaining voters, who would not detect any misuse. Furthermore, in the case that the printing authority accesses verification codes, it could bypass the cast-as-intended verification mechanism, making vote manipulation on the voting client undetectable. These scenarios underscore the printing authority as a potential single point of failure, requiring stringent measures to establish trust in this critical component for enhancing a protocols maturity.

\textbf{The high stakes associated with such operational vulnerabilities are not merely theoretical and have been demonstrated in prominent real-world deployments.} The significant failures of the New South Wales iVote system, which led to its discontinuation~\cite{haldermanNewSouthWales2015}, and the complex evolution of the SwissPost voting system following the discovery of critical flaws~\cite{hainesRunningRaceSwiss2022}, highlight the immense hurdles faced in high-stakes governmental elections. These incidents reinforce the importance of systematically analyzing the trade-offs between system complexity, practical usability, and the underlying trust model. A structured framework like the IVMF is designed to help decision-makers evaluate these dimensions, potentially identifying architectural weaknesses or excessive operational complexity before they contribute to catastrophic failures in a live election environment.

\textbf{Coercion resistance remains one of the most difficult properties to implement effectively.} Coercion resistance in online voting is a complex problem, especially in high-stakes elections where voters might be influenced by external forces. Our analysis reveals that protocols implement a spectrum of mitigation strategies rather than a single, perfect solution. Most protocols that address the issue, such as Decidim or Vecdoni, offer only weak resistance (CRES=1) through a re-voting mechanism, which provides plausible deniability but is not a robust defense against a determined coercer.

A stronger, procedural safeguard is the hybrid model (CRES=2) employed by the \textbf{Estonian i-voting system}, where an in-person paper vote can invalidate a previously cast electronic one. While this is an effective real-world deterrent, we classify it as a procedural mitigation rather than a feature of the protocol itself. The most advanced protocol in our analysis is \textbf{MACI}, which achieves strong cryptographic resistance (CRES=3) through a native key-switching mechanism that relies on a trusted coordinator. This makes it cryptographically difficult for a voter to prove their vote, representing the current state-of-the-art among deployed systems. Achieving fully trustless receipt-freeness remains a significant challenge, as evidenced by the fact that no protocol in our dataset achieved the highest possible score.

\section{Limitations and Future Work}\label{sec_8}
While this study provides a comprehensive analysis of various i-voting protocols, several limitations should be acknowledged, along with multiple avenues for future research. The analysis primarily focused on a select set of key security properties, but several other important aspects were not explored, such as ``Forward Election Integrity'', ``Everlasting Privacy'', and ``Quantum Resistance''. For addressing these aspects, a substantially more nuanced analysis with a quantification of higher granularity is needed, whereas its relevance for practical considerations of decision-makers is unclear. A more detailed examination of these properties in future work could provide a broader understanding of the security landscape in internet voting systems. Furthermore, the quantitative measurement of cryptographic assumptions is ripe for exploration. This could involve comparing the strength and implications of various cryptographic techniques, such as RSA factorization, elliptic curves, and lattice-based cryptography, to better assess the trade-offs between security and efficiency.

In this study, trust models were simplified to consensus protocols using a generic ``majority of many'' assumption. However, different consensus mechanisms, such as Proof-of-Work, Proof-of-Stake, and Distributed Randomness, have distinct trust assumptions that could be analyzed in greater detail in future research to provide a more nuanced view. Moreover, the voting client/device was excluded from the trust analysis, acknowledging its universal challenge across all i-voting solutions. While some protocols attempt to verify the correctness of the voting device using the Benaloh Challenge~\cite{benalohBallotCastingAssurance2007}, future work could incorporate such mechanisms into a more comprehensive trust assessment. In terms of cryptographic mechanisms, generally, systems that rely on techniques such as blind signatures were given the highest score for not depending on a trusted party. Potential attacks to these mechanisms, however, were not considered. These include, statistical or side-channel attacks, which could compromise these mechanisms. Future research should investigate these vulnerabilities to refine trust model scoring, reflecting a wider range of possible attack vectors. Additionally, the financial implications of deploying different internet voting protocols were not part of our analysis. Decision-makers balance security with budgetary constraints. Cost analyses overall and per ballot, thus, should be considered for verifying the maturity concept in e-voting protocol selection. Another aspect that was not addressed is the assumption that a significant number of voters will act as independent auditors by challenging the system’s integrity through mechanisms like ballot spoiling or zero-knowledge proof verification. Quantifying and systematizing the participation of voluntary auditors remains a challenge.

Furthermore, the current application of the IVMF relies on the manual assessment of protocol documentation. While we employed triangulation to mitigate bias, this approach is inherently subjective and labor-intensive. A significant methodological advancement for future work would be to formally model each protocol using a standardized specification language. Once modeled, their security properties and trust assumptions could be quantified through automated analysis, such as formal verification or large-scale simulations against defined adversary models. This approach would drastically improve the replicability and objectivity of the scoring process, eliminate the potential for human error in interpretation, and provide a more rigorous, verifiable foundation for the framework's inputs.

A key avenue for future work also lies in leveraging the inherent flexibility of the IVMF's weighting system. While this paper has focused on 'maturity' by heavily weighting Practical Usability, the framework can be easily adapted to generate alternative rankings based on different priorities. For instance, by setting the weight for Practical Usability to zero ($w_{PU}=0$), the IVMF transforms into a tool for purely theoretical comparison. Such a ranking would disregard real-world adoption and instead highlight protocols with the most robust cryptographic guarantees and elegant architectural designs (i.e., the highest Trust Model scores and lowest Complexity). This theoretical perspective could be invaluable for cryptographic researchers, protocol designers, or organizations looking to adopt a system based on its foundational security strengths rather than its current market presence.

\section{Conclusions}\label{sec_9}
The analysis reveals that many of the differences between i-voting systems lie in the instantiation and management of the bulletin board, a critical component for ensuring transparency and verifiability. Considerations of who controls it and how complex it is constructed were also key influences in protocols' maturity scores. \textbf{Blockchain-based internet voting systems have a distinct advantage in achieving common knowledge natively, thereby eliminating the need for external auditors.} The inherent immutability and transparency of blockchains ensure that every transaction is validated by multiple independent nodes, each of which acts as an auditor. This decentralized verification process is a significant strength of blockchain systems, as it enhances both individual and universal verifiability without relying on a centralized authority. However, the use of blockchain alone does not provide significant benefits in the mentioned properties, because they are achieved assuming that the majority of the network is honest and behaves correctly. A private blockchain is easier to manage, program, and resolve double-spending (double-voting) conflicts, but when the network is small and closed, it does not enhance the security model beyond what is achievable with well-managed traditional distributed systems.

\textbf{Authentication mechanisms vary depending on the use case}, reflecting the diverse requirements of different voting scenarios. Some systems employ decentralized trust models by distributing their components, yet, in practice, most of these components are controlled by a single entity. This centralization can undermine the intended trust model, as the system's security still hinges on the honesty and integrity of that single entity. \textbf{Introducing additional parties and performing secure multi-party computation is a straightforward way to improve a system's trust model.} However, if all components are ultimately controlled by one party, this approach offers little improvement. In such cases, the theoretical benefits of multi-party computation are not realized, as the trust remains concentrated in one entity. Generally, different protocols assume varying degrees of trustworthiness among system components, leading to a range of security solutions. Simpler security measures may suffice when certain parties are trusted, whereas protocols that assume all parties are untrustworthy must implement more robust, complex security mechanisms. While these elements are highly technical, the IVMF serves as an adaptable tool to help experts convey these trade-offs to decision-makers. \textbf{The framework's weighting system is not fixed; it is designed to be modified to reflect different strategic priorities. By adjusting the emphasis on practical usage versus theoretical security, decision-makers can generate rankings tailored to their specific needs, whether that is choosing a battle-tested system for a national election or identifying a cryptographically advanced protocol for future development.} In conclusion, while blockchain-based voting systems offer significant advantages in transparency and verifiability, the overall trust model of any voting system depends heavily on how components are managed and the assumptions made about the trustworthiness of various parties. The challenge for future developments will be to strike a balance between decentralization, usability, and security, ensuring that trust is distributed effectively without introducing unnecessary complexity.

\section*{Acknowledgments}
This work has been partially funded by the EU in the framework of the NGI TRUSTCHAIN project, grant number: \href{https://cordis.europa.eu/project/id/101093274}{101093274}, and funds of the faculty of Electronic Telecommunications and Informatics, Gdańsk University of~Technology.

\printbibliography

@article{baranskiPracticalIVotingStellar2020,
  title = {Practical {{I-Voting}} on {{Stellar Blockchain}}},
  author = {Barański, Stanisław and Szymański, Julian and Sobecki, Andrzej and Gil, David and Mora, Higinio},
  date = {2020-10-28},
  journaltitle = {Applied Sciences},
  shortjournal = {Applied Sciences},
  volume = {10},
  number = {21},
  pages = {7606},
  issn = {2076-3417},
  doi = {10.3390/app10217606},
  url = {https://www.mdpi.com/2076-3417/10/21/7606},
  urldate = {2022-07-19},
  langid = {english}
}

@article{parkGoingBadWorse2021,
  title = {Going from Bad to Worse: From Internet Voting to Blockchain Voting},
  author = {Park, Sunoo and Specter, Michael and Narula, Neha and Rivest, Ronald L},
  date = {2021},
  journaltitle = {Journal of Cybersecurity},
  shortjournal = {Journal of Cybersecurity},
  volume = {7},
  number = {1},
  pages = {tyaa025},
  publisher = {Oxford University Press},
  issn = {2057-2085}
}

@online{buterinBlockchainVotingOverrated2021,
  title = {Blockchain Voting Is Overrated among Uninformed People but Underrated among Informed People},
  author = {Buterin, Vitalik},
  date = {2021},
  url = {https://vitalik.ca/general/2021/05/25/voting2.html},
  urldate = {2022-08-22},
  organization = {Vitalik Buterin's website}
}

@article{jeffersonMythSecureBlockchain2019,
  title = {The {{Myth}} of “{{Secure}}” {{Blockchain Voting}}},
  author = {Jefferson, David},
  date = {2019},
  journaltitle = {Verified Voting. Available online: https://verifiedvoting. org/the-myth-of-secure-blockchain-voting/(accessed on 12 October 2020)},
  shortjournal = {Verified Voting. Available online: https://verifiedvoting. org/the-myth-of-secure-blockchain-voting/(accessed on 12 October 2020)}
}

@inproceedings{specterBallotBustedBlockchain2020,
  title = {The {{Ballot}} Is {{Busted Before}} the {{Blockchain}}: {{A Security Analysis}} of {{Voatz}}, the {{First Internet Voting Application Used}} in \{\vphantom\}{{US}}\vphantom\{\}. {{Federal Elections}}},
  author = {Specter, Michael A and Koppel, James and Weitzner, Daniel},
  date = {2020},
  pages = {1535--1553},
  eventtitle = {29th {{USENIX Security Symposium}} ({{USENIX Security}} 20)},
  isbn = {1-939133-17-3}
}

@online{buterinTrustModels2020,
  title = {Trust {{Models}}},
  author = {Buterin, Vitalik},
  date = {2020},
  url = {https://vitalik.ca/general/2020/08/20/trust.html},
  urldate = {2022-08-26},
  organization = {Vitalik Buterin's website}
}

@incollection{juelsCoercionresistantElectronicElections2010,
  title = {Coercion-Resistant Electronic Elections},
  booktitle = {Towards {{Trustworthy Elections}}},
  author = {Juels, Ari and Catalano, Dario and Jakobsson, Markus},
  date = {2010},
  pages = {37--63},
  publisher = {Springer}
}

@inproceedings{clarksonCivitasSecureVoting2008,
  title = {Civitas: {{Toward}} a Secure Voting System},
  author = {Clarkson, Michael R and Chong, Stephen and Myers, Andrew C},
  date = {2008},
  pages = {354--368},
  publisher = {IEEE},
  eventtitle = {2008 {{IEEE Symposium}} on {{Security}} and {{Privacy}} (Sp 2008)},
  isbn = {0-7695-3168-7}
}

@online{leeBlockchainbasedElectionsWould2018,
  title = {Blockchain-Based Elections Would Be a Disaster for Democracy},
  author = {Lee, Timothy B.},
  date = {2018-06-11},
  url = {https://arstechnica.com/tech-policy/2018/11/blockchain-based-elections-would-be-a-disaster-for-democracy/},
  urldate = {2022-08-30},
  langid = {american},
  organization = {Ars Technica}
}

@online{shanklandNoBlockchainIsn2018,
  title = {No, Blockchain Isn't the Answer to Our Voting System Woes},
  author = {Shankland, Stephen},
  date = {2018},
  url = {https://www.cnet.com/news/privacy/blockchain-isnt-answer-to-voting-system-woes/},
  urldate = {2022-08-30}
}

@online{mearianWhyBlockchainbasedVoting2019,
  title = {Why Blockchain-Based Voting Could Threaten Democracy},
  author = {Mearian, Lucas},
  date = {2019-08-12},
  url = {https://www.computerworld.com/article/3430697/why-blockchain-could-be-a-threat-to-democracy.html},
  urldate = {2022-08-30},
  langid = {english},
  organization = {Computerworld}
}

@software{ethereumfoundationMinimalAntiCollusionInfrastructure2022,
  title = {Minimal {{Anti-Collusion Infrastructure}}},
  author = {{Ethereum Foundation}},
  date = {2022-08-30T21:45:12Z},
  origdate = {2019-06-16T17:01:15Z},
  url = {https://github.com/privacy-scaling-explorations/maci/blob/9b1b1a631090ee89d2bc12f4bcef7763e42caef0/specs/01_introduction.md},
  urldate = {2022-09-02},
  organization = {Privacy \& Scaling Explorations (formerly known as appliedzkp)}
}

@article{halpernKnowledgeCommonKnowledge1990,
  title = {Knowledge and Common Knowledge in a Distributed Environment},
  author = {Halpern, Joseph and Moses, Yoram},
  date = {1990},
  journaltitle = {Journal of the ACM (JACM)},
  shortjournal = {Journal of the ACM (JACM)},
  volume = {37},
  number = {3},
  pages = {549--587},
  publisher = {ACM New York, NY, USA},
  issn = {0004-5411}
}

@incollection{mccorrySmartContractBoardroom2017,
  title = {A {{Smart Contract}} for {{Boardroom Voting}} with {{Maximum Voter Privacy}}},
  booktitle = {Financial {{Cryptography}} and {{Data Security}}},
  author = {McCorry, Patrick and Shahandashti, Siamak F. and Hao, Feng},
  editor = {Kiayias, Aggelos},
  date = {2017},
  volume = {10322},
  pages = {357--375},
  publisher = {Springer International Publishing},
  location = {Cham},
  doi = {10.1007/978-3-319-70972-7_20},
  url = {http://link.springer.com/10.1007/978-3-319-70972-7_20},
  urldate = {2023-01-20},
  isbn = {978-3-319-70971-0}
}

@inproceedings{adidaHeliosWebbasedOpenAudit2008,
  title = {Helios: {{Web-based Open-Audit Voting}}.},
  shorttitle = {Helios},
  booktitle = {{{USENIX}} Security Symposium},
  author = {Adida, Ben},
  date = {2008},
  volume = {17},
  pages = {335--348},
  url = {https://www.usenix.org/legacy/event/sec08/tech/full_papers/adida/adida.pdf},
  urldate = {2023-10-12}
}

@software{mccorryOpenVoteNetwork2023,
  title = {Open {{Vote Network}}},
  author = {McCorry, Patrick},
  date = {2023-10-10T14:03:12Z},
  origdate = {2016-08-23T11:49:13Z},
  url = {https://github.com/stonecoldpat/anonymousvoting},
  urldate = {2023-10-12}
}

@incollection{kiayiasSelftallyingElectionsPerfect2002,
  title = {Self-Tallying {{Elections}} and {{Perfect Ballot Secrecy}}},
  booktitle = {Public {{Key Cryptography}}},
  author = {Kiayias, Aggelos and Yung, Moti},
  editor = {Naccache, David and Paillier, Pascal},
  editora = {Goos, Gerhard and Hartmanis, Juris and Van Leeuwen, Jan},
  editoratype = {redactor},
  date = {2002},
  volume = {2274},
  pages = {141--158},
  publisher = {Springer Berlin Heidelberg},
  location = {Berlin, Heidelberg},
  doi = {10.1007/3-540-45664-3_10},
  url = {http://link.springer.com/10.1007/3-540-45664-3_10},
  urldate = {2024-01-02},
  isbn = {978-3-540-43168-8 978-3-540-45664-3}
}

@online{malavoltaHomomorphicTimeLockPuzzles2019,
  title = {Homomorphic {{Time-Lock Puzzles}} and {{Applications}}},
  author = {Malavolta, Giulio and Thyagarajan, Sri Aravinda Krishnan},
  date = {2019},
  number = {2019/635},
  url = {https://eprint.iacr.org/2019/635},
  urldate = {2024-01-02},
  pubstate = {prepublished}
}

@online{suwitoEvolutionBulletinBoard2021,
  title = {Evolution of {{Bulletin Board}} \& Its Application to {{E-Voting}} – {{A Survey}}},
  author = {Suwito, Misni Harjo and Ueshige, Yoshifumi and Sakurai, Kouichi},
  date = {2021},
  number = {2021/047},
  url = {https://eprint.iacr.org/2021/047},
  urldate = {2024-01-02},
  pubstate = {prepublished}
}

@article{williamsRemoteVotingAge2022,
  title = {Remote {{Voting}} in the {{Age}} of {{Cryptography}}},
  author = {Williams, Nathaniel},
  date = {2022-12-05},
  journaltitle = {MIT Computational Law Report},
  url = {https://law.mit.edu/pub/remotevotingintheageofcryptography/release/1},
  urldate = {2024-02-03},
  langid = {english}
}

@online{VocdoniIntroductionVocdoni,
  title = {Vocdoni Introduction | {{Vocdoni}} Developer Portal},
  url = {https://developer.vocdoni.io/protocol/overview},
  urldate = {2024-02-03},
  langid = {english}
}

@inproceedings{duenas-cidTrustDistrustEDemocracy2022,
  title = {Trust and {{Distrust}} in E-{{Democracy}}},
  booktitle = {{{DG}}.{{O}} 2022: {{The}} 23rd {{Annual International Conference}} on {{Digital Government Research}}},
  author = {Duenas-Cid, David and Janowski, Tomasz and Krimmer, Robert},
  date = {2022-09-14},
  series = {Dg.o 2022},
  pages = {486--488},
  publisher = {Association for Computing Machinery},
  location = {New York, NY, USA},
  doi = {10.1145/3543434.3543637},
  url = {https://doi.org/10.1145/3543434.3543637},
  urldate = {2024-02-16},
  isbn = {978-1-4503-9749-0}
}

@misc{mattbeckerProofVoteEndtoend2018,
  title = {Proof of {{Vote}}®: {{An}} End-to-End Verifiable Digital Voting Protocol Using Distributed Ledger Technology (Blockchain)},
  author = {{Matt Becker} and {Lauren Chandler} and {Liv Stromme} and {Patrick Hayes} and {Wesley Hedrick} and {Kurtis Jensen} and {Srini Kandikattu} and {Pete Martin} and {Scott Meier} and {Leopoldo Peña} and {Kun Peng} and {Aleck Silva-Pinto} and {Jeffrey Stern}},
  date = {2018},
  url = {https://github.com/votem/proof-of-vote/tree/master},
  urldate = {2024-02-20}
}

@incollection{cortierBeleniosSimplePrivate2019,
  title = {Belenios: {{A Simple Private}} and {{Verifiable Electronic Voting System}}},
  shorttitle = {Belenios},
  booktitle = {Foundations of {{Security}}, {{Protocols}}, and {{Equational Reasoning}}},
  author = {Cortier, Véronique and Gaudry, Pierrick and Glondu, Stéphane},
  editor = {Guttman, Joshua D. and Landwehr, Carl E. and Meseguer, José and Pavlovic, Dusko},
  date = {2019},
  volume = {11565},
  pages = {214--238},
  publisher = {Springer International Publishing},
  location = {Cham},
  doi = {10.1007/978-3-030-19052-1_14},
  url = {http://link.springer.com/10.1007/978-3-030-19052-1_14},
  urldate = {2024-03-07},
  isbn = {978-3-030-19051-4},
  langid = {english}
}

@incollection{neumannHolisticFrameworkEvaluation2014,
  title = {A Holistic Framework for the Evaluation of Internet Voting Systems},
  booktitle = {Design, {{Development}}, and {{Use}} of {{Secure Electronic Voting Systems}}},
  author = {Neumann, Stephan and Volkamer, Melanie},
  date = {2014},
  pages = {76--91},
  publisher = {IGI Global},
  url = {https://www.igi-global.com/chapter/a-holistic-framework-for-the-evaluation-of-internet-voting-systems/109229},
  urldate = {2024-03-26}
}

@article{alonsoEvotingSystemEvaluation2018,
  title = {E-Voting System Evaluation Based on the {{Council}} of {{Europe}} Recommendations: {{Helios Voting}}},
  shorttitle = {E-Voting System Evaluation Based on the {{Council}} of {{Europe}} Recommendations},
  author = {Alonso, Luis Panizo and Gasco, Mila and Del Blanco, David Y. Marcos and Alonso, José Á Hermida and Barrat, Jordi and Moreton, Héctor Aláiz},
  date = {2018},
  journaltitle = {IEEE Transactions on Emerging Topics in Computing},
  volume = {9},
  number = {1},
  pages = {161--173},
  publisher = {IEEE},
  url = {https://ieeexplore.ieee.org/abstract/document/8540060/},
  urldate = {2024-03-26}
}

@online{haenniCHVoteProtocolSpecification2017,
  title = {{{CHVote Protocol Specification}}},
  author = {Haenni, Rolf and Koenig, Reto E. and Locher, Philipp and Dubuis, Eric},
  date = {2017},
  number = {2017/325},
  url = {https://eprint.iacr.org/2017/325},
  urldate = {2024-04-08},
  pubstate = {prepublished}
}

@inproceedings{cortierSoKVerifiabilityNotions2016,
  title = {{{SoK}}: {{Verifiability Notions}} for {{E-Voting Protocols}}},
  shorttitle = {{{SoK}}},
  booktitle = {2016 {{IEEE Symposium}} on {{Security}} and {{Privacy}} ({{SP}})},
  author = {Cortier, Véronique and Galindo, David and Küsters, Ralf and Müller, Johannes and Truderung, Tomasz},
  date = {2016-05},
  pages = {779--798},
  issn = {2375-1207},
  doi = {10.1109/SP.2016.52},
  url = {https://ieeexplore.ieee.org/document/7546535},
  urldate = {2024-04-08},
  eventtitle = {2016 {{IEEE Symposium}} on {{Security}} and {{Privacy}} ({{SP}})}
}

@inproceedings{bernhardSoKComprehensiveAnalysis2015,
  title = {{{SoK}}: {{A Comprehensive Analysis}} of {{Game-Based Ballot Privacy Definitions}}},
  shorttitle = {{{SoK}}},
  booktitle = {2015 {{IEEE Symposium}} on {{Security}} and {{Privacy}}},
  author = {Bernhard, David and Cortier, Véronique and Galindo, David and Pereira, Olivier and Warinschi, Bogdan},
  date = {2015-05},
  pages = {499--516},
  issn = {2375-1207},
  doi = {10.1109/SP.2015.37},
  url = {https://ieeexplore.ieee.org/document/7163044},
  urldate = {2024-04-08},
  eventtitle = {2015 {{IEEE Symposium}} on {{Security}} and {{Privacy}}}
}

@inproceedings{canettiUniversallyComposableSecurity2001,
  title = {Universally Composable Security: A New Paradigm for Cryptographic Protocols},
  shorttitle = {Universally Composable Security},
  booktitle = {Proceedings 42nd {{IEEE Symposium}} on {{Foundations}} of {{Computer Science}}},
  author = {Canetti, R.},
  date = {2001-10},
  pages = {136--145},
  issn = {1552-5244},
  doi = {10.1109/SFCS.2001.959888},
  url = {https://ieeexplore.ieee.org/document/959888},
  urldate = {2024-04-08},
  eventtitle = {Proceedings 42nd {{IEEE Symposium}} on {{Foundations}} of {{Computer Science}}}
}

@online{larraiaSVoteControlComponents2022,
  title = {{{sVote}} with {{Control Components Voting Protocol}}. {{Computational Proof}} of {{Complete Verifiability}} and {{Privacy}}.},
  author = {Larraia, Enrique and Finogina, Tamara and Costa, Nuria},
  date = {2022},
  number = {2022/1509},
  url = {https://eprint.iacr.org/2022/1509},
  urldate = {2024-05-21},
  pubstate = {prepublished}
}

@online{GeneralFrameworkElectronic,
  title = {General {{Framework}} for {{Electronic Voting}} | {{Elections}} in {{Estonia}}},
  url = {https://www.valimised.ee/en/internet-voting/more-about-i-voting/general-framework-electronic-voting},
  urldate = {2024-05-22}
}

@inproceedings{chaidosBeleniosRFNoninteractiveReceiptFree2016,
  title = {{{BeleniosRF}}: {{A Non-interactive Receipt-Free Electronic Voting Scheme}}},
  shorttitle = {{{BeleniosRF}}},
  booktitle = {Proceedings of the 2016 {{ACM SIGSAC Conference}} on {{Computer}} and {{Communications Security}}},
  author = {Chaidos, Pyrros and Cortier, Véronique and Fuchsbauer, Georg and Galindo, David},
  date = {2016-10-24},
  pages = {1614--1625},
  publisher = {ACM},
  location = {Vienna Austria},
  doi = {10.1145/2976749.2978337},
  url = {https://dl.acm.org/doi/10.1145/2976749.2978337},
  urldate = {2024-07-16},
  eventtitle = {{{CCS}}'16: 2016 {{ACM SIGSAC Conference}} on {{Computer}} and {{Communications Security}}},
  isbn = {978-1-4503-4139-4},
  langid = {english}
}

@online{FreedomtoolOrg,
  title = {Freedomtool.Org},
  url = {https://freedomtool.com},
  urldate = {2024-07-16},
  langid = {english}
}

@incollection{cortierElectionVerifiabilityHelios2014,
  title = {Election {{Verifiability}} for {{Helios}} under {{Weaker Trust Assumptions}}},
  booktitle = {Computer {{Security}} - {{ESORICS}} 2014},
  author = {Cortier, Véronique and Galindo, David and Glondu, Stéphane and Izabachène, Malika},
  editor = {Kutyłowski, Mirosław and Vaidya, Jaideep},
  date = {2014},
  volume = {8713},
  pages = {327--344},
  publisher = {Springer International Publishing},
  location = {Cham},
  doi = {10.1007/978-3-319-11212-1_19},
  url = {http://link.springer.com/10.1007/978-3-319-11212-1_19},
  urldate = {2024-07-17},
  isbn = {978-3-319-11211-4},
  langid = {english}
}

@online{glaeserCicadaFrameworkPrivate2023,
  title = {Cicada: {{A}} Framework for Private Non-Interactive on-Chain Auctions and Voting},
  shorttitle = {Cicada},
  author = {Glaeser, Noemi and Seres, István András and Zhu, Michael and Bonneau, Joseph},
  date = {2023},
  number = {2023/1473},
  url = {https://eprint.iacr.org/2023/1473},
  urldate = {2024-07-18},
  pubstate = {prepublished}
}

@online{ghangasZkSnapScalableZero,
  title = {{{ZkSnap}}: {{Scalable}}, Zero Cost, End to End Private Voting for {{DAOs}}},
  author = {Ghangas, Rahul and Londhe, Shreyas},
  url = {https://aeriuslabs.org/whitepaper},
  urldate = {2024-07-23}
}

@online{AgoraBringingOur,
  title = {Agora: {{Bringing}} Our Voting Systems into the 21st Century},
    author = {Agora},
    date = {2018},
  url = {https://static1.squarespace.com/static/5b0be2f4e2ccd12e7e8a9be9/t/5f37eed8cedac41642edb534/1597501378925/Agora_Whitepaper.pdf}
}

@book{barandiaranDecidimTechnopoliticalNetwork2024,
  title = {Decidim, a {{Technopolitical Network}} for {{Participatory Democracy}}: {{Philosophy}}, {{Practice}} and {{Autonomy}} of a {{Collective Platform}} in the {{Age}} of {{Digital Intelligence}}},
  shorttitle = {Decidim, a {{Technopolitical Network}} for {{Participatory Democracy}}},
  author = {Barandiaran, Xabier E. and Calleja-López, Antonio and Monterde, Arnau and Romero, Carol},
  date = {2024},
  series = {{{SpringerBriefs}} in {{Political Science}}},
  publisher = {Springer Nature Switzerland},
  location = {Cham},
  doi = {10.1007/978-3-031-50784-7},
  url = {https://link.springer.com/10.1007/978-3-031-50784-7},
  urldate = {2024-07-23},
  isbn = {978-3-031-50783-0},
  langid = {english}
}

@online{plurality2023,
  title={Plurality: The Future of Collaborative Technology and Democracy},
  author={Weyl, E. Glen and Tang, Audrey and {the Plurality Community}},
  year={2023},
  url={https://github.com/pluralitybook/plurality/blob/main/contents/english},
  publisher={GitHub},
}

@misc{interviewer1BlockchainTechnologyNot2024,
  title = {Interview: Blockchain Technology Is Not the Solution to Internet Voting Problem},
author = {{Interviewee 1}},
  namea = {{Interviewee 1}},
  nameatype = {collaborator},
  date = {2024-03-13},
  langid = {english}
}

@online{DecidimUse,
urldate = {2024-09-25},
author = {Decidim},
  title = {Decidim in Use},
  url = {https://decidim.org/usedby/}
}

@online{IntroducingSnapshotsOnchain,
  title = {Introducing {{Snapshot}}'s Onchain Voting Protocol: {{Snapshot X}}},
  shorttitle = {Introducing {{Snapshot}}'s Onchain Voting Protocol},
  url = {https://snapshot.mirror.xyz/F0wSmh8LROHhLYGQ7VG6VEG1_L8_IQk8eC9U7gFwep0},
  urldate = {2024-09-25}
}

@report{harrisonStudyEVotingPractices2023,
  title = {Study on {{E-Voting Practices}} in the {{EU}}},
  author = {Harrison, Sarah and Vives, Elisabet and Gentile, Giulia and Bruter, Michael},
  date = {2023-12},
  url = {https://www.lse.ac.uk/business/consulting/reports/study-on-e-voting-practices-in-the-eu.aspx},
  urldate = {2024-09-30},
  langid = {british}
}

@article{achenbachImprovedCoercionResistantElectronic2015,
  title = {Improved {{Coercion-Resistant}} Electronic Elections through Deniable {{Re-Voting}}},
  author = {Achenbach, Dirk and Kempka, Carmen and Löwe, Bernhard and Müller-Quade, Jörn},
  date = {2015-08},
  journaltitle = {USENIX Journal of Election Technology and Systems (JETS)},
  volume = {3},
  pages = {26--45},
  publisher = {USENIX Association},
  location = {Washington, D.C.},
  issn = {2328-2797},
  url = {https://www.usenix.org/conference/jets15/workshop-program/presentation/achenbach}
}

@article{buterinEthereumWhitePaper2013,
  title = {Ethereum White Paper},
  author = {Buterin, Vitalik},
  date = {2013},
  journaltitle = {GitHub repository},
  volume = {1},
  pages = {22--23},
  url = {https://static.peng37.com/ethereum_whitepaper_laptop_3.pdf},
  urldate = {2024-09-30}
}

@article{nakamotoBitcoinPeerpeerElectronic2008,
  title = {Bitcoin: {{A}} Peer-to-Peer Electronic Cash System},
  author = {Nakamoto, Satoshi},
  date = {2008},
  journaltitle = {Decentralized Business Review},
  shortjournal = {Decentralized Business Review},
  pages = {21260}
}

@online{CybervoteProgrammeResearch2003,
  title = {Cybervote {{Programme}} for Research, Technological Development and Demonstration on a "{{User-friendly}} Information Society},
  date = {2003},
  url = {https://cordis.europa.eu/programme/id/FP5-IST},
  urldate = {2024-09-27},
  langid = {english},
  organization = {CORDIS | European Commission}
}

@inproceedings{maatenRemoteEvotingEstonian2004,
  title = {Towards Remote E-Voting: {{Estonian}} Case},
  shorttitle = {Towards Remote E-Voting},
  booktitle = {Electronic Voting in {{Europe-Technology}}, Law, Politics and Society, Workshop of the {{ESF TED}} Programme Together with {{GI}} and {{OCG}}},
  author = {Maaten, Epp},
  date = {2004},
  pages = {83--90},
  publisher = {Gesellschaft für Informatik eV},
  url = {https://dl.gi.de/items/cc43f6bb-eb9d-473a-bafc-6dcf97db7e46},
  urldate = {2024-09-27}
}

@inproceedings{cranorSensusSecurityconsciousElectronic1997,
  title = {Sensus: A Security-Conscious Electronic Polling System for the {{Internet}}},
  shorttitle = {Sensus},
  booktitle = {Proceedings of the {{Thirtieth Hawaii International Conference}} on {{System Sciences}}},
  author = {Cranor, L.F. and Cytron, R.K.},
  date = {1997-01},
  volume = {3},
  pages = {561-570 vol.3},
  issn = {1060-3425},
  doi = {10.1109/HICSS.1997.661700},
  url = {https://ieeexplore.ieee.org/document/661700/?arnumber=661700},
  urldate = {2024-09-27},
  eventtitle = {Proceedings of the {{Thirtieth Hawaii International Conference}} on {{System Sciences}}}
}

@article{benalohVerifiableSecretBallotElections1987,
  title = {Verifiable {{Secret-Ballot Elections}}},
  author = {Benaloh, Josh},
  date = {1987-09-01},
  url = {https://www.microsoft.com/en-us/research/publication/verifiable-secret-ballot-elections/},
  urldate = {2023-02-09},
  langid = {american}
}

@inproceedings{benalohReceiptfreeSecretballotElections1994,
  title = {Receipt-Free Secret-Ballot Elections (Extended Abstract)},
  booktitle = {Proceedings of the Twenty-Sixth Annual {{ACM}} Symposium on {{Theory}} of Computing  - {{STOC}} '94},
  author = {Benaloh, Josh and Tuinstra, Dwight},
  date = {1994},
  pages = {544--553},
  publisher = {ACM Press},
  location = {Montreal, Quebec, Canada},
  doi = {10.1145/195058.195407},
  url = {http://portal.acm.org/citation.cfm?doid=195058.195407},
  urldate = {2024-09-27},
  eventtitle = {The Twenty-Sixth Annual {{ACM}} Symposium},
  isbn = {978-0-89791-663-9},
  langid = {english}
}

@inproceedings{krimmerBitsPaperComparing2005,
  title = {Bits or {{Paper}}? {{Comparing Remote Electronic Voting}} to {{Postal Voting}}.},
  shorttitle = {Bits or {{Paper}}?},
  booktitle = {{{EGOV}} ({{Workshops}} and {{Posters}})},
  author = {Krimmer, Robert and Volkamer, Melanie},
  date = {2005},
  pages = {225--232},
  publisher = {Citeseer},
  url = {https://citeseerx.ist.psu.edu/document?repid=rep1&type=pdf&doi=b675ceb003b719aaa5bff885347162cb84c4f76d},
  urldate = {2024-09-27}
}

@book{staveleyGreekRomanVoting1972,
  title = {Greek and {{Roman Voting}} and {{Elections}}},
  author = {Staveley, E. S.},
  date = {1972},
  publisher = {Thames \& Hudson},
  location = {[London]}
}

@incollection{fujiokaPracticalSecretVoting1993,
  title = {A Practical Secret Voting Scheme for Large Scale Elections},
  booktitle = {Advances in {{Cryptology}} — {{AUSCRYPT}} '92},
  author = {Fujioka, Atsushi and Okamoto, Tatsuaki and Ohta, Kazuo},
  editor = {Seberry, Jennifer and Zheng, Yuliang},
  editora = {Goos, Gerhard and Hartmanis, Juris},
  editoratype = {redactor},
  date = {1993},
  volume = {718},
  pages = {244--251},
  publisher = {Springer Berlin Heidelberg},
  location = {Berlin, Heidelberg},
  doi = {10.1007/3-540-57220-1_66},
  url = {http://link.springer.com/10.1007/3-540-57220-1_66},
  urldate = {2024-09-27},
  isbn = {978-3-540-57220-6}
}

@online{CaseStudiesMACI,
  title = {Case {{Studies}} | {{MACI}}},
  url = {https://maci.pse.dev/docs/use-cases/public-goods-funding/quadratic-funding/case-studies},
  urldate = {2024-10-02},
  langid = {english}
}

@online{WelcomeSnapshotDocs2024,
  title = {Welcome to {{Snapshot}} Docs! | Snapshot},
  date = {2024-01-18},
  url = {https://docs.snapshot.org},
  urldate = {2024-09-25},
  langid = {english}
}

@incollection{lichtIvoteNotIvote2021,
  title = {To I-Vote or {{Not}} to i-Vote: {{Drivers}} and {{Barriers}} to the {{Implementation}} of {{Internet Voting}}},
  shorttitle = {To I-Vote or {{Not}} to i-Vote},
  booktitle = {Electronic {{Voting}}},
  author = {Licht, Nathan and Duenas-Cid, David and Krivonosova, Iuliia and Krimmer, Robert},
  editor = {Krimmer, Robert and Volkamer, Melanie and Duenas-Cid, David and Kulyk, Oksana and Rønne, Peter and Solvak, Mihkel and Germann, Micha},
  date = {2021},
  volume = {12900},
  pages = {91--105},
  publisher = {Springer International Publishing},
  location = {Cham},
  doi = {10.1007/978-3-030-86942-7_7},
  url = {https://link.springer.com/10.1007/978-3-030-86942-7_7},
  urldate = {2024-02-16},
  isbn = {978-3-030-86941-0},
  langid = {english}
}

@online{electrustAnxietiesInternetVoting2021,
  title = {Anxieties over Internet Voting Reflect Wider Social Concerns},
  author = {{ELECTRUST}},
  date = {2021},
  url = {https://cordis.europa.eu/article/id/448713-anxieties-over-internet-voting-reflect-wider-social-concerns},
  urldate = {2024-10-04},
  organization = {Dynamics of Trust and Distrust Creation in Internet Voting}
}

@article{khoReviewCryptographicElectronic2022,
  title = {A Review of Cryptographic Electronic Voting},
  author = {Kho, Yun-Xing and Heng, Swee-Huay and Chin, Ji-Jian},
  date = {2022},
  journaltitle = {Symmetry},
  volume = {14},
  number = {5},
  pages = {858},
  publisher = {MDPI},
  url = {https://www.mdpi.com/2073-8994/14/5/858},
  urldate = {2024-09-06}
}

@article{neumannSecIVoQuantitativeSecurity2016,
  title = {{{SecIVo}}: A Quantitative Security Evaluation Framework for Internet Voting Schemes},
  shorttitle = {{{SecIVo}}},
  author = {Neumann, Stephan and Volkamer, Melanie and Budurushi, Jurlind and Prandini, Marco},
  date = {2016-08},
  journaltitle = {Annals of Telecommunications},
  shortjournal = {Ann. Telecommun.},
  volume = {71},
  number = {7-8},
  pages = {337--352},
  issn = {0003-4347, 1958-9395},
  doi = {10.1007/s12243-016-0520-0},
  url = {http://link.springer.com/10.1007/s12243-016-0520-0},
  urldate = {2024-09-30},
  langid = {english}
}

@online{Snapshot2024,
  title = {Snapshot {{X}}},
  date = {2024-09-20},
  url = {https://docs.snapshot.box},
  urldate = {2024-09-26},
  langid = {english}
}

@online{VotecasterFarcasterPolls,
  title = {Votecaster - {{Farcaster}} Polls},
  url = {https://farcaster.vote},
  urldate = {2024-09-26},
  langid = {english}
}

@online{teamSwissbasedAgoraRecords2018,
  title = {Swiss-Based {{Agora}} Records First Government Election on Blockchain as Accredited Observer in {{Sierra}}…},
  author = {Team, Agora Blockchain},
  date = {2018-03-22T05:10:05},
  url = {https://medium.com/agorablockchain/swiss-based-agora-powers-worlds-first-ever-blockchain-elections-in-sierra-leone-984dd07a58ee},
  urldate = {2024-10-10},
  langid = {english},
  organization = {AgoraBlockchain}
}

@online{VotemCaseStudies,
  title = {Votem: {{Case Studies}}},
  url = {https://votem.com/case-studies/},
  urldate = {2024-10-10},
  langid = {american},
  organization = {Online Voting with Votem®}
}

@article{adidaElectingUniversityPresident2009,
  title = {Electing a University President Using Open-Audit Voting: {{Analysis}} of Real-World Use of {{Helios}}},
  shorttitle = {Electing a University President Using Open-Audit Voting},
  author = {Adida, Ben and De Marneffe, Olivier and Pereira, Olivier and Quisquater, Jean-Jacques},
  date = {2009},
  journaltitle = {EVT/WOTE},
  volume = {9},
  number = {10},
  url = {https://www.usenix.org/event/evtwote09/tech/full_papers/adida-helios.pdf},
  urldate = {2024-10-10}
}

@inproceedings{cortierFeaturesUsageBelenios2022,
  title = {Features and Usage of {{Belenios}} in 2022},
  booktitle = {The {{International Conference}} for {{Electronic Voting}} ({{E-Vote-ID}} 2022)},
  author = {Cortier, Véronique and Gaudry, Pierrick and Glondu, Stéphane},
  date = {2022},
  publisher = {University of Tartu Press},
  url = {https://inria.hal.science/hal-03791757/},
  urldate = {2024-10-10}
}

@inproceedings{serdultFifteenYearsInternet2015,
  title = {Fifteen Years of Internet Voting in {{Switzerland}} [{{History}}, {{Governance}} and {{Use}}]},
  booktitle = {2015 {{Second International Conference}} on {{eDemocracy}} \& {{eGovernment}} ({{ICEDEG}})},
  author = {Serdult, Uwe and Germann, Micha and Mendez, Fernando and Portenier, Alicia and Wellig, Christoph},
  date = {2015-04},
  pages = {126--132},
  doi = {10.1109/ICEDEG.2015.7114482},
  url = {https://ieeexplore.ieee.org/document/7114482/?arnumber=7114482},
  urldate = {2024-10-14}
}

@online{fchEVoting,
  title = {E-{{Voting}}},
  author = {FCh, Swiss Federal Chancellery},
  url = {https://www.bk.admin.ch/bk/en/home/politische-rechte/e-voting.html},
  urldate = {2024-10-14},
  langid = {english}
}

@incollection{haenniCHVoteSixteenBest2020,
  title = {{{CHVote}}: {{Sixteen Best Practices}} and {{Lessons Learned}}},
  shorttitle = {{{CHVote}}},
  booktitle = {Electronic {{Voting}}},
  author = {Haenni, Rolf and Dubuis, Eric and Koenig, Reto E. and Locher, Philipp},
  editor = {Krimmer, Robert and Volkamer, Melanie and Beckert, Bernhard and Küsters, Ralf and Kulyk, Oksana and Duenas-Cid, David and Solvak, Mihkel},
  date = {2020},
  volume = {12455},
  pages = {95--111},
  publisher = {Springer International Publishing},
  location = {Cham},
  doi = {10.1007/978-3-030-60347-2_7},
  url = {https://link.springer.com/10.1007/978-3-030-60347-2_7},
  urldate = {2024-10-14},
  isbn = {978-3-030-60346-5},
  langid = {english}
}

@online{scytlHowHoldFully2023,
  title = {How to {{Hold Fully Digital Elections}}},
  author = {Scytl},
  date = {2023-10-25T12:24:36},
  url = {https://medium.com/edge-elections/how-to-hold-fully-digital-elections-46e23f4b44c9},
  urldate = {2024-10-10},
  langid = {english},
  organization = {EDGE Elections}
}

@online{ESADECaseScytl2017,
  title = {{{ESADE}} Case on {{Scytl}} Receives {{European}} Recognition - {{Esade}}},
  date = {2017-06-12},
  url = {https://www.esade.edu/en/news/esade-case-on-scytl-receives-european-recognition},
  urldate = {2024-10-10},
  langid = {english}
}

@online{SCYTLELECTIONTECHNOLOGIES,
  title = {{{Scytl Election Technologies SL}}},
  url = {https://ec.europa.eu/info/funding-tenders/opportunities/portal/screen/how-to-participate/org-details/892346166},
  urldate = {2024-10-10}
}

@online{EDemocracyOpenData2024,
  title = {E-{{Democracy}} \& Open Data},
  date = {2024-06-09T10:43:23+03:00},
  url = {https://e-estonia.com/solutions/e-governance/e-democracy/},
  urldate = {2024-10-10},
  langid = {english},
  organization = {e-Estonia}
}

@online{ElectionsEstonia,
  title = {Elections in {{Estonia}}},
  url = {https://www.valimised.ee/en},
  urldate = {2024-10-10}
}

@article{hobbis_voter_2017,
	title = {Voter {Integrity}, {Trust} and the {Promise} of {Digital} {Technologies}: {Biometric} {Voter} {Registration} in {Solomon} {Islands}},
	volume = {27},
	issn = {0066-4677},
	shorttitle = {Voter {Integrity}, {Trust} and the {Promise} of {Digital} {Technologies}},
	url = {https://doi.org/10.1080/00664677.2017.1324287},
	doi = {10.1080/00664677.2017.1324287},
	number = {2},
	urldate = {2024-10-22},
	journal = {Anthropological Forum},
	author = {Hobbis, Stephanie Ketterer and Hobbis, Geoffrey},
	month = apr,
	year = {2017},
	note = {Publisher: Routledge
\_eprint: https://doi.org/10.1080/00664677.2017.1324287},
	pages = {114--134},
}

@misc{freitag_new_2022,
	title = {A new {Privacy} {Preserving} and {Scalable} {Revocation} {Method} for {Self} {Sovereign} {Identity} -- {The} {Perfect} {Revocation} {Method} does not exist yet},
	url = {http://arxiv.org/abs/2211.13041},
	doi = {10.48550/arXiv.2211.13041},
	urldate = {2023-12-21},
	publisher = {arXiv},
	author = {Freitag, Andreas},
	month = nov,
	year = {2022},
	note = {arXiv:2211.13041 [cs]},
}

@misc{european_parliament_regulation_2024,
	title = {Regulation ({EU}) 2024/1183 amending {Regulation} ({EU}) {No} 910/2014 as regards establishing the {European} {Digital} {Identity} {Framework}},
	url = {https://eur-lex.europa.eu/eli/reg/2024/1183/oj},
	language = {en},
	urldate = {2024-05-19},
	author = {{European Parliament} and {European Council}},
	month = apr,
	year = {2024},
	note = {Doc ID: 32024R1183
Doc Sector: 3
Doc Title: Regulation (EU) 2024/1183 of the European Parliament and of the Council of 11 April 2024 amending Regulation (EU) No 910/2014 as regards establishing the European Digital Identity Framework
Doc Type: R
Usr\_lan: en},
}

@article{barry_state---art_2022,
	title = {State-of-the-art literature review methodology: {A} six-step approach for knowledge synthesis},
	volume = {11},
	issn = {2212-277X},
	shorttitle = {State-of-the-art literature review methodology},
	url = {https://pmejournal.org/articles/10.1007/S40037-022-00725-9},
	doi = {10.1007/S40037-022-00725-9},
	language = {en-US},
	number = {5},
	urldate = {2023-12-15},
	author = {Barry, Erin S. and Merkebu, Jerusalem and Varpio, Lara},
	month = sep,
	year = {2022},
	note = {Number: 5
Publisher: Ubiquity Press},
	pages = {1},
}

@article{barry_understanding_2022,
	title = {Understanding {State}-of-the-{Art} {Literature} {Reviews}},
	volume = {14},
	issn = {1949-8349},
	url = {https://www.ncbi.nlm.nih.gov/pmc/articles/PMC9765914/},
	doi = {10.4300/JGME-D-22-00705.1},
	number = {6},
	urldate = {2023-08-20},
	journal = {Journal of Graduate Medical Education},
	author = {Barry, Erin S. and Merkebu, Jerusalem and Varpio, Lara},
	month = dec,
	year = {2022},
	pmid = {36591417},
	pmcid = {PMC9765914},
	pages = {659--662},
}

@misc{tan_open_2023,
	title = {Open {Problems} in {DAOs}},
	url = {http://arxiv.org/abs/2310.19201},
	doi = {10.48550/arXiv.2310.19201},
	urldate = {2024-03-05},
	publisher = {arXiv},
	author = {Tan, Joshua Z. and Merk, Tara and Hubbard, Sarah and Oak, Eliza R. and Pirovich, Joni and Rennie, Ellie and Hoefer, Rolf and Zargham, Michael and Potts, Jason and Berg, Chris and Youngblom, Reuben and De Filippi, Primavera and Frey, Seth and Strnad, Jeff and Mannan, Morshed and Nabben, Kelsie and Elrifai, Silke Noa and Hartnell, Jake and Hill, Benjamin Mako and Maddox, Alexia and Lim, Woojin and South, Tobin and Juels, Ari and Boneh, Dan},
	month = oct,
	year = {2023},
	note = {arXiv:2310.19201 [cs]
version: 1},
}

@article{tammpuu_estonian_2022,
	title = {Estonian {E}-{Residency} and {Conceptions} of {Platformbased} {State}-{Individual} {Relationship}},
	copyright = {© 2022. This work is published under http://www.kirj.ee/13169/ (the “License”). Notwithstanding the ProQuest Terms and Conditions, you may use this content in accordance with the terms of the License.},
	issn = {14060922},
	url = {https://www.proquest.com/docview/2737499954/abstract/4AD688E366784FCAPQ/1},
	doi = {10.3176/tr.2022.1.01},
	language = {English},
	urldate = {2024-01-12},
	author = {Tammpuu, Piia 1 and Masso, Anu 1 and Ibrahimi, Mergime 2 and Abaku, Tam 2 1 University of Tartu 2 Tallinn University of Technology},
	year = {2022},
	note = {Num Pages: 3-21
Publisher: Teaduste Akadeemia Kirjastus (Estonian Academy Publishers)},
	pages = {3--21},
}

@article{fitzpatrick_high_2022,
	title = {“{The} {High} {Mass} of {Democracy}” —{Why} {Germany} {Remains} {Aloof} to the {Idea} of {Electronic} {Voting}},
	volume = {4},
	issn = {2673-3145},
	url = {https://www.frontiersin.org/journals/political-science/articles/10.3389/fpos.2022.876476/full},
	doi = {10.3389/fpos.2022.876476},
	language = {English},
	urldate = {2024-10-22},
	journal = {Frontiers in Political Science},
	author = {Fitzpatrick, Jasmin and Jöst, Paula},
	month = jul,
	year = {2022},
	note = {Publisher: Frontiers},
}

@article{beck_governance_2018,
	title = {Governance in the {Blockchain} {Economy}:  {A} {Framework} and {Research} {Agenda}},
	volume = {19},
	issn = {1536-9323},
	shorttitle = {Governance in the {Blockchain} {Economy}},
	url = {https://aisel.aisnet.org/jais/vol19/iss10/1},
	number = {10},
	journal = {Journal of the Association for Information Systems},
	author = {Beck, Roman and Müller-Bloch, Christoph and King, John},
	month = oct,
	year = {2018},
}

@misc{messias_understanding_2024,
	title = {Understanding {Blockchain} {Governance}: {Analyzing} {Decentralized} {Voting} to {Amend} {DeFi} {Smart} {Contracts}},
	shorttitle = {Understanding {Blockchain} {Governance}},
	url = {http://arxiv.org/abs/2305.17655},
	doi = {10.48550/arXiv.2305.17655},
	urldate = {2024-03-05},
	publisher = {arXiv},
	author = {Messias, Johnnatan and Pahari, Vabuk and Chandrasekaran, Balakrishnan and Gummadi, Krishna P. and Loiseau, Patrick},
	month = jan,
	year = {2024},
	note = {arXiv:2305.17655 [cs]},
}

@article{johansson_roadmap_2019,
	title = {Roadmap for a communication maturity index for organizations—{Theorizing}, analyzing and developing communication value},
	volume = {45},
	issn = {0363-8111},
	url = {https://www.sciencedirect.com/science/article/pii/S0363811118304065},
	doi = {10.1016/j.pubrev.2019.05.012},
	number = {4},
	urldate = {2024-03-03},
	journal = {Public Relations Review},
	author = {Johansson, Catrin and Grandien, Christina and Strandh, Kicki},
	month = nov,
	year = {2019},
	pages = {101791},
}

@misc{ministers_deputies_recommendation_2017,
	title = {Recommendation {CM}/{Rec}(2017)5 on standards for e-voting},
	url = {https://rm.coe.int/0900001680726f6f},
	author = {{Ministers' Deputies}},
	month = jul,
	year = {2017},
}

@book{dener_govtech_2021,
	title = {{GovTech} {Maturity} {Index}: {The} {State} of {Public} {Sector} {Digital} {Transformation}},
	isbn = {978-1-4648-1765-6},
	shorttitle = {{GovTech} {Maturity} {Index}},
	language = {en},
	publisher = {World Bank Publications},
	author = {Dener, Cem and Nii-Aponsah, Hubert and Ghunney, Love E. and Johns, Kimberly D.},
	month = oct,
	year = {2021},
	note = {Google-Books-ID: pvdIEAAAQBAJ},
}

@book{creswell_designing_2017,
	title = {Designing and {Conducting} {Mixed} {Methods} {Research}},
	isbn = {978-1-4833-4698-4},
	language = {en},
	publisher = {SAGE Publications},
	author = {Creswell, John W. and Clark, Vicki L. Plano},
	month = aug,
	year = {2017},
	note = {Google-Books-ID: eTwmDwAAQBAJ},
}

@article{knosp_research_2018,
	title = {Research {IT} maturity models for academic health centers: {Early} development and initial evaluation},
	volume = {2},
	issn = {2059-8661},
	shorttitle = {Research {IT} maturity models for academic health centers},
	url = {https://www.cambridge.org/core/journals/journal-of-clinical-and-translational-science/article/research-it-maturity-models-for-academic-health-centers-early-development-and-initial-evaluation/F4FAF953A4C2ACEC9DE8EB8EFBBBD9F3},
	doi = {10.1017/cts.2018.339},
	language = {en},
	number = {5},
	urldate = {2024-03-03},
	journal = {Journal of Clinical and Translational Science},
	author = {Knosp, Boyd M. and Barnett, William K. and Anderson, Nicholas R. and Embi, Peter J.},
	month = oct,
	year = {2018},
	pages = {289--294},
}

@article{van_ede_assembling_2024,
	title = {Assembling a population health management maturity index using a {Delphi} method},
	volume = {24},
	issn = {1472-6963},
	url = {https://doi.org/10.1186/s12913-024-10572-5},
	doi = {10.1186/s12913-024-10572-5},
	number = {1},
	urldate = {2024-03-03},
	journal = {BMC Health Services Research},
	author = {van Ede, A. F. T. M. and Stein, K. V. and Bruijnzeels, M. A.},
	month = jan,
	year = {2024},
	pages = {110},
}

@article{ding_desci_2022,
	title = {{DeSci} {Based} on {Web3} and {DAO}: {A} {Comprehensive} {Overview} and {Reference} {Model}},
	volume = {9},
	issn = {2329-924X},
	shorttitle = {{DeSci} {Based} on {Web3} and {DAO}},
	url = {https://ieeexplore.ieee.org/document/9906878},
	doi = {10.1109/TCSS.2022.3204745},
	number = {5},
	urldate = {2024-02-11},
	journal = {IEEE Transactions on Computational Social Systems},
	author = {Ding, Wenwen and Hou, Jiachen and Li, Juanjuan and Guo, Chao and Qin, Jirong and Kozma, Robert and Wang, Fei-Yue},
	month = oct,
	year = {2022},
	note = {Conference Name: IEEE Transactions on Computational Social Systems},
	pages = {1563--1573},
}

@article{nolan_managing_1973,
	title = {Managing the computer resource: a stage hypothesis},
	volume = {16},
	issn = {0001-0782},
	shorttitle = {Managing the computer resource},
	url = {https://dl.acm.org/doi/10.1145/362280.362284},
	doi = {10.1145/362280.362284},
	number = {7},
	urldate = {2024-03-04},
	journal = {Communications of the ACM},
	author = {Nolan, Richard L.},
	month = jul,
	year = {1973},
	pages = {399--405},
}

@misc{european_commission_directive_1999,
	title = {Directive 1999/93/{EC} of the {European} {Parliament} and of the {Council} of 13 {December} 1999 on a {Community} framework for electronic signatures},
	volume = {013},
	url = {http://data.europa.eu/eli/dir/1999/93/oj/eng},
	language = {en},
	urldate = {2023-11-30},
	author = {{European Commission}},
	month = dec,
	year = {1999},
}

@article{bowen_document_2009,
	title = {Document {Analysis} as a {Qualitative} {Research} {Method}},
	volume = {9},
	issn = {1443-9883},
	doi = {10.3316/QRJ0902027},
	number = {2},
	journal = {Qualitative Research Journal},
	author = {Bowen, Glenn A.},
	year = {2009},
	pages = {27--40},
}

@article{leibetseder_critical_2011,
	title = {A {Critical} {Review} on the {Concept} of {Social} {Technology}},
	volume = {1},
	journal = {Social Technologies},
	author = {Leibetseder, Bettina},
	month = jan,
	year = {2011},
	pages = {7--24},
}

@book{creswell_designing_2011,
	address = {Los Angeles and London},
	edition = {2nd ed.},
	title = {Designing and conducting mixed methods research: {Advanced} {Nursing} {Practice}},
	isbn = {978-1-4129-7517-9},
	publisher = {SAGE},
	author = {Creswell, John W. and Plano Clark, Vicki L.},
	year = {2011},
}

@incollection{datta_paradigm_2006,
	address = {London [etc.]},
	title = {Paradigm {Wars}: {A} {Basis} for {Peaceful} {Coexistence} and {Beyond}},
	isbn = {978-1-4129-1163-4},
	booktitle = {Mixed methods},
	publisher = {SAGE},
	author = {Datta, Lois-ellin},
	editor = {Bryman, Alan},
	year = {2006},
	pages = {31--52},
}

@article{kitchenham_guidelines_2007,
	title = {Guidelines for performing {Systematic} {Literature} {Reviews} in {Software} {Engineering}},
	volume = {2},
	author = {Kitchenham, Barbara and Charters, Stuart},
	month = jan,
	year = {2007},
}

@misc{biedermann_evaluating_2024,
	title = {Evaluating {Progress} in {Web3} {Grants}: {Introducing} the {Grant} {Maturity} {Index}},
	shorttitle = {Evaluating {Progress} in {Web3} {Grants}},
	url = {http://arxiv.org/abs/2410.19828},
	doi = {10.48550/arXiv.2410.19828},
	urldate = {2024-10-30},
	publisher = {arXiv},
	author = {Biedermann, Ben and Gibrel, Fahima},
	month = oct,
	year = {2024},
	note = {arXiv:2410.19828},
}

@book{oecd_handbook_2008,
	address = {Paris},
	title = {Handbook on {Constructing} {Composite} {Indicators}: {Methodology} and {User} {Guide}},
	shorttitle = {Handbook on {Constructing} {Composite} {Indicators}},
	url = {https://www.oecd-ilibrary.org/economics/handbook-on-constructing-composite-indicators-methodology-and-user-guide_9789264043466-en},
	language = {en},
	urldate = {2024-08-02},
	publisher = {Organisation for Economic Co-operation and Development},
	author = {{OECD}},
	year = {2008},
}

@online{ben-sassonAuroraTransparentSuccinct2018,
  title = {Aurora: {{Transparent Succinct Arguments}} for {{R1CS}}},
  shorttitle = {Aurora},
  author = {Ben-Sasson, Eli and Chiesa, Alessandro and Riabzev, Michael and Spooner, Nicholas and Virza, Madars and Ward, Nicholas P.},
  date = {2018},
  number = {2018/828},
  url = {https://eprint.iacr.org/2018/828},
  urldate = {2024-11-04},
  pubstate = {prepublished}
}

@article{hoffmanTrustSecurityExpanded2006,
  title = {Trust beyond Security: An Expanded Trust Model},
  shorttitle = {Trust beyond Security},
  author = {Hoffman, Lance J. and Lawson-Jenkins, Kim and Blum, Jeremy},
  date = {2006-07},
  journaltitle = {Communications of the ACM},
  shortjournal = {Commun. ACM},
  volume = {49},
  number = {7},
  pages = {94--101},
  issn = {0001-0782, 1557-7317},
  doi = {10.1145/1139922.1139924},
  url = {https://dl.acm.org/doi/10.1145/1139922.1139924},
  urldate = {2024-10-30},
  langid = {english}
}

@article{choSurveyTrustModeling2015,
  title = {A {{Survey}} on {{Trust Modeling}}},
  author = {Cho, Jin-Hee and Chan, Kevin and Adali, Sibel},
  date = {2015-11-21},
  journaltitle = {ACM Computing Surveys},
  shortjournal = {ACM Comput. Surv.},
  volume = {48},
  number = {2},
  pages = {1--40},
  issn = {0360-0300, 1557-7341},
  doi = {10.1145/2815595},
  url = {https://dl.acm.org/doi/10.1145/2815595},
  urldate = {2024-10-30},
  langid = {english}
}

@article{shaikhTrustModelMeasuring2015,
  title = {Trust Model for Measuring Security Strength of Cloud Computing Service},
  author = {Shaikh, Rizwana and Sasikumar, M.},
  date = {2015},
  journaltitle = {Procedia Computer Science},
  volume = {45},
  pages = {380--389},
  publisher = {Elsevier},
  url = {https://www.sciencedirect.com/science/article/pii/S1877050915004081},
  urldate = {2024-10-30}
}

@online{guptaPLUMEECDSANullifier2022,
  title = {{{PLUME}}: {{An ECDSA Nullifier Scheme}} for {{Unique Pseudonymity}} within {{Zero Knowledge Proofs}}},
  shorttitle = {{{PLUME}}},
  author = {Gupta, Aayush and Gurkan, Kobi},
  date = {2022},
  number = {2022/1255},
  url = {https://eprint.iacr.org/2022/1255},
  urldate = {2024-11-08}
}

@article{benalohBallotCastingAssurance2007,
  title = {Ballot {{Casting Assurance}} via {{Voter-Initiated Poll Station Auditing}}.},
  author = {Benaloh, Josh},
  date = {2007},
  journaltitle = {EVT},
  volume = {7},
  pages = {14--14},
  url = {https://www.usenix.org/event/evt07/tech/full_papers/benaloh/benaloh.pdf},
  urldate = {2024-11-19}
}

@inproceedings{ellul2023good,
  title={When is good enough good enough? On software assurances},
  author={Ellul, Joshua and Pace, Gordon J and Revolidis, Ioannis and Schneider, Gerardo},
  booktitle={ERA Forum},
  volume={23},
  number={3},
  pages={337--360},
  year={2023},
  organization={Springer}
}

@article{benalohSimpleVerifiableElections2006,
  title = {Simple Verifiable Elections.},
  author = {Benaloh, Josh},
  date = {2006},
  journaltitle = {EVT},
  volume = {6},
  pages = {5--5},
  url = {https://www.usenix.org/legacy/event/evt06/tech/full_papers/benaloh/benaloh.pdf?ref=https://githubhelp.com},
  urldate = {2024-12-12},
}

@inproceedings{culnanePeeredBulletinBoard2014,
  title = {A {{Peered Bulletin Board}} for {{Robust Use}} in {{Verifiable Voting Systems}}},
  booktitle = {2014 {{IEEE}} 27th {{Computer Security Foundations Symposium}}},
  author = {Culnane, Chris and Schneider, Steve},
  date = {2014-07},
  pages = {169--183},
  issn = {2377-5459},
  doi = {10.1109/CSF.2014.20},
  url = {https://ieeexplore.ieee.org/abstract/document/6957110},
  urldate = {2025-08-26},
  eventtitle = {2014 {{IEEE}} 27th {{Computer Security Foundations Symposium}}}
}

@inproceedings{kiayiasSecurityPropertiesEVoting2018,
  title = {On the {{Security Properties}} of E-{{Voting Bulletin Boards}}},
  booktitle = {Security and {{Cryptography}} for {{Networks}}},
  author = {Kiayias, Aggelos and Kuldmaa, Annabell and Lipmaa, Helger and Siim, Janno and Zacharias, Thomas},
  editor = {Catalano, Dario and De Prisco, Roberto},
  date = {2018},
  pages = {505--523},
  publisher = {Springer International Publishing},
  location = {Cham},
  doi = {10.1007/978-3-319-98113-0_27},
  isbn = {978-3-319-98113-0},
  langid = {english}
}

@online{bokslagEvaluatingEvotingTheory2016,
  title = {Evaluating E-Voting: Theory and Practice},
  shorttitle = {Evaluating E-Voting},
  author = {Bokslag, Wouter and family=Vries, given=Manon, prefix=de, useprefix=true},
  date = {2016-02-08},
  eprint = {1602.02509},
  eprinttype = {arXiv},
  eprintclass = {cs},
  url = {http://arxiv.org/abs/1602.02509},
  urldate = {2024-09-06},
  langid = {english},
  pubstate = {prepublished}
}

@inproceedings{hainesRunningRaceSwiss2022,
  title = {Running the~{{Race}}: {{A Swiss Voting Story}}},
  shorttitle = {Running the~{{Race}}},
  booktitle = {Electronic {{Voting}}},
  author = {Haines, Thomas and Pereira, Olivier and Teague, Vanessa},
  editor = {Krimmer, Robert and Volkamer, Melanie and Duenas-Cid, David and Rønne, Peter and Germann, Micha},
  date = {2022},
  pages = {53--69},
  publisher = {Springer International Publishing},
  location = {Cham},
  doi = {10.1007/978-3-031-15911-4_4},
  isbn = {978-3-031-15911-4},
  langid = {english}
}

@inproceedings{haldermanNewSouthWales2015,
  title = {The {{New South Wales iVote System}}: {{Security Failures}} and {{Verification Flaws}} in a {{Live Online Election}}},
  shorttitle = {The {{New South Wales iVote System}}},
  booktitle = {Proceedings of the 5th {{International Conference}} on {{E-Voting}} and {{Identity}} - {{Volume}} 9269},
  author = {Halderman, J. Alex and Teague, Vanessa},
  date = {2015-09-02},
  series = {{{VoteID}} 2015},
  pages = {35--53},
  publisher = {Springer-Verlag},
  location = {Berlin, Heidelberg},
  doi = {10.1007/978-3-319-22270-7_3},
  url = {https://doi.org/10.1007/978-3-319-22270-7_3},
  urldate = {2025-08-28},
  isbn = {978-3-319-22269-1}
}

@article{fieller_tests_1957,
	title = {Tests for {Rank} {Correlation} {Coefficients}. {I}},
	volume = {44},
	issn = {0006-3444},
	url = {https://www.jstor.org/stable/2332878},
	doi = {10.2307/2332878},
	number = {3/4},
	urldate = {2025-08-28},
	journal = {Biometrika},
	author = {Fieller, E. C. and Hartley, H. O. and Pearson, E. S.},
	year = {1957},
	note = {Publisher: [Oxford University Press, Biometrika Trust]},
	pages = {470--481},
}

@techreport{Mankins2004,
  author = {Mankins, John C.},
  title  = {Technology Readiness Levels: A White Paper},
  institution  = {Advanced Concepts Office, Office of Space Access and Technology, NASA},
  year   = {2004},
  note   = {Originally issued April 6, 1995. Edited version: 22 December 2004},
  url    = {https://www.artemisinnovation.com/images/TRL_White_Paper_2004-Edited.pdf}
}

@inproceedings{olechowski_technology_2015,
	title = {Technology readiness levels at 40: {A} study of state-of-the-art use, challenges, and opportunities},
	shorttitle = {Technology readiness levels at 40},
	url = {https://ieeexplore.ieee.org/document/7273196/},
	doi = {10.1109/PICMET.2015.7273196},
	urldate = {2025-08-28},
	booktitle = {2015 {Portland} {International} {Conference} on {Management} of {Engineering} and {Technology} ({PICMET})},
	author = {Olechowski, Alison and Eppinger, Steven D. and Joglekar, Nitin},
	month = aug,
	year = {2015},
	note = {ISSN: 2159-5100},
	pages = {2084--2094},
}

@online{landquistOvercomingBottlenecksHomomorphic2025,
  title = {Overcoming {{Bottlenecks}} in {{Homomorphic Encryption}} for the 2024 {{Mexican Federal Election}}},
  author = {Landquist, Eric and Sawhney, Nimit and Sawhney, Simer},
  date = {2025-04-14},
  eprint = {2504.13198},
  eprinttype = {arXiv},
  eprintclass = {cs},
  doi = {10.48550/arXiv.2504.13198},
  url = {http://arxiv.org/abs/2504.13198},
  urldate = {2025-08-27},
  pubstate = {prepublished}
}

@online{OleInsightsRemote2024,
  title = {¡{{Olé}}! {{Insights}} from {{Remote Voting}} in the 2024 {{Mexican Federal Election}}},
  date = {2024},
  urldate = {2025-08-27},
  langid = {english},
  organization = {NASS}
}

@misc{voatzVoatzMobileVoting2019,
  title = {Voatz {{Mobile Voting Platform}}: {{An Overview}}: {{Security}}, {{Identity}}, {{Auditability}}},
  author = {{Voatz}},
  date = {2019},
  url = {https://voatz.com/wp-content/uploads/2020/07/voatz-security-whitepaper.pdf},
  langid = {english},
}

@article{yatskovskaya_integrated_2018,
  author = {Yatskovskaya, Ekaterina and Srai, Jagjit Singh and Kumar, Mukesh},
  doi = {10.3390/su10030896},
  journal = {Sustainability},
  month = mar,
  note = {Number: 3 Publisher: Multidisciplinary Digital Publishing Institute},
  number = {3},
  pages = {896},
  title = {Integrated {Supply} {Network} {Maturity} {Model}: {Water} {Scarcity} {Perspective}},
  url = {https://www.mdpi.com/2071-1050/10/3/896},
  volume = {10},
  year = {2018},
}

@article{kucinska-landwojtowicz_organizational_2023,
  author = {Kuci{\'n}ska-Landw{\'o}jtowicz, Aneta and Czabak-G{\'o}rska, Izabela Dagmara and Domingues, Pedro and Sampaio, Paulo and Ferradaz de Carvalho, Carolina},
  doi = {10.1108/IJQRM-12-2022-0360},
  journal = {International Journal of Quality \& Reliability Management},
  month = jan,
  note = {Publisher: Emerald Publishing Limited},
  number = {1},
  pages = {60--83},
  title = {Organizational maturity models: the leading research fields and opportunities for further studies},
  url = {https://doi.org/10.1108/IJQRM-12-2022-0360},
  volume = {41},
  year = {2023},
}

\newpage
\appendix

\section{Trust Model Sub-Indicator Scores (TM(p))}\label{app:tmp}

\begin{table}[htbp]
\centering
\footnotesize
\caption{Trust Model Scores $TM(p)$ ranked under the inductively defined default weight set ($w_1$).}
\label{tab:TM_scores_ranked}
\begin{tabular}{lrrr}
\toprule
Name Internet Voting System & Rank & Normalized $TM(p)_{w_1}$ & $TM(p)_{w_1}$ \\
\midrule
Snapshot X              & 1  & 1.0000 & 5.5000 \\
Stellot                 & 1  & 1.0000 & 5.5000 \\
Cicada                  & 3  & 0.9777 & 5.4000 \\
MACI                    & 4  & 0.9465 & 5.2600 \\
Open Vote Network       & 5  & 0.8797 & 4.9600 \\
Vocdoni                 & 6  & 0.8107 & 4.6500 \\
CHVote                  & 7  & 0.8051 & 4.6250 \\
Scytl                   & 8  & 0.6982 & 4.1450 \\
Estonian e-voting system & 9  & 0.6648 & 3.9950 \\
Agora                   & 10 & 0.6548 & 3.9500 \\
Votem, Proof of Vote    & 11 & 0.6528 & 3.9411 \\
Belenios                & 12 & 0.4767 & 3.1506 \\
zkSnap                  & 13 & 0.4655 & 3.1000 \\
Helios                  & 14 & 0.4065 & 2.8350 \\
Decidim                 & 15 & 0.3842 & 2.7350 \\
Snapshot                & 16 & 0.1537 & 1.7000 \\
Voatz                   & 17 & 0.0000 & 1.0100 \\
\bottomrule
\end{tabular}
\end{table}

\newpage

\section{Complexity (CMPX) and Practical Usage (PU) Data}\label{app:cmpx_and_pu}

\begin{table}[htbp]
\centering
\footnotesize

\caption{Comparison Complexity (CMPX) and Practical Usage (PU) for the analyzed e-voting systems (sorted by CMPX, then PU).}

\begin{tabular}{|p{2.9cm}|p{6.2cm}|c|p{1.8cm}|c|}
\hline
\textbf{Name p} & \textbf{Components and their CMPX scores} & \textbf{CMPX(p)} & \textbf{PU sources} & \textbf{PU(p)} \\
\hline
Votem, Proof of Vote~\cite{mattbeckerProofVoteEndtoend2018} & Election Authority (1), DKG/Decryption Trustees (4), Authentication Authorities (4), Authorization Authorities (4), Ballot Distribution Servers (4), Mix-Network Nodes (4), PBFT Private Blockchain (5), Public Verifiers (3) & 29 & \cite{VotemCaseStudies} & 1 \\
\hline
Vocdoni~\cite{williamsRemoteVotingAge2022} & Voting Organiser (1), Census Service (1), Scrutinizers (4), Keykeepers (4), Oracles (5), Decentralised Storage (2), Gateways (4), Vochain Private BC (5), Public blockchain (2) & 28 & \cite{VotecasterFarcasterPolls} & 1 \\
\hline
Agora~\cite{AgoraBringingOur} & Election Authority (1), Private Bulletin Board Blockchain (5), Cothority (4), Cotena (4), Bitcoin Blockchain (2), Valeda Auditors (4) & 20 & \cite{teamSwissbasedAgoraRecords2018} & 1 \\
\hline
Scytl~\cite{larraiaSVoteControlComponents2022} & Election Administrators (1), Voting Server (1), Print Office (1), Control Components (CCR and CCM) (4), Board Members (Key-Holders) (4), Auditors (3) & 14 & \cite{SCYTLELECTIONTECHNOLOGIES,ESADECaseScytl2017,scytlHowHoldFully2023,serdultFifteenYearsInternet2015} & 2 \\
\hline
Voatz~\cite{voatzVoatzMobileVoting2019} & Admin (1), Voatz Backend Servers (1), Identity Verification Service (Jumio) (2), Hyperledger Fabric (5), Auditors (3) & 12 & \cite{OleInsightsRemote2024,landquistOvercomingBottlenecksHomomorphic2025} & 2 \\
\hline
Estonian e-voting system~\cite{GeneralFrameworkElectronic} & Organiser (1), Collector (1), Processor (1), Registration Service (1), Talliers (4), Auditors (3) & 11 & \cite{ElectionsEstonia, EDemocracyOpenData2024} & 3 \\
\hline
Belenios~\cite{cortierBeleniosSimplePrivate2019} & Server administrator (1), Credential authority (1), Voting Server (1), Trustees (4), Auditors (3) & 10 & \cite{cortierFeaturesUsageBelenios2022} & 1 \\
\hline
CHVote~\cite{haenniCHVoteProtocolSpecification2017} & Administrator (1), Printing Authority (1), Election Authorities (4), Auditors (3) & 9 & \cite{serdultFifteenYearsInternet2015, fchEVoting, haenniCHVoteSixteenBest2020} & 3 \\
\hline
Helios~\cite{adidaHeliosWebbasedOpenAudit2008} & Administrator (1), Helios Server (1), Key-holders (4), Auditors (3) & 9 & \cite{adidaElectingUniversityPresident2009} & 1 \\
\hline
Decidim~\cite{barandiaranDecidimTechnopoliticalNetwork2024} & Administrator (1), Decidim Server (1), Key-Holders (4), Monitoring Committee (3) & 9 & \cite{DecidimUse} & 1 \\
\hline
Snapshot~\cite{WelcomeSnapshotDocs2024} & Admin (1), Snapshot-hub (1), Shutter network (2), IPFS (2) & 6 & \cite{IntroducingSnapshotsOnchain} & 2 \\
\hline
Snapshot X~\cite{Snapshot2024} & Admin (1), Mana (1), Shutter network (2), Starknet Blockchain (2) & 6 & \cite{Snapshot2024} & 1 \\
\hline
zkSnap~\cite{ghangasZkSnapScalableZero} & Organiser (1), Trusted Coordinator (1), DRAND (2), Public Blockchain (2) & 6 & -- & 0 \\
\hline
MACI~\cite{ethereumfoundationMinimalAntiCollusionInfrastructure2022} & Organiser (1), Trusted Coordinator (1), Public Blockchain (2) & 4 & \cite{CaseStudiesMACI} & 2 \\
\hline
Stellot~\cite{baranskiPracticalIVotingStellar2020} & Organiser (1), Token Distribution Server (1), Public Blockchain (2) & 4 & -- & 0 \\
\hline
Cicada~\cite{glaeserCicadaFrameworkPrivate2023} & Administrator (1), Off-chain solver (1), Public BC (2) & 4 & -- & 0 \\
\hline
Open Vote Network~\cite{mccorryOpenVoteNetwork2023} & Organiser (1), Public Blockchain (2) & 3 & -- & 0 \\
\hline
\end{tabular}
\end{table}
\FloatBarrier

\newpage
\section{Trust Models}\label{app:trust_models}

\newcolumntype{L}{>{\raggedright\arraybackslash}p{.125\textwidth}}
\newcolumntype{M}{>{\raggedright\arraybackslash}p{.25\textwidth}} 
\newcolumntype{N}{>{\raggedright\arraybackslash}p{0.63\textwidth}} 

\subsection{Estonian e-voting system (IVXV)}\label{sec_A.1}

\begin{table}[htbp]
\footnotesize
\centering
\caption{Trust model evaluation for Estonian e-voting system (IVXV)~\cite{GeneralFrameworkElectronic}}
\begin{tabular}{|L|M|N|}
\hline
\textbf{Property} & \textbf{Trust Model} & \textbf{Justification} \\
\hline
Voting secrecy & Collector AND All of Talliers (1/n + 1/1) & Attacker receives ballots from Collector, and then decrypts them with the help of Talliers. \\\hline
Voter anonymity & Collector AND Processor (few/1) & Collector has the encrypted and anonymized votes. Processor has the capacity to re-establish links between votes and voter id. Since the Collector plays also the role of Authentication service, it can connect the voter id with the real identity thus breaking the voter anonymity property. "The Voter Application and the Collection Service serve as mediators between the Voter and the Identification Service in a suitable manner. As a result of the process, the Collection Service learns the identity of the Voter." (\cite{GeneralFrameworkElectronic}, page 11). \\\hline
Individual Verifiability & Processor AND Auditor (1/N + 1/1) & If the Collector colludes with the Registration Service, they can respond positively to voter requests, but then filter out the vote when it is passed to the Processor. Or the Processor may discard the votes passed by the Collector, so the Auditor must also collude. \\\hline
Universal Verifiability & Processor AND Auditor (1/N + 1/1) & The processor can filter out any votes if not watched by auditor. Organisers can provide arbitrary annulment list if not verified by the auditor. \\\hline
Eligibility verifiability & (Collector AND Auditor) OR Organiser (1/1) & The Organiser can include a list with an extended list of eligible voters. Or the collector may decide to include ballots from voters that are not included in the Eligibility list provided by the Organiser. This has to be checked by the Auditor. \\\hline
Coercion Resistance & Auditor AND Processor (N/A) & The protocol achieves coercion resistance using a hybrid model, meaning that there is parallel physical voting which overrides online voting. While the online protocol itself is not receipt-free, the overall election process allows a voter to cast a ballot in-person at a physical polling station, which overrides any previously cast i-vote. This is a robust, officially sanctioned method for a coerced voter to cast their true vote. \\
\hline
\end{tabular}
\label{tab:IVXV}
\end{table}
\FloatBarrier
\newpage

\subsection{Scytl}\label{sec_A.2}

\begin{table}[htbp]
\footnotesize
\centering
\caption{Trust model evaluation for Scytl~\cite{larraiaSVoteControlComponents2022}}
\begin{tabular}{|L|M|N|}
\hline
\textbf{Property} & \textbf{Trust Model} & \textbf{Justification} \\
\hline
Voting secrecy & Voting Server AND Print Office AND All of Mixing and Decryption Control Components [CCM] (1/n + few/1) & Voting Server shares the encrypted votes with the attacker. The Print Office colludes with CCMs to recover the decryption key and shares them with the attacker. Having the decryption key, the attacker can decrypt the votes. \\
\hline
Voter anonymity & (Voting Server OR Control Components) AND Print Office (few/1) & Print Office is responsible for creating credential material for the voters. The credentials are attached to the encrypted ballot sent by the voter (\cite{larraiaSVoteControlComponents2022}, Figure 18). So if the Voting Server or Control Components which process ballots can correspond them with credentials, then, by colluding with Print Office they can deduce the voter identity. \\
\hline
Individual Verifiability & Voting Server AND Print Office AND All Control Components (CCR) AND Auditors (1/N + 1/n + few/1) & After successful CRC and VCC verification by Control Components, the Voting Server could discard the vote and record an altered one. The Print Office can ensure that the false codes provided to the voter appear legitimate. Auditors can provide false verification results, making it appear that all votes have been correctly recorded and included in the final tally. \\
\hline
Universal Verifiability & Voting Server AND Print Office AND Control Components (CCR) AND Auditors (1/N + 1/n + few/1) & Individual voters are convinced about their votes being recorded correctly as described in Individual Verifiability. Control Components generate false CRCs that are liked to manipulated votes generated by the adversary. The Voting Server stores altered votes. Then the Control Components manipulate the decryption and mixing process to ensure the altered votes are counted. Auditors collude and validated manipulated process, making it appear that no one voted twice and the tally is accurate. The final tally is announced based on the manipulated vote count, ensuring it reflects the false data. \\
\hline
Eligibility verifiability & Print Office (1/1) & The Print Office could generate additional fake credentials that Voting Server and Control Components would accept. \\
\hline
Coercion Resistance & None (N/A) & The protocol's "vote encryption challenge," a core feature of its individual verifiability, is fundamentally incompatible with coercion resistance. It explicitly allows the voter to access the secret random parameters used for encryption, which serve as a perfect receipt to prove the content of their cast ballot to a coercer. \\
\hline
\end{tabular}
\label{tab:scytl}
\end{table}
\FloatBarrier
\newpage

\subsection{CHVote}\label{sec_A.3}

\begin{table}[htbp]
\footnotesize
\centering
\caption{Trust model evaluation for CHVote~\cite{haenniCHVoteProtocolSpecification2017}}
\begin{tabular}{|L|M|N|}
\hline
\textbf{Property} & \textbf{Trust Model} & \textbf{Justification} \\
\hline
Voting secrecy & Administrator AND All of Election Authority (1/n + 1/1) & Administrator together with Election Authorities can reconstruct the decryption key. Having the decryption key, the attacker can decrypt the votes. \\
\hline
Voter anonymity & Printing Authority AND All Election Authority (1/n + 1/1) & The printing authority knows the private voting and confirmation credentials (X and Y) of each voter. If the printing authority shared this information with the election authorities, they could link these credentials to the voter's identity, compromising the voter's anonymity. With access to private credentials, election authorities could break the shuffle process and link submitted ballots and confirmations back to specific voters, undermining the protocol's core principle of anonymity. \\
\hline
Individual Verifiability & Administrator AND Print Authority AND All of A Few Election Authority AND Auditors (1/N + 1/n + few/1) & Administrator and Printing Authority collude to print manipulated verification codes on ballot cards. Election Authorities prepare to intercept and alter votes during the counting process. Administrator ensures that the voting system returns manipulated verification codes that match those on the ballot card. Election Authorities intercept encrypted votes. Administrator and Election Authorities update voting records to include fraudulent votes. Auditors collude to create false audit reports that falsely certify the integrity of the election. \\
\hline
Universal Verifiability & Administrator AND Election Authority AND Auditors (1/N + 1/n + few/1) & Administrator together with Election Authorities would have to produce wrong results. Auditors, in collusion, generate false verification reports that falsely affirm the integrity of the election. \\
\hline
Eligibility verifiability & Printing authority (1/1) & "the printing authority receives very sensitive information from the election authorities, for example the credentials for submitting a vote or the verification codes for the candidates. In principle, knowing this information allows the submission of votes on behalf of eligible voters. Exploiting this knowledge would be noticed by the voters when trying to submit a ballot, but obviously not by voters abstaining from voting. Even worse, if check is given access to the verification codes, it can easily bypass the cast-as-intended verification mechanism, i.e., voters can no longer detect vote manipulations on the voting client" \\
\hline
Coercion Resistance & None (N/A) & The protocol provides no mechanism for coercion resistance. The official protocol specification explicitly states that the design goal is only to ensure the risk is "not significantly higher compared to voting by postal mail," effectively defining strong receipt-freeness as out-of-scope. \\
\hline
\end{tabular}
\label{tab:CHVote}
\end{table}
\FloatBarrier
\newpage

\subsection{Voatz}\label{sec_A.4}

\begin{table}[htbp]
\footnotesize
\centering
\caption{Trust model evaluation for Voatz~\cite{voatzVoatzMobileVoting2019}}
\begin{tabular}{|L|M|N|}
\hline
\textbf{Property} & \textbf{Trust Model} & \textbf{Justification} \\
\hline
Voting secrecy & Voatz Backend Server (1/1) & The central backend server necessarily decrypts or has access to the decrypted ballot content before it's processed. \\
\hline
Voter anonymity & Jumio OR Backend server generates Anonymous ID (1/1) & Voatz claims ballots are doubly anonymized and use an anonymous ID. However, the security analysis found that a Voatz administrator with backend access could observe API requests when ballots are submitted in real time and deanonymize votes. The IDs from registration and the ballot submission are linked on the central server. \\
\hline
Individual Verifiability & Voatz Server (1/1) & A voter receives a PDF receipt of their choices via email. They can verify that this receipt matches their intent. However, this is not a cryptographic proof they can use to independently verify their ballot was included correctly in the final tally. They must trust that the server sent them an honest receipt and that this receipt corresponds to a ballot that was correctly recorded and later printed for tabulation. \\
\hline
Universal Verifiability & Auditors AND (Private Blockchain OR Backend) (1/N + 2/n) & The blockchain provides an immutable audit log, but it is a permissioned chain. The public cannot independently verify the cryptographic proofs of a correct tally. Verifiability is delegated to a set of designated auditors who are granted access by the system operator (Voatz/election officials). \\
\hline
Eligibility verifiability & Organiser OR Jumio OR Voatz Server (1/1) & The election authority may provide an incorrect list of eligible voters. Jumio may use different list, or incorrectly verify non eligible users. \\
\hline
Coercion Resistance & None (N/A) & The system provides a PDF receipt of the voter's choices. This receipt can be shown to a coercer as proof of how they voted. This fundamentally violates the principle of receipt-freeness, which is essential for strong coercion resistance. While some pilots allowed for spoiling and re-voting, this only provides a weak mitigation and does not solve the receipt problem. \\
\hline
\end{tabular}
\label{tab:Voatz}
\end{table}
\FloatBarrier

\subsection{Belenios}\label{sec_A.5}

\begin{table}[htbp]
\footnotesize
\centering
\caption{Trust model evaluation for Belenios~\cite{cortierBeleniosSimplePrivate2019}}
\begin{tabular}{|L|M|N|}
\hline
\textbf{Property} & \textbf{Trust Model} & \textbf{Justification} \\
\hline
Voting secrecy & All Keyholders (1/n) & Keyholders can decrypt votes that are publicly available to everyone on the Message Board. \\
\hline
Voter anonymity & Voting Server AND Registrar (few/1) & The registrar is responsible for generating and sending the signing keys to voters. The voting server receives and stores the corresponding verification keys. If both the registrar and the voting server collude, they can link the verification key (used in the recorded ballot) back to the voter’s real identity. This is because the registrar knows how voters are mapped to their verification keys, and the voting server can match the verification keys to the ballots cast. \\
\hline
Individual Verifiability & Belenios Server AND Auditors (1/N + 1/1) & The voting server would have to manipulate the public bulletin board to display a fake entry for the voter's ballot while excluding it from the actual tally. Therefore the actual tally must be audited and controlled by the auditors who ensure the common knowledge of the actual tally. \\
\hline
Universal Verifiability & Belenios AND Auditors (1/N + 1/1) & The voting server would have to manipulate the public bulletin board to display a fake entry for the voter's ballot while excluding it from the actual tally. Therefore the actual tally must be audited and controlled by the auditors who ensure the common knowledge of the actual tally. \\
\hline
Eligibility verifiability & Voting Server AND Registrar (few/1) & Registrator guarantees that only ballots from eligible voters are recorded on message board. If the registrar issues credentials to ineligible entities and the voting server accepts ballots from these entities, observers can be misled into believing that only eligible voters participated. \\
\hline
Coercion Resistance & Voter AND Belenios Server (N/A) & As explicitly stated in its documentation, the standard Belenios protocol is not coercion-resistant. A voter can prove their vote by revealing the encryption randomness. We evaluate the base protocol; a separate variant, BeleniosRF~\cite{chaidosBeleniosRFNoninteractiveReceiptFree2016}, uses re-randomizable signed encryption where the randomness is shared between voter and collector so it can not be revealed/proved by voter without collusion with the Belenios Server. \\
\hline
\end{tabular}
\label{tab:Belenios}
\end{table}
\FloatBarrier

\subsection{Helios}\label{sec_A.6}
\begin{table}[htbp]
\footnotesize
\centering
\caption{Trust model evaluation for Helios~\cite{adidaHeliosWebbasedOpenAudit2008}}
\begin{tabular}{|L|M|N|}
\hline
\textbf{Property} & \textbf{Trust Model} & \textbf{Justification} \\
\hline
Voting secrecy & All Keyholders (1/n) & Keyholders can decrypt votes that are publicly available to everyone on the Message Board. \\
\hline
Voter anonymity & Organiser OR Helios Server (1/1) & The message board is public and everyone knows who has posted which ballot. The real identity of the voter is known to the Organiser who created the census and the Helios server who sends the emails. Therefore, only the organiser and the Helios server can link the real identity of the voter to the ballot cast. \\
\hline
Individual Verifiability & Helios Server AND Auditors (1/N + 1/1) & The voting server would have to manipulate the public bulletin board to display a fake entry for the voter's ballot while excluding it from the actual tally. Therefore the actual tally must be audited and controlled by the auditors who ensure the common knowledge of the actual tally. \\
\hline
Universal Verifiability & Helios Server AND Auditors (1/N + 1/1) & The voting server would have to manipulate the public bulletin board to display a fake entry for the voter's ballot while excluding it from the actual tally. Therefore the actual tally must be audited and controlled by the auditors who ensure the common knowledge of the actual tally. \\
\hline
Eligibility verifiability & Helios Server (1/1) & Helios Serve can include additional votes at will unless Organiser audit the message board against the list of eligible voters. \\
\hline
Coercion Resistance & None (N/A) & The system allows easy coercion by revealing the encrypted ballot along with randomness. \\
\hline
\end{tabular}
\label{tab:Helios}
\end{table}
\FloatBarrier

\subsection{Decidim}\label{sec_A.7}

\begin{table}[htbp]
\footnotesize
\centering
\caption{Trust model evaluation for Decidim~\cite{barandiaranDecidimTechnopoliticalNetwork2024}}
\begin{tabular}{|L|M|N|}
\hline
\textbf{Property} & \textbf{Trust Model} & \textbf{Justification} \\
\hline
Voting secrecy & All Keyholders (1/n) & Keyholders can decrypt votes that are publicly available to everyone on the Message Board. \\
\hline
Voter anonymity & Organiser OR Decidim Server (1/1) & The Decidim supports both Decidim Authentication and Census, where the access codes are sent to the voters by the Organiser or his delegate (e.g. by post). Therefore, the link between the access code and the real identity can only be reconstructed by either the Decidim server (Decidim Authentication) or the Organiser (External Census). \\
\hline
Individual Verifiability & Decidim Server AND Monitoring Comittee (1/N + 1/1) & Decidim can store each vote sent by the voter in a separate ballot box; each time the voter requests a ballot audit, the Decidim response is returned with the original. However, during the counting process, the vote can be skipped.  Therefore, the actual count must be audited and controlled by the auditors, who ensure the common knowledge of the actual tally. \\
\hline
Universal Verifiability & Decidim AND Monitoring Comittee (1/N + 1/1) & Decidim can store every vote sent by the voter in separate ballot box, each time the voter request ballot audit, the Decidim response with the original one. However, during the tally, the vote can be skipped.  Therefore the actual tally must be audited and controlled by the auditors who ensure the common knowledge of the actual tally. \\
\hline
Eligibility verifiability & Decidim Server (1/1) & Decidim Server can skip or include additional votes at will unless the Organiser or Auditors validate the message board against the list of eligible voters. \\
\hline
Coercion Resistance & None (N/A) & The system achieves weak coercion resistance by allowing votes to vote many times. This is a weak form of coercion resistance, as a coercer could observe the voter until polls close to ensure the final vote is not changed. However, it provides more protection than a single, irrevocable vote. \url{https://docs.decidim.org/bulletin-board/admin/virtual-voting-booth.html#_vote_again} \\
\hline
\end{tabular}
\label{tab:Decidim}
\end{table}
\FloatBarrier
\newpage

\subsection{Votem, Proof of Vote}\label{sec_A.8}

\begin{table}[htbp]
\footnotesize
\centering
\caption{Trust model evaluation for Votem, Proof-of-Vote~\cite{mattbeckerProofVoteEndtoend2018}}
\begin{tabular}{|L|M|N|}
\hline
\textbf{Property} & \textbf{Trust Model} & \textbf{Justification} \\
\hline
Voting secrecy & Threshold-of-Taillers (few/n) & Votem uses ElGamal threshold encryption, so a threshold of talliers must collude to reconstruct the decryption key. An attacker would have to obtain all the encrypted votes recorded on the blockchain, which can be obtained from any node on the blockchain. \\
\hline
Voter anonymity & All of A Few Mixers (1/n) & "By utilizing several mixers, we guarantee vote anonymity as long as at least one mixer is honest." (~\cite{mattbeckerProofVoteEndtoend2018}, Page 33) \\
\hline
Individual Verifiability & Majority of Private Bulletin Board Blockchain + Auditors (1/N + 2/n) & Majority of the private blockchain nodes incorectly confirms that the ballot was recorded, then all the auditors confirm that the processes were executed correctly. \\
\hline
Universal Verifiability & Majority of Private Bulletin Board Blockchain + Auditors (1/N + 2/n) & As long as the majority of private blockchain nodes are honest and follow the protocol the voter can be sure that his vote is included.  \\
\hline
Eligibility verifiability & Authorizers (1/n) & "How does the protocol ensure that only eligible voters cast votes? Proof of Vote guarantees this as described by the Authentication Process." (~\cite{mattbeckerProofVoteEndtoend2018}, page 35) "Authorizers are entities that can match a voter's identifying information to a Ballot Style. In many cases there is only a single authorizer (the Election Authority itself, in possession of the voters list). However, Proof of Vote implements a multi-signature scheme for authorizers to allow for distributed and redundant authorization in cases where Election Authorities want to distribute trust and responsibility for the voters list management. (~\cite{mattbeckerProofVoteEndtoend2018}, page 10) \\
\hline
Coercion Resistance & None (N/A) & A voter can demonstrate their vote to a coercer by revealing the inputs to the encryption function. The coercer can then verify on the public blockchain that the corresponding encrypted ballot was cast by the voter, confirming the successful coercion. \\
\hline
\end{tabular}
\label{tab:Votem}
\end{table}
\FloatBarrier
\newpage

\subsection{Agora}\label{sec_A.9}

\begin{table}[htbp]
\footnotesize
\centering
\caption{Trust model evaluation for Agora~\cite{AgoraBringingOur}}
\begin{tabular}{|L|M|N|}
\hline
\textbf{Property} & \textbf{Trust Model} & \textbf{Justification} \\
\hline
Voting secrecy & Majority of Keyholders/Cothority (Talliers) (few/n) & The attacker needs to get the list of all the encrypted votes recorded on the blockchain, which they can get from any node on the blockchain. Each node in the Cothority has a piece of the decryption key, and only together can they decrypt the encrypted ballots to count the votes. \\
\hline
Voter anonymity & Election Authority AND All of a Few Mix-Nodes (1/n + 1/1) & All mixing nodes would have to collude to deanonymise the ballots. The Election Authority would have access to voter identities and could potentially link these identities to the issued ballots. \\
\hline
Individual Verifiability & Majority of Private Bulletin Board Blockchain + Valenda auditors (1/N + 2/n) & Majority of the private blockchain nodes confirm that the ballot was recorded, then all the auditors confirm that the correct hash of a ballot-box was commited to the Bitcoin network. \\
\hline
Universal Verifiability & Majority of Private Bulletin Board Blockchain + Valenda auditors (1/N + 2/n) & Majority of the private blockchain nodes compile a modified ballot-box, which is then moved forward to the tally phase, if all of the auditors collude and confirm to the modified ballot-box then the incident passes unnoticed. \\
\hline
Eligibility verifiability & Election Authority (1/1) & Election Authority controls the list of eligable voters. \\
\hline
Coercion Resistance & None (N/A) & Voters can vote only once and they can prove how they voted. \\
\hline
\end{tabular}
\label{tab:Agora}
\end{table}
\FloatBarrier

\subsection{Snapshot}\label{sec_A.10}

\begin{table}[htbp]
\footnotesize
\centering
\caption{Trust model evaluation for Snapshot~\cite{WelcomeSnapshotDocs2024}}
\begin{tabular}{|L|M|N|}
\hline
\textbf{Property} & \textbf{Trust Model} & \textbf{Justification} \\
\hline
Voting secrecy & Keypers (1/n) & Snapshot uses the Shutter network to delegate threshold decryption so only a majority of all Keypers have to collude to reconstruct the decryption key. \\
\hline
Voter anonymity & None (N/A) & Votes are sent directly via the voters' public addresses. \\
\hline
Individual Verifiability & Snapshot-hub (1/1) & The Snapshot-hub may skip the voter's ballot at the tally stage, no one could prove the accident as the blockchain hasn't seen the transaction on-chain. \\
\hline
Universal Verifiability & Snapshot-hub (1/1) & The Snapshot-hub may skip the voter's ballot at the tally stage, no one could prove the accident as the blockchain hasn't seen the transaction on-chain. \\
\hline
Eligibility verifiability & Crypto (0/N) & Everyone who can pass the specified predicate like “I’ve been part of the DAO for 1k blocks.” is elligable to vote. \\
\hline
Coercion Resistance & None (N/A) & Votes are typically public or pseudonymous, but there is no mechanism to prevent a voter from proving their vote. \\
\hline
\end{tabular}
\label{tab:Snapshot}
\end{table}
\FloatBarrier
\newpage

\subsection{Snapshot X}\label{sec_A.11}

\begin{table}[htbp]
\footnotesize
\centering
\caption{Trust model evaluation for Snapshot X~\cite{Snapshot2024}}
\begin{tabular}{|L|M|N|}
\hline
\textbf{Property} & \textbf{Trust Model} & \textbf{Justification} \\
\hline
Voting secrecy & Keypers (1/n) & Snapshot uses the Shutter network to delegate threshold decryption so only a majority of all Keypers have to collude to reconstruct the decryption key. \\
\hline
Voter anonymity & None (N/A) & Votes are sent directly via the voters' public addresses. \\
\hline
Individual Verifiability & Public blockchain (2/N) & As long as all nodes correctly execute the smart contract and accept voters' ballots. Voters can verify that thier votes are included in the final tally. \\
\hline
Universal Verifiability & Public blockchain (2/N) & As long as all nodes correctly execute the smart contract and accept voters' ballots. Voters can verify that thier votes are included in the final tally. \\
\hline
Eligibility verifiability & Crypto (0/N) & Everyone who can pass the specified predicate like “I’ve been part of the DAO for 1k blocks.” is elligable to vote. \\
\hline
Coercion Resistance & None (N/A) & As an on-chain version of Snapshot, it provides greater verifiability and censorship resistance but does not add coercion resistance mechanisms. \\
\hline
\end{tabular}
\label{tab:SnapshotX}
\end{table}
\FloatBarrier

\subsection{Vocdoni}\label{sec_A.12}

\begin{table}[htbp]
\footnotesize
\centering
\caption{Trust model evaluation for Vocdoni~\cite{williamsRemoteVotingAge2022}}
\begin{tabular}{|L|M|N|}
\hline
\textbf{Property} & \textbf{Trust Model} & \textbf{Justification} \\
\hline
Voting secrecy & Threshold-Of-Keykeepers (few/n) & There are some validators on the blockchain who are also keykeepers. When a new encrypted (secret until the end) election is created, each keykeeper creates an encryption key. They publish the public part via a transaction (setProcessKeysTx). The voters select up to N (at least one) of the keys and encrypt the vote with the N keys, in onion mode. Only if all keykeeper validators collide, they can decrypt the votes. \\
\hline
Voter anonymity & Crypto (0/N) & Anonymity is protected by the use of nullifiers and proof of inclusion. \\
\hline
Individual Verifiability & Private Blockchain Vecdoni (2/n) & Private Blockchain can store every vote sent by the voter in separate ballot box, each time the voter request ballot audit, the Blockchain nodes response with the original one. However, during the tally, the vote can be skipped.  \\
\hline
Universal Verifiability & Private Blockchain Vecdoni (2/n) & The majority of the private blockchain nodes collude and execute malicious version of the protocol, dropping or modifying votes. The voters would not notice as they are not auditing the whole process and the results are not validated by the external auditors or public blockchains. \\
\hline
Eligibility verifiability & Crypto (0/N) & "A Merkle Tree is generated with a snapshot of the census and published to IPFS. The census can be generated in one of many ways, including: i) The organization creates a CSV spreadsheet containing voter information (i.e. name, ID number, etc), ii) An Ethereum Storage Proof, iii) Each user creates a key pair (self-sovereign identity) and signs up to a Registry DB." All of the methods except "Ethereum Storage Proof" are based on the trust to the organiser who controls the list of eligable voters. \\
\hline
Coercion Resistance & None (N/A) & The protocol allows for re-voting, which is considered a weak mechanism as the final vote remains provable to a coercer who can monitor the voter's activity until the election concludes. \\
\hline
\end{tabular}
\label{tab:Vocdoni}
\end{table}
\FloatBarrier
\newpage

\subsection{zkSnap}\label{sec_A.13}

\begin{table}[htbp]
\footnotesize
\centering
\caption{Trust model evaluation for zkSnap~\cite{ghangasZkSnapScalableZero}}
\begin{tabular}{|L|M|N|}
\hline
\textbf{Property} & \textbf{Trust Model} & \textbf{Justification} \\
\hline
Voting secrecy & Trusted Coordinator OR (Trusted Coordinator AND DRAND) (1/1) & The votes submitted are encrypted, but the trusted coordinator knows the decryption keys so can decrypt the ballots at any time. Or the DRAND consortium (https://drand.love/) colludes and solves the puzzle and reconstructs the decryption key before the deadline. \\
\hline
Voter anonymity & Crypto (0/N) & Anonymity is protected by the use of Pseudonymously Linked Unique Message Entity (PLUME)~\cite{guptaPLUMEECDSANullifier2022}. \\
\hline
Individual Verifiability & Trusted Coordinator (1/1) & The Trusted Coordinator may skip the voter's ballot at the tally stage, no one could prove the accident as the blockchain hasn't seen the transaction on-chain. \\
\hline
Universal Verifiability & Trusted Coordinator (1/1) & The Trusted Coordinator may skip voters' ballots at the tally stage, no one could prove the accident as the blockchain hasn't seen the transactions on-chain. \\
\hline
Eligibility verifiability & Crypto (0/N) & Instead of using a snapshot of eligible voters, each voter generates storage proofs that attest to a predicate like “I’ve been part of the DAO for 1k blocks.” using storage proofs, \url{https://github.com/aragon/evm-storage-proofs}. \\
\hline
Coercion Resistance & None (N/A) & Voters can vote only once and they can prove how they voted. \\
\hline
\end{tabular}
\label{tab:zkSnap}
\end{table}
\FloatBarrier

\subsection{Stellot}\label{sec_A.14}

\begin{table}[htbp]
\footnotesize
\centering
\caption{Trust model evaluation for Stellot~\cite{baranskiPracticalIVotingStellar2020}}
\begin{tabular}{|L|M|N|}
\hline
\textbf{Property} & \textbf{Trust Model} & \textbf{Justification} \\
\hline
Voting secrecy & Token Distribution Server (1/1) & The published votes are encrypted. However, the Token Distribution Server has the keys to decrypt them. \\
\hline
Voter anonymity & Crypto (0/N) & Anonymity is protected by the use of blind signatures. \\
\hline
Individual Verifiability & Public blockchain (2/N) & As long as the majority of public blockchain nodes are honest and follow the protocol the voter can be sure that his vote is included. \\
\hline
Universal Verifiability & Public blockchain (2/N) & As long as the majority of public blockchain nodes are honest and follow the protocol the voter can be sure that his vote is included. \\
\hline
Eligibility verifiability & IdP OR Token Distribution Server (1/1) & IdP guarantee the voter's correctness and uniqueness, if malicious can allow one person to issue many certificates. TDS is the one who authorise the voter by blindly signing the access-tokens, if malicious can sing any token, without correct authorisation.  \\
\hline
Coercion Resistance & None (N/A) & Voters can vote only once and they can prove how they voted. \\
\hline
\end{tabular}
\label{tab:Stellot}
\end{table}
\FloatBarrier
\newpage

\subsection{MACI}\label{sec_A.15}

\begin{table}[htbp]
\footnotesize
\centering
\caption{Trust model evaluation for MACI~\cite{ethereumfoundationMinimalAntiCollusionInfrastructure2022}}
\begin{tabular}{|L|M|N|}
\hline
\textbf{Property} & \textbf{Trust Model} & \textbf{Justification} \\
\hline
Voting secrecy & Trusted Coordinator (1/1) & "No one except a trusted coordinator should be able to decrypt a vote." \\
\hline
Voter anonymity & Organiser (1/1) & The organiser knows the addresses of the voters that will be used to cast their votes. \\
\hline
Individual Verifiability & Public blockchain (2/N) & "No one — not even the trusted coordinator, should be able to censor a vote." ~ \url{https://maci.pse.dev/docs/introduction#features} \\
\hline
Universal Verifiability & Public blockchain (2/N) & As long as the majority of public blockchain nodes are honest and follow the protocol the voter can be sure that his vote is included. \\
\hline
Eligibility verifiability & Organiser (1/1) & The system is based on the publicly known list of eligible voters. However in general, there must be some announced list of eligable voters, and the list is controlled externally. \\
\hline
Coercion Resistance & Voter AND Trusted Coordinator (1/1) & MACI is specifically designed for coercion resistance using a protocol-native cryptographic mechanism (key-switching). A voter cannot prove how they voted to an external party, but this guarantee relies on the coordinator not colluding with the voter. \\
\hline
\end{tabular}
\label{tab:MACI}
\end{table}
\FloatBarrier

\subsection{Cicada}\label{sec_A.16}

\begin{table}[htbp]
\footnotesize
\centering
\caption{Trust model evaluation for Cicada~\cite{glaeserCicadaFrameworkPrivate2023}}
\begin{tabular}{|L|M|N|}
\hline
\textbf{Property} & \textbf{Trust Model} & \textbf{Justification} \\
\hline
Voting secrecy & N/A & "The basic Cicada framework does not guarantee long-term ballot privacy. Submissions are public after the Open stage. This is because users publish their HTLPs on-chain: once public, the votes contained in the HTLPs are only guaranteed to be hidden for the time it takes to compute T sequential steps, after which point it is plausible that someone has computed the solution." \\
\hline
Voter anonymity & Crypto (0/N) & Anonymity is protected by the use of nullifiers and proof of inclusion. \\
\hline
Individual Verifiability & Public blockchain (2/N) & As long as the majority of public blockchain nodes are honest and follow the protocol the voter can be sure that his vote is included. \\
\hline
Universal Verifiability & Public blockchain (2/N) & As long as the majority of public blockchain nodes are honest and follow the protocol the voter can be sure that his vote is included. \\
\hline
Eligibility verifiability & Organiser (1/1) & The system is based on the publicly known list of eligible voters. However in general, there must be some announced list of eligable voters, and the list is controlled externally. \\
\hline
Coercion Resistance & None (N/A) & "However, we consider receipt-freeness as a non-goal in our protocol design." (~\cite{glaeserCicadaFrameworkPrivate2023}, sec 1.1) \\
\hline
\end{tabular}
\label{tab:Cicada}
\end{table}
\FloatBarrier
\newpage

\subsection{OpenVoteNetwork (OVN)}\label{sec_A.17}

\begin{table}[htbp]
\footnotesize
\centering
\caption{Trust model evaluation for OpenVoteNetwork ~\cite{mccorryOpenVoteNetwork2023}}
\begin{tabular}{|L|M|N|}
\hline
\textbf{Property} & \textbf{Trust Model} & \textbf{Justification} \\
\hline
Voting secrecy & Crypto (0/N) & The voting can be decrypted only once all the voters cast their votes. Also "full collusion of the remaining voters is required to reveal an individual vote" \\
\hline
Voter anonymity & Organiser (1/1) & The organiser knows the addresses of the voters that will be used to cast their votes. \\
\hline
Individual Verifiability & Public blockchain (2/N) & As long as all nodes correctly execute the smart contract and accept voters' ballots. Voters can verify that thier votes are included in the final tally. \\
\hline
Universal Verifiability & Public blockchain (2/N) & As long as all nodes correctly execute the smart contract and accept voters' ballots. Voters can verify that thier votes are included in the final tally. \\
\hline
Eligibility verifiability & Organiser (1/1) & The system is based on the publicly known list of eligible voters. However in general, there must be some announced list of eligable voters, and the list is controlled externally. \\
\hline
Coercion Resistance & None (N/A) & Everyone can prove to anyone how he voted by providing randomness to decrypt the vote. The identities are publicly known so the proof is easily verified. \\
\hline
\end{tabular}
\label{tab:OVN}
\end{table}
\FloatBarrier

\end{document}